\documentclass[prx,aps,twocolumn,superscriptaddress,longbibliography]{revtex4-1} 

\usepackage{graphicx}
\usepackage{bbold}
\usepackage{bbm}
\usepackage[utf8]{inputenc}
\usepackage[T1]{fontenc}

\usepackage[utf8]{inputenc}
\usepackage[T1]{fontenc}
\usepackage[usenames,dvipsnames,svgnames,table]{xcolor}
\usepackage{amstext}
\usepackage{amssymb}
\usepackage{graphicx}
\usepackage{amsmath}
\usepackage{graphicx}
\usepackage{txfonts}
\usepackage{bbm}
\usepackage{latexsym}
\usepackage{dcolumn}

\usepackage[usenames,dvipsnames,svgnames,table]{xcolor}
\definecolor{myurlcolor}{rgb}{0,0,0.7}
\definecolor{myrefcolor}{rgb}{0.8,0,0}
\usepackage[unicode=true,pdfusetitle, bookmarks=true,bookmarksnumbered=false,bookmarksopen=false, breaklinks=false,pdfborder={0 0 0},backref=false,colorlinks=true, linkcolor=myrefcolor,citecolor=myurlcolor,urlcolor=myurlcolor]{hyperref}

\newcommand{\eref}[1]{\eqref{#1}}
\newcommand{\eqnref}[1]{Equation~(\ref{#1})}
\newcommand{\eqnsref}[2]{Equations~\eqref{#1} and \eqref{#2}}
\newcommand{\figref}[1]{Figure~\ref{#1}}
\newcommand{\figsref}[1]{Figures~\ref{#1}}

\newcommand{\secref}[1]{Section~\ref{#1}}
\newcommand{\appref}[1]{\ref{#1}}

\newcommand{\refcite}[1]{reference~\cite{#1}}
\newcommand{\refscite}[1]{references~\cite{#1}}

\newcommand{\BE}{\begin{equation}}
\newcommand{\EE}{\end{equation}}
\newcommand{\BEA}{\begin{eqnarray}}
\newcommand{\EEA}{\end{eqnarray}}

\newcommand{\ket}[1]{\left| {#1} \right\rangle}
\newcommand{\bra}[1]{\left\langle {#1}\right|}
\newcommand{\ketbra}[2]{\ket{#1}\!\bra{#2}}
\newcommand{\proj}[1]{\ketbra{#1}{#1}}

\newcommand{\eins}{\mathbbm{1}}

\renewcommand{\t}[1]{\textrm{#1}}
\newcommand{\tr}{\mathrm{tr}}
\renewcommand{\sp}{{\sigma}}
\newcommand{\Bvec}{{\vec{r}}}
\newcommand{\Cang}{{\vartheta}}
\renewcommand{\Re}[1]{\textrm{Re}\!\left\{#1\right\}}
\renewcommand{\Im}[1]{\textrm{Im}\!\left\{#1\right\}}
\newcommand\hsprod[2]{{\langle #1,  #2 \rangle}}
\newcommand\dynmat{\mathsf{D}}
\newcommand\hsf[1]{\Theta\!\left(#1\right)}
\renewcommand{\vec}[1]{\boldsymbol{#1}}

\begin{document}

\title[]{Fundamental limits to frequency estimation: \\A comprehensive  microscopic perspective}

\author{J. F. Haase}
\email{jan.frhaase@gmail.com}
\affiliation{Institut für Theoretische Physik, Albert-Einstein-Allee 11, Universität
Ulm, D-89069 Ulm, Germany}
\author{A. Smirne}
\email{andrea.smirne@uni-ulm.de}
\affiliation{Institut für Theoretische Physik, Albert-Einstein-Allee 11, Universität
Ulm, D-89069 Ulm, Germany}
\author{J. Ko\l{}ody\'{n}ski}
\affiliation{ICFO-Institut de Ciènces Fotòniques, The Barcelona Institute of Science and Technology, 08860 Castelldefels (Barcelona), Spain}
\author{R. Demkowicz-Dobrza\'{n}ski}
\affiliation{Faculty of Physics, University of Warsaw, 02-093 Warszawa, Poland}
\author{S.F. Huelga}
\email{susana.huelga@uni-ulm.de}
\affiliation{Institut für Theoretische Physik, Albert-Einstein-Allee 11, Universität
Ulm, D-89069 Ulm, Germany}
%
%
%
%
%
%
%
\begin{abstract}
We consider a metrology scenario in which qubit-like probes are used to sense an external field that 
affects their energy splitting in a linear fashion. Following the frequency estimation approach in which one optimizes the state and sensing time 
of the probes to maximize the sensitivity, we provide a systematic study of the attainable precision under the impact of noise 
originating from independent bosonic baths. Specifically, we invoke an explicit microscopic derivation of the 
probe dynamics using the spin-boson model with weak coupling of arbitrary geometry.
We clarify how the secular approximation leads to a phase-covariant dynamics, where the 
noise terms commute with the field Hamiltonian, while the inclusion of 
non-secular contributions 
breaks the phase-covariance.
Moreover, unless one restricts to a particular (i.e., Ohmic) spectral density 
of the bath modes, the noise terms may contain relevant information about the frequency to be estimated. Thus, by considering general 
evolutions of a single probe, we study regimes in which these two effects have a non-negligible impact on the achievable
precision. We then consider baths of Ohmic spectral density yet fully accounting for the lack of phase-covariance, in order 
to characterize the ultimate attainable scaling of precision when $N$ probes are used in parallel.
Crucially, we show that beyond the semigroup (Lindbladian) regime
the Zeno limit imposing the $1/N^{3/2}$ scaling of the mean squared error, recently derived assuming 
phase-covariance, generalises to any dynamics of the probes, unless the latter are coupled to the baths in the
direction perfectly transversal to the frequency encoding---when a novel scaling of $1/N^{7/4}$ arises. 
As our microscopic approach covers all classes of dissipative dynamics,
from semigroup to non-Markovian ones (each of them potentially non-phase-covariant),
it provides an exhaustive picture, in which all the different asymptotic scalings of precision naturally emerge.
\end{abstract}
\maketitle
\section{Introduction}
Quantum metrology is a rapidly evolving research field with a potential to soon 
become a commercial technology \cite{Dowling2015,Schleich2016}. Over the last decades, 
it has developed in many different directions encompassing
a broad spectrum of settings in which quantum systems are employed to precisely sense, 
measure or track physical parameters \cite{Toth2014,Demkowicz2015,Pezze2016,Degen2017}.
Despite other important quantum phenomena enhancing precision measurements \cite{Braun2017}, 
its major part has been devoted to scenarios with multiple probes, whose inter-entanglement  
allows to surpass precision limits typical to classical statistics, i.e., 
the \emph{Standard Quantum Limit} (SQL) \cite{Giovannetti2004}. 
As a result, the precision of sensing a parameter (either intrinsic 
or externally imprinted, e.g., by a field) encoded in each of the probes 
dramatically improves with the probe number. In optical interferometry \cite{Demkowicz2015}, in which the SQL is dictated by the photon shot noise, 
the use of squeezed light has allowed for ultrasensitive phase measurements \cite{Schnabel2017},
with a spectacular application in gravitational-wave detectors \cite{LIGO2011,LIGO2013}.
Similarly, in experiments involving multiple atoms \cite{Pezze2016}, in which the atomic projection noise 
defines the SQL, thanks to preparation of spin-squeezed \cite{Ma2011} or maximally 
entangled \cite{Greenberger1989} states, novel standards of atomic transition-frequency 
have been proposed \cite{Wineland1992,Wineland1994,Leibfried2004}. On the other hand, 
large atomic ensembles \cite{Budker2007} and nitrogen-vacancy centres \cite{Taylor2008} have 
become most sensitive magnetometers to date \cite{Degen2017};~optomechanical devices have 
lead to state-of-art displacement measurements \cite{Clerk2010}, while trapped-ion and optical-lattice atomic clocks  
have achieved both highest stability and accuracy in time-keeping \cite{Ludlow2015}.

In parallel, novel theoretical methods have been developed in order to 
quantify the ultimate performance of quantum metrology protocols and 
supplement the optimisation of their implementations. In particular, the techniques of estimation 
theory \cite{Helstrom1967} and statistical inference \cite{Barndorff2003} have been generalised 
into the quantum realm, introducing quantum notions of, e.g.:~Fisher Information \cite{Braunstein1994}, 
filtering \cite{Bouten2007} or waveform estimation \cite{Tsang2011};~%
which then had to be adapted in order to account for the inevitable 
quantum noise processes occurring in real-life experiments \cite{Huelga1997,Escher2011,Demkowicz2012,Tsang2013}.
\begin{figure}[t!]
\includegraphics[width=\columnwidth]{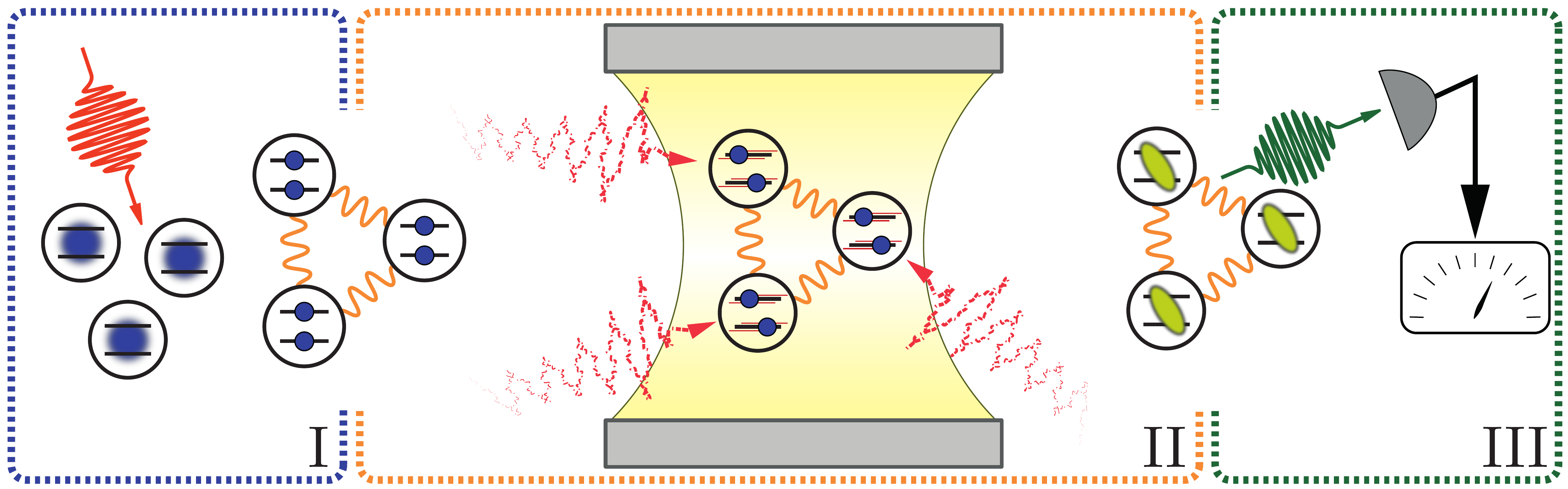}
\caption{
\textbf{Quantum metrology scheme} -- its consecutive stages with 
sensing probes being disturbed by bosonic baths. (I, \textit{blue}):~The probes modelled by
two-level particles are prepared in a desired entangled state.  (II, \textit{orange}):~The 
probes evolve for time $t$ during which each of them is driven by an external 
field and decoheres due to interactions with the bath surrounding it. 
(III, \textit{green}):~A global measurement is performed on all the probes, so that 
a fluctuation of the field may be most precisely resolved based on measurement outcomes.}  
\label{fig:metro_scheme}
\end{figure}

The main motivation of our work is to provide a \emph{microscopic model} that, on the one hand, 
allows for an explicit derivation of the up-to-date \emph{noisy quantum metrology} results while giving 
them a clear connection with the microscopic details of the probe-environment interaction;~but, on the other, is capable to go beyond what is known 
thanks to its rich structure that, however, has an indisputable physical interpretation. 
In order to do so, we resort to the canonical qubit-based metrology scheme depicted in \figref{fig:metro_scheme}, 
in which we set the probes to also be weakly interacting with bosonic baths during the sensing process. 
As a result, the dissipative dynamics of each of them is then described by the \emph{spin-boson model}  \cite{Leggett1987}
in the weak-coupling regime---a model commonly used in describing dynamics of open 
quantum systems, also beyond light-atom interactions, e.g, to model charge transfer \cite{Garg1985},
tunneling in materials \cite{Golding1992} or magnetic flux in SQUIDs \cite{Makhlin2001}.
Importantly, depending on the coupling geometry and the spectral density of bath modes, 
the model induces dissipative probe dynamics encompassing common noise descriptions,
whose use has been previously motivated in the metrological context either 
phenomenologically \cite{Escher2011,Demkowicz2012,Kolodynski2013},
or by considering the classical stochastic-fluctuations approach \cite{Szankowski2014}. 
It then not only provides a unifying picture, but also gives novel microscopic derivations to some 
noise-types, e.g., rank-one Pauli noise \cite{Sekatski2017} that includes transversal noise \cite{Chaves2013}.
Moreover, stemming from the microscopic picture, it allows to take into account the 
effect of the dissipative dynamics being dependent on the parameter being sensed---which
we demonstrate to significantly improve the attainable sensing precision at a single-probe level.
Last but not least, it gives a clear interpretation of the \emph{phase-covariance assumption} 
\cite{Holevo1993,Holevo1996,Vacchini2010}, which forces the noise terms to commute with the parameter-encoding Hamiltonian,
as it is then naturally guaranteed by the \emph{secular approximation} within which one discards fast oscillating 
terms in the master equation \cite{BreuerPetruccione}. Hence, by considering the model yielding
non-secular dynamics induced by the baths with Ohmic spectral densities, we are able to 
explicitly show that it is the \emph{Zeno limit} (see \refscite{Matsuzaki2011,Chin2012})
that dictates the asymptotic precision scaling also when the phase-covariance is broken. To this point, this limit was shown to be universal only in the case of secular dynamics \cite{Smirne2016}, hence this recent result is generalized for the considered model. Yet it can even be breached when the coupling of each bath is 
perfectly transversal.

The present manuscript has the following structure: \secref{sec:nqfe} contains an extensive introduction to the field of frequency estimation, illustrates the considered setup and recalls necessary tools for its analysis. 
The notion of phase covariance and its characterization in open quantum systems is established in \secref{sec:blv}, along with the corresponding form in terms of a master equation. Subsequently, the microscopic model of choice is illustrated in \secref{sec:spinbosonMOD}, where we demonstrate its capability to realize both phase-covariant and non-phase-covariant dynamics. The final sections deal with the metrologic properties of the model. We clarify the effect of non-phase-covariant dynamics by using a single probe in \secref{sec:sqFI}, using a short time expansion of the dynamics, independent of the environmental spectral density one chooses to be realized by the model. \secref{sec:QFIBounds} contains a thorough study of the asymptotic scalings in the  regime of large number of probes.
%
%
%
%
%
%
%
%
%
%
%
%
%
%
%
%
\section{Noisy quantum frequency estimation}
\label{sec:nqfe}
In all quantum metrology schemes employing multiple probes, as the one depicted 
in \figref{fig:metro_scheme}, the parameter to be determined---e.g., the external field in sensing 
\cite{Degen2017}, the photon path-difference in optical interferometry \cite{Demkowicz2015}, or the 
atomic internal transition frequency in spectrocopy \cite{Wineland1992,Wineland1994,Leibfried2004}---%
is crucially encoded onto each of the probes in an independent manner. As a result, by exploring the 
quantum entanglement in between them, the \emph{Standard Quantum Limit} (SQL) can be breached. 
In the classical setting, the SQL forces the mean squared error of estimation to decrease according
to the central limit theorem \cite{Kay1993}---at most as $\sim\!1/N$  with the number of probes---as the 
growth of $N$ can then be effectively interpreted as an overall increase in the size of the measurement 
data available. However, when the probes are prepared in an entangled state, such an intuition must be abandoned. 
In particular, by entangling all the probes with one another, e.g., by preparing them in
a GHZ state \cite{Greenberger1989}, the mean squared error may drop even as $\sim\!1/N^2$---%
attaining the fundamental \emph{Heisenberg Limit} (HL) on precision \cite{Giovannetti2004}.

In this work, we consider the task of \emph{frequency estimation} 
that is directly motivated by the atomic spectroscopy experiments
\cite{Wineland1992,Wineland1994}. However, it applies to any sensing scenario in which the 
duration of each experimental repetition  should be treated as a resource, while still 
operating in the regime of large statistics of the measurement data gathered  
\footnote{For instance, when maximising the sensitivity of slope detection 
in external-field sensing scenarios \cite{Degen2017}.}.
In such a case, the estimated parameter, $\omega_0$,
corresponds to the effective magnitude of a Hamiltonian $H_{\omega_0}$
inducing a unitary transformation on each of the probes:
\BE
\mathcal{U}_{\omega_0}(t)=U_{\omega_0}(t) \bullet U_{\omega_0}^\dagger(t) 
\quad {\rm with}\quad 
U_{\omega_0}(t) = e^{-i H_{\omega_0} t},
\label{eq:unitary_encoding}
\EE
where $H_{\omega_0}=\omega_0 \,\hat{h}$ and $\hat h$ is some fixed operator 
\footnote{Without loss of generality, we also require for convenience that 
maximal variance of $\hat h$ is $\max_{\psi} \left.\Delta^2 \hat h\right|_\psi=1/4$.}. 
The parameter $\omega_0$ can thus be 
naturally interpreted as the atomic transition frequency 
in spectroscopy experiments \cite{Bollinger1996} or, equivalently, the strength of
an external field being sensed, e.g., the magnetic field in atomic \cite{Budker2007} or 
NV-centre-based \cite{Taylor2008} magnetometry setups. Let us emphasize that within
frequency estimation tasks the encoding Hamiltonian, $H_{\omega_0}$, is assumed to 
be fixed, what contrasts the sensing scenarios in which either the parameter $\omega_0$ 
varies in time and must be tracked \cite{Tsang2011}, or $\hat h$ itself 
is a time-dependent operator \cite{Pang2017}. 

Importantly, in contrast to phase estimation tasks in optical interferometry \cite{Demkowicz2015}, 
in frequency estimation one must explicitly account for the finite time-scale over which $\omega_0$ 
is imprinted on the probes. In particular, $t$ in \eqnref{eq:unitary_encoding} 
that constitutes the encoding time specifies also the duration of a single 
round (repetition) of the protocol---we assume throughout this work that both the preparation 
and measurement stages in \figref{fig:metro_scheme} take negligible durations 
(see \cite{Dooley2016} for a generalisation). As a result, when optimising the protocol 
to maximise the precision attained, one must take into account the fact that, although the total duration 
of an experiment, $T$, can always be assumed to be significantly larger than the duration 
of a single protocol round ($T\gg t$), by decreasing $t$ the total number of repetitions, 
$\nu=T/t$, is increased. Such a possibility can have a positive impact on the achieved precision, 
as the mean squared error improves then at a classical, $\sim\!1/\nu$, rate due to 
more measurement data being gathered over the total experimental time $T$.

\subsection{Frequency estimation task as a quantum channel estimation protocol}
\label{sec:freq_est_prot}
\begin{figure}[t!]
\includegraphics[width=\columnwidth]{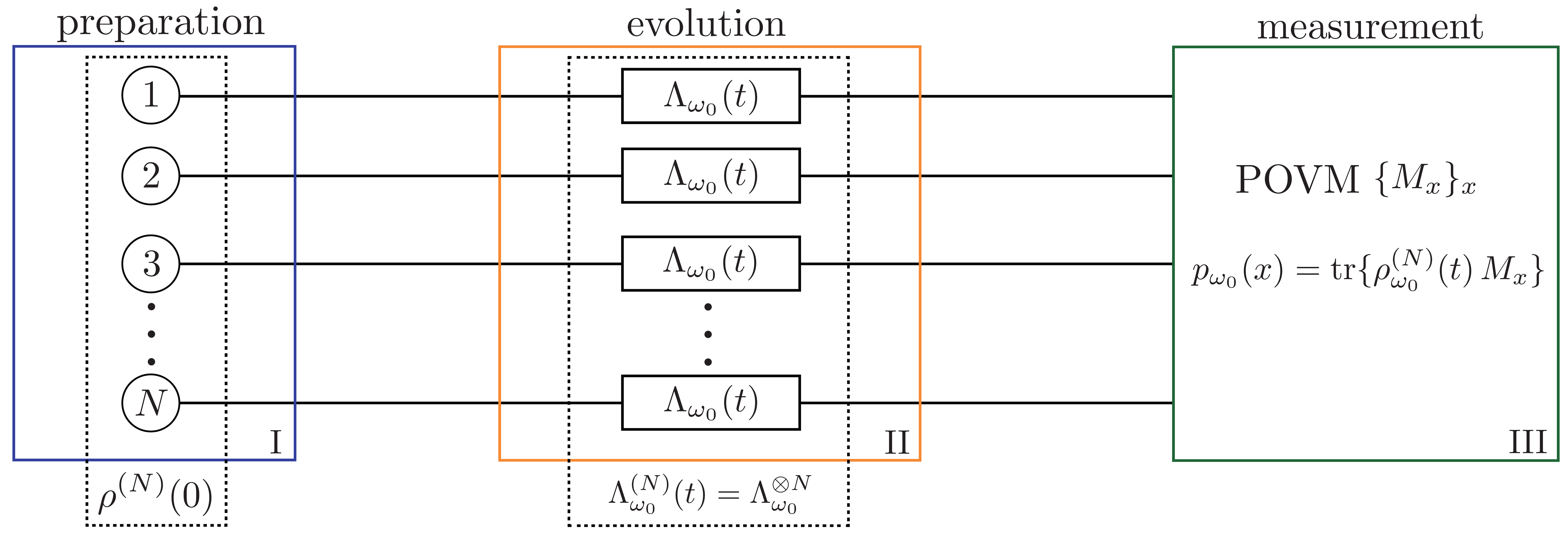}
\caption{
\textbf{Quantum channel estimation protocol} -- a formal interpretation of the quantum metrology scheme 
of \figref{fig:metro_scheme}. 
(I, \textit{blue}):~$\rho^{(N)}(0)$ represents the initial collective state of the $N$ probes. 
(II, \textit{orange}):~As probes evolve independently, the quantum channel representing 
global dynamics factorises, $\Lambda_{\omega_0}^{(N)}=\Lambda_{\omega_0}^{\otimes N}$,
into tensor product of probe channels, $\Lambda_{\omega_0}$, which encode the parameter, $\omega_0$, 
to be estimated and incorporate the impact of local noise. 
(III, \textit{green}):~A global measurement performed on the probes is formally represented 
by a POVM, $\{M_x\}_x$, whose elements determine the probability distribution of the measurement outcomes
labelled by $x$. 
}
\label{fig:paraEstimScheme}
\end{figure}
Any frequency estimation task, and more generally any metrology scheme of \figref{fig:metro_scheme}, can be 
viewed at the abstract level as a \emph{quantum channel estimation} protocol \cite{Fujiwara2001,Fujiwara2008}, depicted 
in \figref{fig:paraEstimScheme}. In particular, the consecutive stages described in \figref{fig:metro_scheme} 
can be formalised in the following way. Initially, the $N$ probes are prepared in a potentially entangled quantum
state $\rho^{(N)}(0)$. Subsequently, the frequency $\omega_0$ is encoded onto each of the probes 
via the action of a completely positive and trace preserving (CPTP) map 
\cite{Nielsen2010,Bengtsson2006}---a \emph{quantum channel} $\Lambda_{\omega_0}(t)$---%
that specifies each probe dynamics. As a result, the global state
of the probes after the frequency encoding stage of duration $t$ reads
\BE
\rho^{(N)}_{\omega_0}(t)=\Lambda_{\omega_0}(t)^{\otimes N}\,[\rho^{(N)}(0)].
\label{eq:final_state}
\EE
In the absence of noise the probe channel corresponds to the unitary $\omega_0$-encoding 
introduced in \eqnref{eq:unitary_encoding}, i.e., $\Lambda_{\omega_0}(t)=\mathcal{U}_{\omega_0}(t)$.
However, in general, it may incorporate the impact of any type of local noise that affects 
each probe in an uncorrelated fashion.

Note that in this work we do not consider the effects of global decoherence mechanisms that disturb all the 
probes in a correlated manner. Such noise processes are known to impose \emph{fixed} lower bounds 
on the achievable precision in metrology tasks that cannot be circumvented by any choice of probe states 
and measurements, even in the asymptotic limit of $N\to\infty$ \cite{Dorner2012,Jeske2014,Knysh2013,Knysh2014}.
As our motivation here is the investigation of the asymptotic precision scaling in frequency estimation---in particular,
its potential quantum enhancement beyond SQL despite dissipative dynamics---we want to consider schemes in which the error
asymptotically vanishes with $N$ and it is this scaling that unambiguously quantifies the performance. 

The measurement stage of any metrology scheme of \figref{fig:metro_scheme} corresponds in \figref{fig:paraEstimScheme}
to a generalised quantum measurement, either local or global, that is performed on the final state \eref{eq:final_state} 
yielding an outcome $x$. It is formally defined by a \emph{positive operator valued measure} (POVM), $\{M_x\}_x$, 
whose elements constitute positive-semidefinite operators, $M_x\geq0$, that sum to identity, 
$\sum_x M_x =\eins$ \cite{Nielsen2010}.  The measurement outcome $x$ is then associated with its corresponding  POVM element, 
so that the outcomes are distributed according to $p_{\omega_0}(x)=\tr\lbrace \rho^{(N)}_{\omega_0}(t) \,M_x\rbrace$.
Given that the protocol is repeated $\nu=T/t$ times over the total experiment
duration $T$, a dataset of measurement outcomes $\vec{x}_{\nu}=\lbrace x_1,  \ldots, x_\nu \rbrace$ is collected.
Then, based on the data, an \emph{estimator} $\tilde \omega(\vec{x}_{\nu})$ is constructed
whose value is aimed to most accurately reproduce the estimated frequency $\omega_0$. 
Moreover, as the experiment is assumed to last much longer than a single protocol round, $T\gg t$,
the measurement data collected can always be taken to be sufficiently large for 
the asymptotic ($\nu\to\infty$) statistical analysis to apply. 

Notice that the measurement outcomes are independently distributed with $p_{\omega_0}(\vec{x}_{\nu}) = \prod_{i=1}^{\nu}p_{\omega_0}(x_i)$,
as we have, for simplicity, disregarded the possibility of conducting adaptive strategies in which one 
adjusts the measurement, i.e., the POVM, in each protocol round based on the outcomes previously 
collected \cite{Berry2009}.  We are allowed to do so, as all the precision bounds discussed in the following 
sections are guaranteed to be saturated without the need of adaptive measurements in the so-called 
``local estimation regime'' \cite{Kay1993}, i.e., when sensing deviations of $\omega_0$ from a known value 
\footnote{A situation that naturally applies in the so-called ``slope detection''' scenarios 
(e.g., in Ramsey spectroscopy) of quantum sensing \cite{Degen2017}.}. As such a scenario is the most optimistic one, 
the precision limits it provides can be considered fundamental---being applicable to \emph{all} the more conservative approaches
as $\nu\to\infty$ \cite{Guta2007}. However, let us stress that the above requirement of ``estimation locality'' 
may, indeed, be relaxed by allowing for the measurements to be adaptive, given the promise that the true value 
of $\omega_0$ lies within a fixed, yet narrow enough window \cite{Fujiwara2006}. Nevertheless, if one was to 
consider the value of $\omega_0$ to be largely unpredictable, one must explicitly follow Bayesian inference 
approaches to frequency estimation \cite{Macieszczak2014} in which the notions of SQL and HL must 
also be redefined \cite{Jarzyna2015}.

Finally, the performance of the estimation protocol is quantified by the \emph{mean squared error} (MSE)
of the estimator constructed, i.e, 
\BE
\Delta^2\tilde{\omega} := 
\sum_{\vec{x}_{\nu}} 
p_{\omega_0}(\vec{x}_{\nu}) \left(\tilde \omega(\vec{x}_{\nu}) - \omega_0 \right)^2,
\label{eq:MSE}
\EE
which must be minimised by optimising the initial state, $\rho^{(N)}(0)$, and the measurement POVM, $\{ M_x\}_x$,
used in each round of the protocol of \figref{fig:paraEstimScheme}.

\subsection{Ultimate precision attained in quantum frequency estimation}
\label{sec:qcrb}
The minimal MSE \eref{eq:MSE} which can be attained by any consistent and unbiased 
estimator is determined by the \emph{Cram{\'e}r-Rao bound} (CRB) \cite{Kay1993}:
\BE
\Delta^2\tilde{\omega} \geq \frac{1}{\nu\, F_{\rm{cl}}[p_{\omega_0}]},
\quad\textrm{where}\quad 
F_{\rm{cl}}[p_{\omega_0}]:=\sum_x\frac{\dot{p}_{\omega_0}(x)^2}{p_{\omega_0}(x)}
\label{eq:CRB}
\EE
is the \emph{Fisher Information} (FI) that is fully defined by 
the probability distribution of measurement outcomes, $p_{\omega_0}(x)$, and its dependence on 
the estimated $\omega_0$. Here, and throughout the manuscript, we use the dot symbol 
to denote the derivative with respect to the estimated parameter, so that 
$\dot{\bullet}=\frac{\mathrm{d}\bullet}{\mathrm{d} \omega_0}$.
As $p_{\omega_0}(x)=\tr\lbrace \rho^{(N)}_{\omega_0}(t) \,M_x\rbrace$
with the measured state given in \eqnref{eq:final_state}, the CRB constitutes 
the ultimate limit on the precision attained by the protocol of \figref{fig:paraEstimScheme}
given a particular:~initial state $\rho^{(N)}(0)$, POVM $\{M_x\}_x$, and 
protocol duration $t$.

Importantly, the optimization of \eqnref{eq:CRB} over measurements can be completely avoided
in the quantum setting, as one may first explicitly maximise the FI over all POVMs 
by defining the \textit{Quantum-Fisher-Information} (QFI) as \cite{Braunstein1994}:
\BE
F_Q[\rho^{(N)}_{\omega_0}(t)]:=\max_{\{M_x\}_x} F_{\rm{cl}}[p_{\omega_0}]=\tr\left\{\rho_{\omega_0}^{(N)}(t) L_{{\omega_0}}^2\right\},
\label{eq:QFI}
\EE
which is now fully determined by the state $\rho_{\omega_0}^{(N)}(t)$ of \eqnref{eq:final_state}
with $L_{\omega_0}$ being its symmetric logarithmic derivative (SLD) satisfying
$\dot{\rho}_{\omega_0}^{(N)}(t)= \frac{1}{2}\left(L_{\omega_0} \rho_{\omega_0}^{(N)}(t)+\rho_{\omega_0}^{(N)}(t) L_{\omega_0}\right)$.
In general, the evaluation of the SLD and, hence, the QFI \eref{eq:QFI} requires the
explicit eigendecomposition of the state $\rho_{\omega_0}^{(N)}(t)$, which becomes rapidly 
intractable due to its dimension growing exponentially with the probe number $N$. 
However, in the absence of noise this is not the case, as the evolution of the probes 
is fully dictated by their Hamiltonians. Recalling \eqnsref{eq:unitary_encoding}{eq:final_state} we may then write:
\begin{equation}
\Lambda^{\otimes N}_{\omega_0}(t)=U_{\omega_0}^{(N)}(t) \bullet U_{\omega_0}^{(N)\dagger}(t) 
\quad\textrm{with}\quad 
U_{\omega_0}^{(N)}(t)=e^{-i H_{\omega_0}^{(N)} t},
\end{equation}
where, $H_{\omega_0}^{(N)}=\sum_{n=1}^{N}H_{\omega_0}^{[n]}=\omega_0 \sum_{n=1}^{N} \hat{h}^{[n]}$
is the effective global frequency-encoding Hamiltonian with $n$ indexing the probes.
Thus, when considering pure initial states $\rho^{(N)}(0)=\proj{\psi^{(N)}}$ in the protocol \footnote{
Pure initial states may always be considered optimal due convexity of the 
QFI \eref{eq:QFI} on states, $F_Q[\sum_i p_i\rho_{\omega}^{(i)}]\le\sum_i p_i F_Q[\rho_{\omega}^{(i)}]$ \cite{Alipour2015}.}, 
the QFI \eref{eq:QFI} simplifies to \cite{Demkowicz2015}:
\BE
F_Q\!\left[\mathcal{U}_{\omega_0}^{\otimes N}(t)[\psi^{(N)}]\right]= 4t^2 \,\Delta^2 H_{\omega_0}^{(N)}|_{\psi^{(N)}},
\label{eq:QFI_pure}
\EE
where $\Delta^2 H_{\omega_0}^{(N)}|_{\psi^{(N)}}$ is just variance of the 
frequency-encoding Hamiltonian for the state $\ket{\psi^{(N)}}$.

Combining \eqnsref{eq:CRB}{eq:QFI}, we arrive at the \textit{quantum Cram{\'e}r-Rao bound} (QCRB) \cite{Helstrom1967} 
that we utilise throughout this work as the benchmark dictating the ultimate achievable  precision:
\begin{eqnarray}
\Delta^2\tilde{\omega} \geq \min_{t\in[0,T]} \frac{1}{\nu F_Q[\rho^{(N)}_{\omega_0}(t)]} =\frac{1}{T}\,\min_{t\in[0,T]} \frac{t}{F_Q[\rho^{(N)}_{\omega_0}(t)]}.
\label{eq:QCRB}
\end{eqnarray}
However, in the setting of frequency estimation, as indicated above, it must also be optimized 
over the duration time $t$ of each protocol repetition. Then, as long as $T\gg t$, the QCRB \eref{eq:QCRB} 
sets the fundamental limit on precision for a given initial state $\rho^{(N)}(0)$, 
which is utilized in each round of the protocol of \figref{fig:paraEstimScheme}, while
the probes evolve according to particular dynamics specified by \eqnref{eq:final_state}.

\subsection{Realistic bounds on precision in the presence of local noise}
\label{sec:nfe}

Now, stemming from \eqnsref{eq:QFI_pure}{eq:QCRB}, we can formally define the notions of SQL and HL 
in frequency estimation when the noise is absent as, respectively: 
\begin{eqnarray}
\Delta^2\tilde{\omega}_{\rm{SQL}}T = \frac{1}{t} \, \frac{1}{N}
\quad\t{and}\quad
\Delta^2\tilde{\omega}_{\rm{HL}}T =  \frac{1}{t}\, \frac{1}{N^2}.
\label{eq:SQL_HL}
\end{eqnarray}
The above MSEs correspond to the minimal values of the QCRB \eref{eq:QCRB}
attained when optimising the protocol over all separable and entangled initial states 
$\rho^{(N)}(0)$, respectively. In particular, $\Delta^2\tilde{\omega}_{\rm{SQL}}$ 
is achieved by preparing the probes in a product $\ket{\phi}^{\otimes N}$ with 
$\ket{\phi}=\t{argmax}_\psi \left.\Delta^2 \hat h\right|_\psi$ , while 
$\Delta^2\tilde{\omega}_{\rm{HL}}$ is attained with $\ket{\psi^{(N)}}=\frac{1}{\sqrt{2}}(\ket{\lambda_{\rm{min}}}+\ket{\lambda_{\rm{max}}})$, 
where $\ket{\lambda_{\rm{min/max}}}$ are the eigenvectors corresponding to minimal/maximal eigenvalues of the Hamiltonian 
$H_{\omega_0}^{(N)}$ in \eqnref{eq:QFI_pure} \cite{Giovannetti2006}.

However, things change quite drastically if noise is taken into account. The first results in this direction were obtained in 
\refcite{Huelga1997}, which deals with a purely dephasing noise acting at rate $\gamma$ independently and identically on each of the probes, 
such that the resulting evolution is given by a quantum dynamical semigroup, i.e., it is fixed by a Lindblad equation \cite{Lindblad1976,Gorini1976}. 
The probes are described as qubits, as we will do from now on.
Even with the preparation of entangled probes, one unavoidably recovers the SQL scaling, 
no matter how weak the dephasing is,
with at most a constant factor of improvement. While this analysis was dedicated to a specific 
initial preparation and measurement (a generalized Ramsey scheme), it was afterwards
extended to arbitrary preparations and measurements and other kinds of
semigroup dynamics \cite{Escher2011,Demkowicz2012,Kolodynski2013}: 
In the presence of pure dephasing, spontaneous emission, depolarization and loss,
if one considers independent and identical noise and the dynamics is given by a semigroup,
the asymptotic 
scaling is unavoidably bounded to the SQL.

Importantly, all the dynamics for which such limitation was proven are characterized by the fact that the action of the noise commutes with 
the unitary encoding of the parameter. In other terms, the dynamics of the probes, besides being independent and identical, is \textit{phase-covariant} (PC) \cite{Holevo1993,Holevo1996,Vacchini2010}, which means that at any time $t$ the quantum channel $\Lambda_{\omega_0}(t)$ can be decomposed 
into the unitary encoding term and a noise term, and these two commute.
More precisely, the dynamics of a two-level system is said to be 
PC, if for any rotation by an angle $\phi$, $\mathcal{U}_{\phi} = \exp\lbrace i \phi \sp_z \rbrace \bullet \exp\lbrace -i \phi \sp_z\rbrace$, 
it holds
\begin{eqnarray}\label{eq:pc}
\left[\mathcal{U}_{\phi}, \Lambda_{\omega_0}(t)\right] = 0, \qquad \forall \phi, t.
\end{eqnarray}
This is easily shown to be equivalent to 
\footnote{Note that in \cite{Demkowicz2012, Kolodynski2013}, as well as in \cite{Smirne2016}, an $\omega_0$-independent
noise term $\Gamma(t)$ was considered. Here, instead, we will take into account a possible dependence on $\omega_0$
also in $\Gamma_{\omega_0}(t)$.}~$\Lambda_{\omega_0}(t) = \mathcal{U}_{\omega_0}(t) \circ\Gamma_{\omega_0}(t) = \Gamma_{\omega_0}(t) \circ \mathcal{U}_{\omega_0}(t)$, where $\Gamma_{\omega_0}(t)$ is a noise term, in other words the quantum channel acting on the probe that can associated purely with the noise.

It has been demonstrated that by going beyond the assumptions 
of the above no-go theorems, i.e., the semigroup and phase-covariance properties,
asymptotic precision scalings beyond SQL can be observed despite uncorrelated noise.
On the one hand, by breaking the phase-covariance and considering noise that is perfectly transversal to
the frequency encoding, the ultimate lower bound on precision has been derived \cite{Chaves2013}:
\BE
\Delta^2\tilde{\omega}_{\perp\&\rm{semi}}T \;\gtrsim\; \frac{1}{N^{5/3}}
\label{eq:trans_semi}
\EE 
and shown to be asymptotically attainable up to a constant factor (denoted by $\gtrsim$).
On the other, by circumventing the semigroup assumption
and considering pure dephasing noise fixed by a time dependent dephasing rate $\gamma(t)$, a scaling
$\propto 1/N^{3/2}$ has been found \cite{Matsuzaki2011,Chin2012,Macieszczak2015}. 
The super-classical $1/N^{3/2}$ scaling was named \textit{Zeno limit} due to it being dictated by the 
quadratic decay of the survival probabilities for short times, analogously to the Zeno effect \cite{Misra1977,Facchi2008}.

Recently \cite{Smirne2016}, an achievable lower bound to the estimation error for the whole class of PC dynamics, including both semigroup and non-semigroup evolutions was derived. 
The maximal estimation precision is fixed by the power-law decay of the short-time expansion of
the noise parameters and
it goes beyond the SQL if and only if the semigroup composition law is violated at short times.
Note that memory effects in the probes dynamics, i.e., non-Markovianity \cite{Rivas2014,Breuer2016}, do not provide any improvement of the estimation precision (apart from the unrealistic case of a full revival of coherences).
In particular, for any PC dynamics with linear decay of the noise parameters, which corresponds to the semigroup evolution, one gets
\BE
\Delta^2\tilde{\omega}_{\rm{PC\&semi}}T \;\gtrsim\; \frac{1}{N},
\label{eq:pc_semi}
\EE
while a quadratic decay yields the Zeno scaling
\begin{eqnarray}
\Delta^2\tilde{\omega}_{\rm{PC\&Zeno}}T \;\gtrsim\; \frac{1}{N^{3/2}}.
\label{eq:ZenoLimit}
\end{eqnarray}
These two scaling behaviors, along with that in \eref{eq:trans_semi} for purely transversal noise, provide the optimal asymptotic estimation precision achievable in the presence of different kinds of noise. Here, we will connect these results to the microscopic description of the probe-environment interaction, but we will also go beyond them by treating the cases of non-phase-covariant and non-semigroup noise.

\section{Phase-covariant vs. non-phase-covariant dynamics}
\label{sec:blv}
From the above discussion, it should be clear that the PC, or 
otherwise \emph{non-phase-covariant} (NPC), nature of the noise strongly influences the metrological bounds on the 
achievable precision in the frequency estimation which are set by the interaction with the environment.
A PC dynamics limits the estimation precision to the SQL for semigroup dynamics, or to the more favorable Zeno limit for non semigroup dynamics.
Hence, it is worth presenting explicitly an
intuitive way to differentiate between PC and NPC dynamics, which we will exploit throughout the paper.
We  use a representation of qubit quantum channels, which relies on the Hilbert-Schmidt scalar
product on the Hilbert space of the linear operators on finite-dimensional Hilbert spaces,
and which is directly linked to the action of the channels on the Bloch sphere. 
For further details the reader is referred to \cite{King2001,Bengtsson2006,Andersson2007, Asorey2009, Chruscinski2010, Chruscinski2010_2, Smirne2010}.

Recall that the Hilbert-Schmidt scalar product among two linear operators $\xi$ and $\chi$ is defined as
\begin{eqnarray}
\hsprod{\xi}{\chi}= \tr\left\{\xi^{\dag} \chi\right\}.
\end{eqnarray}
Hence, given the orthonormal basis of operators $\lbrace \tau_{\alpha}\rbrace_{\alpha = 0, \ldots,3} = \lbrace\mathbbm{1}/\sqrt{2},\sp_j/\sqrt{2}\rbrace_{j=x,y,z}$
acting on $\mathbbm{C}^2$, with $\sp_j$ the Pauli matrices, any qubit state $\rho$ can be represented as
\begin{eqnarray}
\rho=\sum^3_{\alpha=0}  \hsprod{\tau_{\alpha}}{\rho} \,\tau_{\alpha} =  \frac{1}{2}\left(\mathbbm{1}+\Bvec\cdot\vec{\sp}\right).
\label{eq:blsp}
\end{eqnarray} 
Here, $\vec{\sp}$ is the vector of Pauli matrices and $\Bvec$ is the Bloch vector associated with the state $\rho$, which has components $r_j=\tr\lbrace \sp_j \,\rho\rbrace$ for $j=x,y,z$ and must fulfill $|\Bvec|\leq 1$ to guarantee positivity. As well-known, any qubit state is in one-to-one correspondence with a vector inside of a unit sphere centered at the origin, i.e., the \textit{Bloch sphere}. 

In the same way, any linear map $\Xi$ acting on the qubit operators can be represented as a $4 \times 4$ matrix by means of the relation
\begin{eqnarray}
\Xi[\rho] = \sum^3_{\alpha \beta=0} \dynmat^{\Xi}_{\alpha \beta} \hsprod{\tau_{\beta}}{\rho} \,\tau_{\alpha} 
\qquad 
\dynmat^{\Xi}_{\alpha\beta} = \hsprod{\tau_{\alpha}}{\Xi[\tau_{\beta}]}.
\label{eq:matr}
\end{eqnarray}
Thus, given the CPTP dynamical map $\Lambda$, the most general form of the matrix $\dynmat^{\Lambda}$ associated with it reads
\begin{eqnarray}
\dynmat^\Lambda = \left( \begin{array}{cc} 
1 & \vec{0}^\mathrm{T} \\
\vec{v} & V 
\end{array}\right),
\label{eq:generalMap}
\end{eqnarray}
where $\vec{v}$ is a real 3 dimensional column-vector, $\vec{0}^\mathrm{T}$ is a 3 dimensional row-vector of zeros and $V$ a real 3-by-3 matrix. The first row guarantees the preservation of the trace, the real coefficients guarantee the Hermiticity preservation, while the general conditions for the CP can be found in \refcite{King2001}. Using Eqns.~(\ref{eq:blsp}-\ref{eq:generalMap}), one can easily see that the action of the dynamical map $\Lambda$ on a state $\rho$ associated with a Bloch vector $\Bvec$ simply corresponds to the affine transformation
\begin{eqnarray}
\Bvec \longrightarrow \vec{v}+V\Bvec,
\label{eq:aff}
\end{eqnarray}
where $\vec{v}$ describes translations of the Bloch sphere, while $V$ describes rotations, reflections and contractions. The latter point can be shown via the singular value decomposition, which allows us to write the $3 \times 3$ real matrix $V$ as \cite{King2001} 
\begin{eqnarray}
 V = R^{\varphi_1}_{n_1} D R^{\varphi_2}_{n_2},
 \label{eq:svd}
\end{eqnarray}
where $R^{\varphi_1}_{n_1}$ and $R^{\varphi_2}_{n_2}$ are two rotation matrices, about the axis $\hat{n}_k$ by the angle $\varphi_k$ for $k=1,2$, while $D$ is the diagonal matrix $D= \mbox{diag}\left\{d_x,d_y, d_z\right\}$. Then $|d_j|$ describes the contraction along the $j$-axis ($|d_j|\leq1$ to guarantee the positivity of the dynamics), and $d_j<0$ implies a reflection with respect to the plane perpendicular to the $j$-axis.

Such a representation of the dynamical maps allows us to easily detect PC
dynamical maps out of all the possible transformations of the Bloch sphere: for any fixed time, a dynamical map satisfies \eqnref{eq:pc} if and only if its matrix representation reads
\begin{eqnarray}
\dynmat^\Lambda_{\mathrm{PC}} = \left( \begin{array}{cccc} 
1 & 0 & 0 & 0 \\
0 & d \cos \xi & -d \sin \xi & 0 \\
0 & d \cos \xi & d \sin \xi & 0 \\
v_z & 0 & 0 & d_z
\end{array}\right).
\label{eq:PCmap}
\end{eqnarray}
With reference to the general form of a qubit dynamical map in \eqnref{eq:generalMap} and the decomposition in \eqnref{eq:svd}, we see that PC maps are identified by: 
equal contractions along the $x$ and $y$ axes ($D=\mbox{diag}\left\{d,d,d_z\right\}$), a translation only along the $z$-axis ($\vec{v} = \left\{0,0,v_z\right\}$) 
and a rotation only about the $z$-axis, which we get by setting $\hat{n}_1 = \hat{z}$ and $\varphi_1=\xi$, while $R^{\varphi_2}_{n_2} = \mathbbm{1}$ 
(other completely equivalent choices can be made, since $D$ commutes with the rotations about the $z$ axis). 
Of course, PC maps include only the affine transformations of the Bloch sphere commuting with the rotation about the $z$-axis
\footnote{The global rotation about $z$ will be given by the encoding rotation by $\omega*t$
plus possibly a further contribution, i.e., $\phi = \omega t + \Cang$.}, while
NPC maps include also rotations about any axis different from the $z$-axis, translations with non-zero components along the $x$ and $y$ axes 
and unequal contractions along the $x$ and $y$ axes, see \figref{fig:dynamicCompare}(\textbf{a}-\textbf{b)}.

Finally and crucially for our purposes, let us recall that given a PC dynamics,
the functional form of the corresponding master equation can be univocally characterized and it reads \cite{Smirne2016}
\begin{eqnarray}
\frac{\mathrm{d}\rho(t)}{\mathrm{d}t}&=& -i\left[(\omega_0+h(t)) \sp_z ,\rho(t)\right] \notag \\
&& + \gamma_{+}(t)\left(\sp_+ \rho(t) \sp_- -\frac{1}{2}\left\lbrace\sp_- \sp_+,\rho(t)\right\rbrace\right) \notag \\ 
&& + \gamma_{-}(t)\left(\sp_- \rho(t) \sp_+ -\frac{1}{2}\left\lbrace\sp_+ \sp_-,\rho(t)\right\rbrace\right) \notag \\ 
&& + \gamma_{z}(t)\left(\sp_z\rho(t) \sp_z - \rho(t)\right),
\label{eq:pcme}
\end{eqnarray}
for some, possibly time dependent, real coefficients $\ell(t), \gamma_{+}(t), \gamma_{-}(t), \gamma_{z}(t)$.
Equivalently, in the other direction, any master equation of the form as in \eqnref{eq:pcme} will give rise to a PC dynamics.
Formally, any the time-local generator $\mathcal{L}(t)$ of a master equation
\begin{equation}
\frac{\mathrm{d}\rho(t)}{\mathrm{d}t} = \mathcal{L}(t)[\rho(t)],
\end{equation}
is related to the the corresponding dynamical map by the Dyson expansion,
\begin{align}
\Lambda(t) = T_{\leftarrow}e^{\int_0^t \mathrm{d} \tau \mathcal{L}(\tau)} =\sum^{\infty}_{k=0} \int_0^t \mathrm{d}t_1 \int_0^{t_1} \mathrm{d}t_{2} \ldots \int_0^{t_k} \mathcal{L}(t_1)\ldots \mathcal{L}(t_k),
\label{eq:mtl}
\end{align}
where $T_{\leftarrow}$ is the chronological time-ordering operator;~%
and, hence, to the matrix representation of the map, $\dynmat^\Lambda$, 
defined via \eqnref{eq:matr}. In particular, one can show that 
starting from the master equation \eref{eq:pcme} the affine representation of the map 
must take form \eref{eq:PCmap} \cite{Smirne2016}.

%
%
\section{Spin-boson model: weak-coupling master equation and secular approximation}
\label{sec:spinbosonMOD}
We can now move on and introduce the general model we will exploit to investigate the difference between PC and NPC dynamics from a microscopic viewpoint
and, in the following Sections, how they determine different optimal precisions in frequency estimation. 

As emphasized before, we assume that the probes are affected identically and independently by their environments, 
so that the global dynamics is fixed by the one-particle dynamics, see \figref{fig:paraEstimScheme} and \eqnref{eq:final_state}. 
Therefore, we focus on the microscopic derivation of the open-system dynamics of one probe, 
which we present in the following. In particular, we model our sensing qubit with the widely used spin-boson model for quantum dissipation \cite{Leggett1987}. Within this model the environment corresponds to a set of non-interacting harmonic oscillators linearly coupled to the system, which may be directly interpreted as interactions with a radiation field or a phononic (crystal lattice) background. This model provides us with the most general description of the corresponding open two-level system dynamics, including special cases such as pure dephasing \cite{BreuerPetruccione} or purely transversal noise \cite{Clos2012}. The Hamiltonian of the spin-boson model consists of the two-level system Hamiltonian $H_0$, the free Hamiltonian $H_B$ of the environment and the interaction Hamiltonian $H_I$, which sum up to $(\hbar=1)$
\begin{eqnarray}
H &=& H_0 + H_B + H_I = \frac{\omega_0\sp_z}{2} + \sum_n \omega_n a^\dagger_n a_n \notag \\
&&+\left(\cos\Cang \frac{\sp_x}{2}+\sin\Cang \frac{\sp_z}{2}\right)\otimes \sum_n \left(g_n a_n+ g^*_n a_n^\dagger \right).
\label{eq:Hamiltonian}
\end{eqnarray}
The system's frequency $\omega_0$ represents the encoded frequency, while $a_n$ and $a^{\dagger}_n$ are the bosonic annihilation and creation operators of the bath mode $n$ of frequency $\omega_n$, which is coupled to the two-level system with the strength $g_n$. The parameter $\Cang$ defines the \emph{coupling angle}, i.e., the angle between the $x$-axis
and the direction of the coupling operator (in the $xz$-plane): for $\Cang=\pi/2$ we have pure dephasing (or parallel, with respect to $H_0$) interaction, 
while for $\Cang=0$ we have purely transversal (or perpendicular) interaction. 

Finally, note that the Hamiltonian is physically equivalent to a transformed Hamiltonian where the system only couples via $\sp_z$ to the environment, but both $\sp_z$ and $\sp_x$ are included in the system Hamiltonian; e.g., this would describe an experimental realization where the system is driven by the application of an off-axis magnetic field (see \appref{app:HamEngine}).

\subsection{Second-order TCL master equation}\label{sec:som}
To obtain a closed form of the master equation ruling the evolution of the probe subject to the noise fixed by \eqnref{eq:Hamiltonian},
we exploit a perturbative approach, assuming that the system is weakly coupled to the environment. In particular, we use the \textit{time-convolutionless} (TCL) master equation up to the second order \cite{Breuer2001, BreuerPetruccione}. Its general form in the interaction picture is given by (denoting as $\tilde{\rho}(t)$ the system state in the interaction picture with respect to $H_0+H_B$)
\begin{eqnarray}
\frac{\mathrm{d}\tilde{\rho}(t)}{\mathrm{d}t}=-\int_0^t \mathrm{d}\tau \, \tr_E\left\lbrace\left[H_I(t),\left[H_I(\tau),\tilde{\rho}(t)\otimes \rho_E \right]\right]\right \rbrace,
\label{eq:TCLmasterEquation}
\end{eqnarray}
where $H_I(t)$ is the interaction Hamiltonian $H_I$ in the interaction picture. In \appref{app:1}, we describe how to get the desired master equation for the reduced system density matrix starting from the \eqnref{eq:TCLmasterEquation}. At this point, let us just briefly introduce the main required quantities to define such a master equation, along with their physical meaning. First, the interaction Hamiltonian in the interaction picture is given by
\begin{eqnarray}
H_I(t)&=&e^{i H_0 t}\left(\cos\Cang \frac{\sp_x}{2}+\sin\Cang \frac{\sp_z}{2}\right)e^{-i H_0 t} \otimes B(t),
\label{eq:HamiltonianInteractionPicture}
\end{eqnarray}
where
\begin{eqnarray}
B(t) &=& \sum_n  \left(g_ne^{-i\omega_nt}a_n+g^*_ne^{i\omega_nt}a_n^\dagger \right)
\label{eq:BathOperators}
\end{eqnarray}
is the interaction picture of the environmental operator appearing in the interaction Hamiltonian, see \eqnref{eq:Hamiltonian}. The partial trace over the environment introduces the two-time correlation function $\tr_E \left[B(t)B(\tau)\rho_E\right]$ of the environment under its free dynamics, along with its complex conjugate $\tr_E \left[B(\tau)B(t)\rho_E\right]$. This function encompasses the whole relevant information about the environment needed to characterize the open-system evolution in the weak coupling regime: as we will see, it fixes each coefficient of the master equation. In addition, if the initial state of the bath is thermal, i.e.
\begin{eqnarray} 
\rho_E(0) = \frac{\exp\lbrace-\beta H_B \rbrace}{Z},
\end{eqnarray} 
with the inverse temperature $\beta$ and $Z = \tr\left\{\exp\lbrace-\beta H_B\rbrace\right\}$, since $[H_B,\rho_E(0)]=0$ the correlation function only depends on the difference of its time arguments $t-\tau$. Therefore we can define the correlation function $C(t)$ via
\begin{eqnarray}
\tr_E \left[B(t)B(\tau)\rho_E\right]&=&\tr_E \left[B(t-\tau)B(0)\rho_E\right] \equiv C(t-\tau).
\end{eqnarray}
Using the definition of $B(t)$ in \eqnref{eq:BathOperators}, this expression can be written as
\begin{eqnarray}
C(t) = \sum_n g_n^2 \left[N(\omega_n) e^{i\omega_n t}+(N(\omega_n)+1)e^{-i\omega_nt}\right],
\label{eq:CorrFunc}
\end{eqnarray}
where $N(\omega_n)=\tr_E\left\{a_n^\dagger a_n\rho_E\right\}$ represents the average number of excitations in the bath mode $n$. For the considered thermal state it is given by
\begin{eqnarray}
N(\omega)=\frac{1}{e^{\beta \omega}-1} = \frac{1}{2} \left[\coth\left(\frac{\beta \omega}{2}\right) -1\right]. 
\label{eq:nomega}
\end{eqnarray}
The bath correlation function $C(t)$ is conveniently expressed in terms of the \textit{spectral density} of the environment, which is defined by
\begin{eqnarray}
J(\omega) = \sum_n g_n^2 \delta(\omega-\omega_n).
\label{eq:spectralDensityDef}
\end{eqnarray}
This quantity describes the density of the bath modes weighted with the square of their individual coupling strength to the system. In fact, the bath correlation function \eqref{eq:CorrFunc} can be written as
\begin{eqnarray}
C(t) &=& \int_{0}^{\infty}\mathrm{d}\omega J(\omega) \left[N(\omega) e^{i\omega t}+(N(\omega)+1)e^{-i\omega t}\right] \notag\\
&=& \int_{-\infty}^{\infty}\mathrm{d}\omega e^{i\omega t} N(\omega)\left[J(\omega)\hsf{\omega}-J(-\omega)\hsf{-\omega}\right] \notag \\
&\equiv& \int_{-\infty}^{\infty}\mathrm{d}\omega e^{i\omega t} j(\omega).
\label{eq:ctt}
\end{eqnarray}
In the second line we used the formal identity $-N(-\omega)=N(\omega)+1$ (see \eqnref{eq:nomega}) in order to introduce the function $j(\omega)$, i.e., the anti-Fourier transform of the bath correlation function. 
The Heaviside stepfunction  $\hsf{\omega}$ keeps track of the fact that $J(\omega)$ is defined only for positive frequencies. 
Finally, the relation in \eqnref{eq:ctt} allows us to perform the continuum limit straightforwardly by replacing the 
spectral density in \eqnref{eq:spectralDensityDef} with a smooth function of the frequency bath modes \cite{BreuerPetruccione}.
 
As said, the bath correlation function $C(t)$ or, equivalently, the bath spectral density $J(\omega)$ along with the initial state of the bath fix the reduced master equation in the weak coupling regime: since we are dealing with the second order perturbative (TCL) expansion, only the two-time correlation function $C(t)$ is involved, while the bath multi-time correlation functions would only be involved in higher order terms (see also the recent \cite{Gasbarri2017}). As shown in \appref{app:1}, the master equation (back in the Schr{\"o}dinger picture) is then given by
\begin{eqnarray}
\frac{\mathrm{d}\rho (t)}{\mathrm{d}t}&=& \mathcal{L}(t)[\rho(t)] = -i\left[H_0 +H^{\text{LS}}(t) ,\rho(t)\right] \notag \\
&&+\sum_{j,k=\pm,z} b_{kj}(t)\left(\sp_k \rho(t) \sp_j^\dagger -\frac{1}{2}\left\lbrace\sp_j^\dagger \sp_k,\rho(t)\right\rbrace\right),
\label{eq:masterEquation}
~\end{eqnarray}
where we introduced the function
\begin{eqnarray}
\Gamma(\varsigma, t) = \int^t_{0} \mathrm{d} \tau e^{i \varsigma \tau} C(\tau)
\label{eq:gammmain}
\end{eqnarray}
for $\varsigma= \pm \omega_0, 0$, such that
\begin{eqnarray}
b_{zz}(t)&=&\frac{\sin^2\Cang}{2}\Re{\Gamma(0,t)} \nonumber\\
b_{++}(t)&=&\frac{\cos^2\Cang}{2}\Re{\Gamma(-\omega_0,t)} \notag \\
b_{--}(t)&=&\frac{\cos^2\Cang}{2}\Re{\Gamma(\omega_0,t)}\notag \\
b_{+-}(t)&=&b_{-+}^*(t)= \frac{\cos^2\Cang}{4}\left(\Gamma(-\omega_0,t)+\Gamma^*(\omega_0,t)\right)  \notag \\
b_{z+}(t)&=&b_{+z}^*(t)=\frac{\sin\Cang\cos\Cang}{4}\left(\Gamma(0,t)+\Gamma^*(-\omega_0,t)\right) \notag \\
b_{z-}(t)&=&b_{-z}^*(t)=\frac{\sin\Cang\cos\Cang}{4}\left(\Gamma(0,t)+\Gamma^*(\omega_0,t)\right),
\label{eq:coeffmain}
\end{eqnarray}
while the Hamiltonian correction is fixed by the elements:
\begin{eqnarray}
H^{\text{LS}}_{11}(t) &=&  
\frac{\cos^2\Cang}{4} \Im{\Gamma(\omega_0,t)} \nonumber\\
H^{\text{LS}}_{10}(t) &=& H^{\text{LS} *}_{01}(t) =
-\frac{i \cos\Cang \sin\Cang}{4} \nonumber\\
&& \times \left(\Re{\Gamma(0,t)}-\frac{1}{2}\left(\Gamma^*(-\omega_0,t)+\Gamma(\omega_0,t)\right)\right) \nonumber\\   H^{\text{LS}}_{00}(t) &=&  \frac{\cos^2\Cang}{4} \Im{\Gamma(-\omega_0,t)},
\label{eq:hhhmain}
\end{eqnarray}
where $H^{\rm{LS}}_{ij}(t) = \bra{i} H^{\rm{LS}}(t) \ket{j}$ for $i,j = 1,2$.

Let us stress that we did not invoke the Born-Markov approximation \cite{BreuerPetruccione} in our derivation---%
the above time-local master equation includes fully general non-Markovian effects and it will provide us 
with a satisfactory description of the noisy evolution of the probes as long as the interaction with 
the environment is weak enough (i.e., the higher orders of the TCL expansion can be neglected).
In addition, we are taking into account the dependence of the coefficients of the dissipative part of the master equation on the free system frequency $\omega_0$, see \eqnref{eq:coeffmain}, i.e., on the parameter to be estimated. This is a natural consequence of the detailed microscopic derivation of the system dynamics \cite{BreuerPetruccione},
in contrast with the phenomenological approaches, where the master equation
is postulated on the basis of the noise effects to be described. Let us emphasize that only in the case of pure dephasing, for which $\Cang=\pi/2$ and all dissipative terms in \eref{eq:coeffmain} apart from $b_{zz}(t)$ vanish, the dissipative part of the master equation can be assured not to depend on $\omega_0$. Otherwise, this is not generally the case unless a special choice of $J(w)$ is made (e.g., discussed later in \secref{sec:ohm}).

\subsection{Secular approximation}
Finding an explicit solution to the master equation in \eqnref{eq:masterEquation} is in general a complicated task, even after fixing the explicit form of the spectral density of the bath modes. 
On the other hand, the structure of the dynamics can be simplified considerably by making the so-called \textit{secular approximation} \cite{BreuerPetruccione,Breuer2001,Maniscalco2004,Fleming2010}, 
which relies on a time-scale separation between the system free-evolution time $\tau_0$ 
and the relaxation time $\tau_R$ of the system subject to the interaction with the environment. 
Whenever the free dynamics is much faster than the dissipative one, i.e., $\tau_0 \sim \omega_0^{-1} \ll \tau_R$, one can neglect terms oscillating with $e^{\pm i \omega_0 t}$ 
because they will be averaged out to 0 over a time interval of the order of $\tau_R$. 
If we apply this approximation to the weak coupling master equation, see in particular \eqnref{eq:3132}, all off-diagonal coefficients in \eqnref{eq:masterEquation}
and the off-diagonal elements of the Hamiltonian in \eqnref{eq:hhhmain}
 vanish, so that one is left with the master equation
\begin{eqnarray}
\frac{\mathrm{d}\rho(t)}{\mathrm{d}t}&=& -i\left[H_0 +\frac{H^{\text{LS}}_{11}(t)}{2} \sp_z ,\rho(t)\right] \notag \\
&& + b_{++}(t)\left(\sp_+ \rho(t) \sp_- -\frac{1}{2}\left\lbrace\sp_- \sp_+,\rho(t)\right\rbrace\right) \notag \\ 
&& + b_{--}(t)\left(\sp_- \rho(t) \sp_+ -\frac{1}{2}\left\lbrace\sp_+ \sp_-,\rho(t)\right\rbrace\right) \notag \\ 
&& + b_{zz}(t)\left(\sp_z\rho(t) \sp_z - \rho(t)\right),
\label{eq:SecMasterEquation}
\end{eqnarray}
where all the non-zero coefficients are still those of \eqnref{eq:coeffmain}.
This master equation can be explicitly solved for generic coefficients $b_{++}(t), b_{--}(t), b_{zz}(t)$ and $H^{\rm{LS}}_{11}(t)$
(see, e.g., \cite{Smirne2016, Lankinen2016}).

Crucially, we see how the secular master equation in \eqnref{eq:SecMasterEquation} precisely
corresponds to the most general form of a master equation associated with a PC qubit dynamics recalled in \secref{sec:blv},
see \eqnref{eq:pcme}.
Hence the difference between secular and non-secular dynamics provides us with a direct physical explanation
of the difference between PC and NPC dynamics.
The complete (weak-coupling) dynamics described by the master equation in \eqnref{eq:masterEquation} 
will generally lead to NPC dynamical maps, represented by generic matrices $\dynmat^{\Lambda(t)}$ as in \eqnref{eq:generalMap}
and corresponding to a completely general affine transformation of the Bloch sphere.
Instead, if one applies the secular approximation, thus getting the master equation in \eqnref{eq:SecMasterEquation}, 
the resulting dynamics is PC and will be then characterized by dynamical maps with a structure as in \eqnref{eq:PCmap}. 
In other words, within this framework, the distinction between PC and NPC dynamics precisely
corresponds to the distinction between dynamics within or outside the secular regime, i.e., the regime $\tau_0 \ll \tau_R$ where the secular approximation is well-justified. Needless to say, and as we will see explicitly in the next sections, the two kinds of dynamics describe also qualitatively different open-system evolutions. As a paradigmatic example, one can easily see how for any secular master equation the populations and coherences are decoupled, while the inclusion of non-secular terms leads to a coupling between them. The latter can be relevant for different phenomena, such as exitonic transport \cite{Oviedo2016,Jeske2015}, or the speed of the evolution in non-Markovian dynamics \cite{Sun2015,Zhang2015}.
Finally, note that general constraints on the variation of the coherences for a given variation of the populations in the presence of a generic completely positive phase covariant map have been recently derived in \cite{Lostaglio2017}.
\begin{figure*}[t!]
\includegraphics[width=\textwidth]{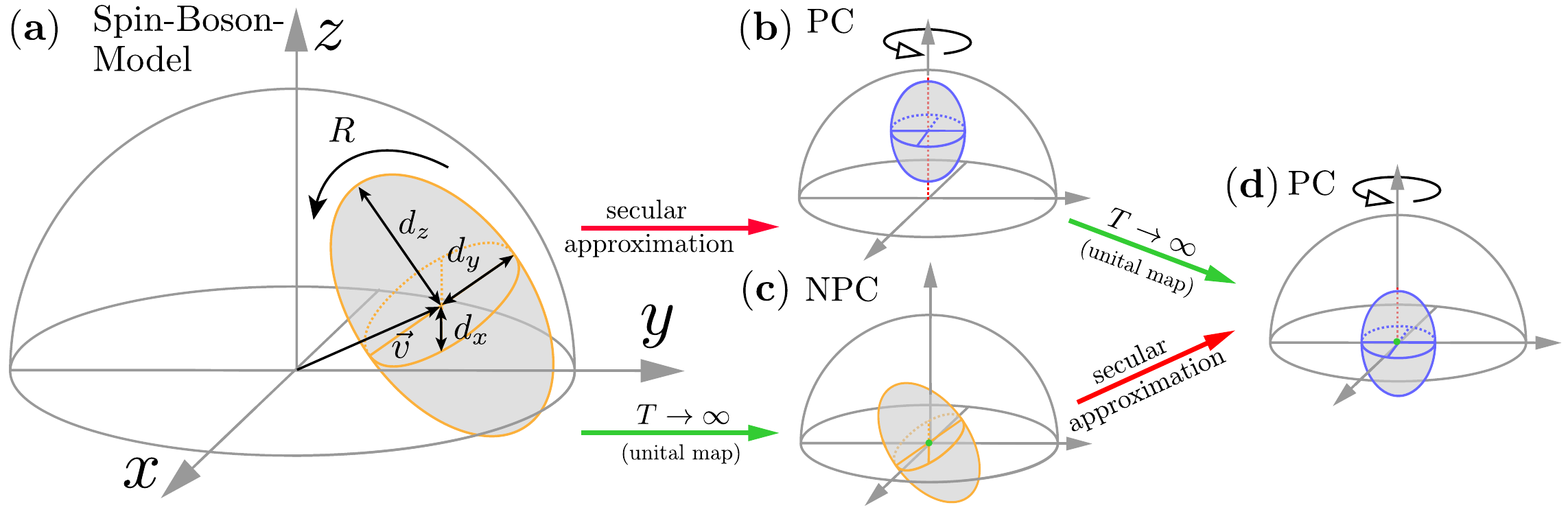}
\caption{
\textbf{Various dynamical regimes of the weak-coupling spin-boson model in the Bloch sphere picture.}
(\textbf{a}):~The NPC dynamics specified by Eqs.~(\ref{eq:masterEquation}-\ref{eq:hhhmain})
can be viewed as a general affine transformation \eref{eq:aff} of the Bloch sphere (see \secref{sec:blv}) 
contracting it to an ellipsoid that is parametrised by (in order):~a rotation $R^{\varphi_1}_{n_1}$, 
contractions $d_x, d_y, d_z$,  along the three axes, a reflection, a second rotation $R^{\varphi_2}_{n_2}$ and a 
translation by a vector $\vec{v}$ (for simplicity, we denote the two rotations by a single $R$ above).
(\textbf{b}):~PC dynamics \eref{eq:SecMasterEquation} is then obtained by applying the 
secular approximation that forces the \emph{cylindrical symmetry} (indicated by a circular arrow) of the ellipsoid 
around the $z$ axis. 
(\textbf{c}):~High-temperature limit of the spin-boson model forces a general map of the NPC dynamics to be \emph{unital}, i.e., 
the translation, $\vec{v}=0$, to vanish. 
(\textbf{d}):~When both high-temperature and secular approximations apply, the resulting quantum map in the Bloch representation 
is both cylindrically symmetric and unital.}
\label{fig:dynamicCompare}
\end{figure*}
%

\section{Solutions in the high-temperature regime}
\label{sec:htr}
In order to get analytic solutions for the NPC dynamics, which will also be useful to compare the different impact of NPC and PC dynamics on the metrological properties of the probes, let us restrict to the case of a bath at a high temperature. Because of that, we can treat the function $j(\omega)$ in the bath correlation function $C(t)$, see \eqnref{eq:ctt}, as a symmetric function of $\omega$: for large values of the temperature, i.e., small values of $\beta$, one has that $N(\omega)\approx 1/(\beta \omega)$, see \eqnref{eq:nomega}, and therefore $j(\omega)\approx j(-\omega)$. Looking at the correlation function in \eqnref{eq:ctt}), we see that in this regime $C(t) \approx C^*(t)$ thus we have $\Gamma(-\omega_0, t) \approx \Gamma^*(\omega_0,t)$, see \eqnref{eq:gammmain}. Together with \eqnref{eq:coeffmain}, we then obtain
\begin{eqnarray}
b_{zz}(t) &=& \frac{\sin^2\Cang}{2}\Gamma(0,t) \label{eq:coefhight}\\
b_{+-}(t)&=&b^*_{-+}(t)=  \frac{\cos^2\Cang}{2}\Gamma(-\omega_0,t) \nonumber\\
b_{++}(t) &=& b_{--}(t) = \Re{b_{+-}(t)}  \nonumber\\
b_{z+}(t)&=&b_{+z}^*(t) = b^*_{z-}(t) = b_{-z}(t) \nonumber\\
&=& \frac{\sin\Cang\cos\Cang}{4}\left(\Gamma(0,t)+\Gamma^*(-\omega_0,t)\right),\nonumber
\end{eqnarray}
while the Hamiltonian correction is given by
\begin{eqnarray}
H^{\text{LS}}(t)_{11}&=&  
\frac{\cos^2\Cang}{4} \Im{\Gamma(\omega_0,t)} \nonumber\\
H^{\text{LS}}(t)_{10}(t) &=& H^{\text{LS}*}(t)_{01}(t)\nonumber\\
&=&\frac{i \cos\Cang \sin\Cang}{4}\left(\Re{\Gamma(0,t)}-\Gamma(-\omega_0,t))\right) \nonumber\\   
H^{\text{LS}}(t)_{00}(t) &=&\frac{\cos^2\Cang}{4} \Im{\Gamma(-\omega_0,t)}.
\end{eqnarray}
These identities can be exploited to simplify the structure of the master equation, and hence of the corresponding dynamical map. 
In \ref{app:htl}, we show explicitly that the
constraints in \eqnref{eq:coefhight} imply the matrix form
\begin{eqnarray}
\dynmat^{\Lambda(t)} = \left( \begin{array}{cc} 
1 & \vec{0}^\mathrm{T} \\
\vec{0} & V(t) 
\end{array}\right),
\label{eq:matrixGenerator}
\end{eqnarray}
so that the translations of the Bloch sphere can be neglected and thus the dynamics can be described by unital maps,
i.e., such that $\Lambda(t)[\mathbbm{1}]=\mathbbm{1}$.
Note that the unitality of the reduced map is a general consequence
of the high temperature limit $T \rightarrow \infty$, in which
the initial state of the bath becomes maximally mixed \cite{Zyczkowski2004}.
By further applying the singular value decomposition to the matrix $V(t)$
one can get the geometrical picture associated with the dynamical map, 
in terms of rotations and contractions of the Bloch sphere, see \secref{sec:blv}. 
Indeed, an analogous result holds if we start from the PC master equation, see \eqnref{eq:SecMasterEquation}, and in the figure \ref{fig:dynamicCompare}(\textbf{c}-\textbf{d}) one can see a graphical representation of the corresponding
transformations of the Bloch sphere.

We will present, in particular, two different solutions of the high-temperature master equations (the PC and NPC ones);
namely,
for short times and a generic spectral density, as well as for an Ohmic spectral density at any time.
\\

\subsection{The short-time evolution}\label{sec:st}
As explained in \appref{app:htl}, using the Dyson series of \eqnref{eq:mtl} we obtain the short-time solution
of master equation \eref{eq:masterEquation} as:
\begin{widetext}
\begin{eqnarray}
\dynmat^{\Lambda(t)}_{(3)}= 
\left(
\begin{array}{cccc}
 1 & 0 & 0 & 0 \\
 0 & 1 -\frac{\omega_0^2t^2}{2}-\frac{1}{2} \alpha  t ^2 \sin ^2\Cang 
 &
-\omega_0  t  +  q
& \frac{1}{2} \alpha  t ^2 \cos\Cang \sin\Cang 
\\
 0 & 
\omega_0  t  - q
 & 1 -\frac{\omega_0^2t^2}{2}-\frac{\alpha  t^2}{2}& \frac{1}{3}\alpha  \omega_0 t^3  \cos\Cang  \sin\Cang\\
 0 & \frac{1}{2} \alpha  t ^2 \cos \Cang  \sin \Cang  & -\frac{1}{3}\alpha  \omega_0 t^3\cos\Cang  \sin\Cang & 1-\frac{1}{2} \alpha  t ^2 \cos ^2\Cang \\
\end{array}
\right)
\label{eq:highTMap}
\end{eqnarray}
\end{widetext}
where
\begin{eqnarray}
\alpha &=& \int_{-\infty}^{\infty}\mathrm{d}\omega j(\omega) \approx \int_{0}^{\infty} \mathrm{d}\omega \frac{2J(\omega)}{\beta\omega}\quad \t{and} \\  q&=& \frac{\omega_0 t^3}{6} \left[\alpha(1 +2\sin^2\Cang)+\omega_0^2 \right].
\end{eqnarray}
Truncating the Dyson series is justified due to the the weak-coupling approximation, while we have kept the terms up to 
the third order (and \emph{not} only to the second order) for a reason which will become clear when we evaluate the QFI 
of the corresponding evolved state in \secref{sec:stl}.  
The short time dynamical maps do not depend on the specific form of spectral density, but only on the global parameter $\alpha$. Furthermore, evaluating the eigenvalues of the Choi matrix reveals that the map is CP \cite{Choi1975}.

Repeating the same calculations for the PC master equation in \eqnref{eq:SecMasterEquation} in the secular approximation, we arrive at
\begin{widetext}
\begin{eqnarray}
\dynmat^{\Lambda(t)}_{(3),\rm{PC}}= \left(
\begin{array}{cccc}
 1 & 0 & 0 & 0 
 \\
 0 &   1- \frac{\omega_0^2t^2 }{2} -\frac{\alpha t^2}{4} \left(1+\sin^2\Cang\right) &-\omega_0 t+q& 0 
 \\
 0 & \omega_0 t-q &1- \frac{\omega_0^2 t^2 }{2}-\frac{\alpha t^2}{4} \left(1+\sin^2\Cang\right) & 0 
 \\
 0 & 0 & 0 &1-\frac{1}{2} \alpha  t ^2 \cos ^2\Cang
 \\
\end{array}
\right). 
\label{eq:highTMapSec}
\end{eqnarray}
\end{widetext}
%

\subsection{Finite-time evolution for an Ohmic spectral density}
\label{sec:ohm}
Here, in order to be able to characterize the reduced dynamics at any time $t$, we focus 
on a particular specific spectral density of the bath---the Ohmic spectral density:
\begin{eqnarray}
J(\omega) = \lambda \omega e^{-\omega/\omega_c},
\label{eq:jom}
\end{eqnarray}
where $\lambda$ quantifies the global strength of the system-environment interaction, while $\omega_c$ sets the cut-off frequency which defines the relevant environmental frequencies in the open-system dynamics. 
We further assume that $\omega_c \gg \omega_0$, so that the dependence of $\Gamma(\varsigma,t)$ on $\varsigma$ can be neglected
and
$\Gamma(\omega_0, t) \approx \Gamma(-\omega_0, t) \approx \Gamma(0, t)$,
as then, see \eqnsref{eq:ctt}{eq:gammmain}:
\begin{eqnarray}
\Gamma(\pm\omega_0,t) &\approx&  \frac{\lambda}{\beta} \int^t_{0} \mathrm{d} \tau \int_{-\infty}^{\infty}\mathrm{d}\omega e^{i\left(\pm \omega_0+\omega\right)\tau} \label{eq:lon}\\
&&\quad\times\;
\left(e^{-\omega/\omega_c}\hsf{\omega}+e^{\omega/\omega_c}\hsf{-\omega}\right) \nonumber\\
&=&\frac{2\lambda}{\beta}\int^t_{0} \mathrm{d} \tau e^{ \pm i \omega_0 \tau} \int_{0}^{\infty}\mathrm{d}\omega e^{-\frac{\omega}{\omega_c}} \cos(\omega \tau) \nonumber\\
&=& \frac{2\lambda}{\beta}\int^t_{0} \mathrm{d} \tau e^{ \pm i \omega_0 \tau} \frac{\omega_c}{1+\omega_c^2 \tau^2} \approx
\frac{2\lambda}{\beta}\int^t_{0} \mathrm{d} \tau \frac{\omega_c}{1+\omega_c^2 \tau^2},
\nonumber
\end{eqnarray}
where in the first and last approximated equalities we used the high-temperature condition and $\omega_0/\omega_c \ll 1$, respectively. 
The coefficients of the master equation in \eqnref{eq:coeffmain} then simplify to
\begin{eqnarray}
b_{zz}(t) &=& \frac{\sin^2\Cang}{2}\Gamma(0,t) \nonumber\\
b_{++}(t) &=& b_{+-}(t) = b_{-+}(t)=b_{--}(t) = \frac{\cos^2\Cang}{2}\Gamma(0,t)  \nonumber\\
b_{z+}(t) &=& b_{+z}(t) = b_{z-}(t)=b_{-z}(t) = \frac{\sin\Cang \cos\Cang}{2}\Gamma(0,t),\label{eq:ohmc}
\end{eqnarray}
while the Hamiltonian correction, $H^{\mathrm{LS}}(t)$, 
vanishes.
We stress that it is the specific choice of Ohmic spectral density
that assures the coefficients of the master equation
to be independent of $\omega_0$---a fact, typically
taken for granted in quantum metrology scenarios \cite{Kolodynski2013,Chaves2013,Dur2014,Brask2015,Demkowicz2015,Smirne2016,Zhou2017}.

Now, using \eqnref{eq:ohmc} one can easily see (e.g., by diagonalizing the matrix with elements given by the coefficients 
$b_{jk}(t)$) that the time-local master equation can be written as
\begin{eqnarray}
\frac{\mathrm{d}\rho(t)}{\mathrm{d}t}= -i\left[H_0 ,\rho(t)\right] + \gamma(t)\left(\bar{\sp} \rho(t) \bar{\sp}^\dagger -\rho(t)\right),
\label{eq:finiteME}
\end{eqnarray}
where the rate $\gamma(t)$ and the dissipative operator $\bar{\sp}$ are given by
\begin{eqnarray}
  \gamma(t) &=& \frac{1}{2}\Gamma(0,t) 
= \frac{\lambda}{\beta} \arctan\left(\omega_c t\right) \nonumber\\
\bar{\sp} &=& \cos\Cang \sp_x + \sin\Cang \sp_z.
\label{eq:ohmf}
\end{eqnarray}

It is worth noting that the dissipative part of the master equation is fixed by one single operator $\bar{\sp}$, i.e.,
we have the most general qubit rank-one Pauli noise, recently proved 
to be correctable in the semigroup case ($\gamma(t)=const$) in quantum metrology by ancilla-assisted error-correction \cite{Sekatski2017},
which has been demonstrated experimentally for transversal coupling, $\Cang=0$, in \refcite{Unden2016}.
In addition, the only noise rate $\gamma(t)$ is a positive function of time, which guarantees not only the CP of the dynamics, but also that the dynamical maps can be always split into CP terms. In this case, 
one speaks of (CP)-divisible dynamics, which coincides with the definition of Markovian quantum dynamics put forward in \cite{Rivas2010}; see also \cite{Rivas2014}. As expected, in the limit of an infinite cut-off, $\omega_c \rightarrow \infty$, the rate goes to a positive constant value, $\gamma(t) \rightarrow \pi \lambda/(2 \beta)$, so that we recover a Lindblad time-homogeneous (semigroup) dynamics \cite{Breuer2001}; see \appref{sec:appc2}, where we also give the explicit
form of the corresponding dynamical maps for $\Cang=0, \pi/2$, i.e., transversal and pure dephasing noise-types, respectively. 

Finally, note that a purely transversal interaction Hamiltonian ($\Cang=0$) yields a purely transversal master equation equation, i.e., the only dissipative operator $\bar{\sp}=\sp_x$ in \eqnref{eq:finiteME} is orthogonal to $H_0$, which is generally not guaranteed for arbitrary spectral densities. $\bar{\sp}=\sp_x$ characterizes what is usually known in the literature \cite{Chaves2013,Brask2015} as (and what we will here denote as) transversal noise.

\begin{figure*}[t!]
\begin{centering}
\vspace{0cm}
\hspace{-0.3cm} \includegraphics[width=2\columnwidth]{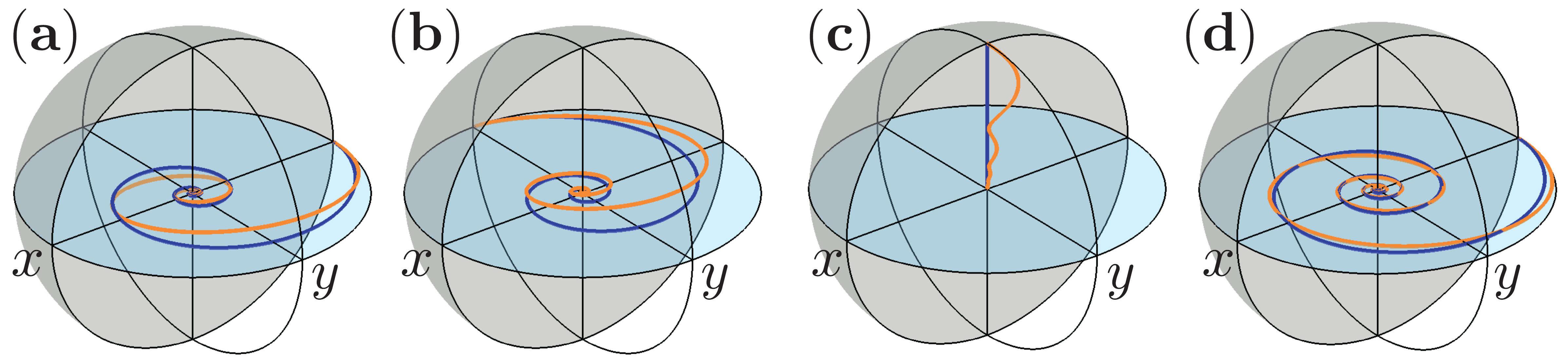}
\par\end{centering}
\caption{%
\textbf{Qubit evolution in the Bloch sphere picture for the resulting NPC \eref{eq:finiteME} and PC 
\eref{eq:masterEquationohmht} dynamics} (orange and blue, respectively). 
In (\textbf{a}-\textbf{c}) the evolution parameter $\Cang=\pi/4$ is chosen, 
so that when starting from an equator state, (\textbf{a}-\textbf{b}), the NPC dynamics
clearly differs from PC leading to a rotation around an axis that is tilted away 
from $z$. Initialising the qubit in an excited state, (\textbf{c}), the PC dynamics 
yields just a decay to a completely mixed state, while for NPC the rotational behaviour 
is still manifested. In (\textbf{d}), perfectly transversal ($\Cang = 0$) coupling 
is considered to illustrate that even though for both NPC and PC an equitorial state 
evolves in the $xy$ plane, the secular approximation of PC strongly modifies the speed 
of contraction.}
\label{fig:Blochsphere}
\end{figure*}
Let us now consider the corresponding dynamics under the secular approximation that provides us with a PC dynamics. 
The coefficients in the third line of \eqnref{eq:ohmc} along with $b_{+-}(t)$ are set to 0 and we are thus left with the PC master equation
\begin{align}
 \frac{\mathrm{d}\rho(t)}{\mathrm{d}t}= -i\left[H_0,\rho(t)\right] +\gamma(t)\sum_{j=\pm,z} d_{j}\left(\sp_j \rho(t) \sp_j^\dagger -\frac{1}{2}\left\lbrace\sp_j^\dagger \sp_j,\rho(t)\right\rbrace\right)
\label{eq:masterEquationohmht}
\end{align}
with $d_+ = d_- = \cos^2\Cang$, while $d_z = \sin^2\Cang$. Once again, the dynamics is CP and due to the positivity of $\gamma(t)$ and the $d_j$s it is even CP-divisible. Despite having now three different dissipative operators, these claims hold because there is only one single time-dependent function which defines all the rates.

In \figref{fig:Blochsphere} we illustrate the different dynamics geometrically by comparing the different evolutions of the open system for the NPC dynamics described by \eqnref{eq:ohmf} and the PC dynamics fixed by \eqnref{eq:masterEquationohmht}, respectively, see \appref{sec:appc2}. In \figsref{fig:Blochsphere}(\textbf{a}-\textbf{c}), we report the evolution for the same dynamics (i.e., the same $\omega_0, \omega_c, \lambda, \beta$ and $\Cang$), for the three different initial conditions which correspond to the three canonical orthogonal axes in the Bloch sphere. Of course, this is enough to detect all the possible linear transformations that the set of states undergoes during the evolution. In the PC dynamics we have contractions and rotations about the $z$-axis, as well as equal contractions along the $x$- and $y$-axes. These are all transformations commuting with the unitary rotation about the $z$-axis, as recalled in \secref{sec:blv}. On the other hand, in the non-secular dynamics we can observe a rotation about an axis with components in the plane perpendicular to the $z$-axis, which clearly breaks the phase covariance of the dynamics. \figref{fig:Blochsphere}(\textbf{d}) is devoted to illustrate another NPC effect, which is already present in the dynamics in \figsref{fig:Blochsphere}(\textbf{a}-\textbf{c}), but is not clearly observable due to the other transformations of the Bloch sphere. We consider a dynamics where $\Cang=0$, thus excluding any rotation apart from that about the $z$-axis \footnote{Once again, this could be shown by exploiting a block-diagonal structure of the generator $\mathcal{L}(t)$
and thus of the resulting dynamical maps; compare with \appref{app:htl}.}. As we see, the NPC dynamics introduces different contractions along the $x$ and $y$ directions, contrary to the PC case. The effects on parameter estimation of the rotations about the $x$- and $y$-axes, as well as the different contractions along them will be investigated in \secref{sec:ft}. 

Finally, note that although the non-secular terms introduce a transient behavior, which departs from the secular (i.e., PC) evolution, the system relaxes, in any case, to the fully mixed state. When probe systems can be interrogated within the transient dynamics, metrological advances may arise, as discussed in the following sections.

%

\section{Single-qubit quantum Fisher information}
\label{sec:sqFI}
We are now in a position to study the precision that can be reached in frequency estimation under the general dynamics considered here. 
We start by addressing the case of a single probe, which already enables us to point out some relevant differences in the behavior of the QFI under a PC and a NPC dynamics, respectively. In the next section, we focus on the asymptotic scaling with the number $N$ of probes.

As recalled in \secref{sec:nqfe}, the QFI fixes the maximum achievable precision via the QCRB in \eqnref{eq:QCRB}. For a single qubit probe, one can directly evaluate the QFI by diagonalizing the state $\rho_{\omega_0}(t)$ at time $t$, see \secref{sec:qcrb}. Here, instead, we use a different and equivalent formulation of the QFI \cite{Zhong2013}, which directly connects it to the Bloch sphere picture of the probe dynamics. Given the Bloch vector $\Bvec(0)$ associated with the initial state $\rho(0)$ and recalling that we are dealing only with unital dynamics, see \secref{sec:htr}, so that the affine transformation of the Bloch sphere in \eqnref{eq:aff} reduces at any time $t$ and for any $\omega_0$ to $\Bvec(0) \rightarrow V_{\omega_0}(t) \Bvec(0)$, the QFI at time $t>0$ can be expressed as
\begin{equation}
F_Q\left[\rho_{\omega_0}(t)\right] = \left| \dot{V}_{\omega_0}(t) \Bvec(0)\right|^2 + 
\frac{\left(V_{\omega_0}(t) \Bvec(0)\cdot  \dot{V}_{\omega_0}(t) \Bvec(0)\right)^2}{1-\left| V_{\omega_0}(t) \Bvec(0)\right|^2};
\label{eq:QFInori}
\end{equation}
the second term is set to 0 for pure states at time $t$, i.e., for $| V_{\omega_0}(t) \Bvec(0)|=1$. Note that we mark the derivative with respect to the parameter by a dot, i.e., $\dot{V}_{\omega_0}\equiv\partial V_{\omega_0}/\partial \omega_0$. In the following, we focus on initially pure states, i.e., $| \Bvec(0)|=1$, since any mixture would decrease the QFI as a consequence of its convexity \cite{Demkowicz2015}. It is then convenient to move to spherical coordinates and adequately parametrise pure states by 
$\Bvec = \lbrace \sin\theta \cos\phi, \sin\theta \sin\phi, \cos\theta \rbrace$.

\subsection{Short-time limit}
\label{sec:stl}
Thus, let us start by looking at the short-time expansion of the QFI in \eqnref{eq:QFInori}. The spherical parametrisation provides us with a clear relation among 
the short-time QFI for the NPC and PC dynamics, see \eqnsref{eq:highTMap}{eq:highTMapSec}, respectively. As a matter of fact, the 
first non-trivial term (i.e., the first contribution to $F_Q$ which is induced by the noise and therefore the first contribution where $F_Q$ differs between NPC and PC dynamics) in the QFI is of the order $t^4$ and it is fixed by those terms up to $
t^3$ in $V_{\omega_0}(t)$ and $\dot{V}_{\omega_0}(t)$. After a straightforward calculation, we arrive in fact at
\begin{eqnarray}
&&F_{Q,\text{PC}}^{(4)}=\sin^2\theta\,\,  t^2  -\frac{1}{3} \alpha \sin^2 \theta \left(1+ \sin^2 \Cang\right) t^4, \label{eq:QFI_sT_pc}  \label{eq:QFIst}\\
&&F_{Q}^{(4)}=F_{Q,\text{PC}}^{(4)}+\alpha t^4 \sin\theta \left(\frac{1}{3} \cos\theta   \sin2\Cang\cos\phi \right.\nonumber\\ 
&&  \left.+\frac{\sin^2\phi\left(\sin\Cang\cos\theta+\cos\Cang \cos\phi \sin\theta\right)^2/4}
  {\cos\theta \cot\theta- 2\tan\Cang \cos\phi \cos\theta+(\cos^{-2}\Cang - \cos^2\phi)\sin\theta}
  \right).\nonumber
\end{eqnarray}
The maximum value of the QFI for a PC dynamics is obtained for $\Cang=0$, i.e., for a pure transversal Hamiltonian \cite{Chaves2013} and for $\theta=\pi/2$, i.e., for a state lying on the equator of the Bloch sphere; 
moreover, the dephasing noise, i.e., $\Cang=\pi/2$, is the most detrimental in this regime.
Although the expression for $F_Q^{(4)}$ in the NPC case is too cumbersome to yield a comprehensible analytical solution for a state which maximizes the QFI in the short time limit,  even for a fixed value of the parameter $\Cang$, we report an approximated evaluation in \appref{app:QFIan}.

As can be directly inferred comparing the two formulas in \eqnref{eq:QFIst}, 
a crucial difference between PC and NPC dynamics is that
in the former case the QFI only depends on the initial distance of the Bloch vector from the $z$-axis and hence on the angle $\theta$, 
while the NPC terms introduce a dependence of the QFI on the direction of the Bloch vector itself and therefore on the angle $\phi$. Such a dependence is
a consequence of the non-commutativity of the encoding Hamiltonian with the action of the noise. For any PC dynamical map $\Lambda_{\omega_0, \mathrm{PC}}$,
if we rotate the state $\rho$ about the $z$-axis by a certain angle $\phi$, we have
that 
\begin{equation}\label{eq:fq}
F_Q\left[\Lambda_{\omega_0, \rm{PC}} \mathcal{U}_{\phi}[\rho]\right] = F_Q\left[\mathcal{U}_{\phi}\Lambda_{\omega_0, \mathrm{PC}}[\rho]\right]
=F_Q\left[\Lambda_{\omega_0, \rm{PC}}[\rho]\right],
\end{equation}
by virtue of \eqnref{eq:pc} the invariance of the QFI under rotations independent from the parameter to be estimated.

Now, the contributions due to the NPC terms are able to enhance the QFI,
as can be seen in \figref{fig:QFIplot}({\textbf{a}), where we illustrate 
the behavior of the difference $\Delta F_Q \equiv F_{Q}^{(4)}-F_{Q,\rm{PC}}^{(4)}$ as a function of the initial conditions. 
Besides the dependence on the initial phase $\phi$, one can clearly observe the presence of several areas where the NPC terms do increase the QFI. 
Moreover, there are two maxima of the increment, one in the neighborhood of $\phi=0$ and one in the neighborhood of $\phi=\pi$; 
we plot $\Delta F_Q$ for values $\phi\in[0,\pi]$, since it is a symmetric function under the reflection $\phi\rightarrow 2\pi-\phi$, see \eqnref{eq:QFIst}. In the plot, we fixed $\Cang = \pi/4$ but the behavior is qualitatively the same for different values of $\Cang$. Indeed, $\Delta F_Q$ goes to 0 for $\Cang$ going to $\pi/2$ since for a pure dephasing Hamiltonian the secular approximation has no effect, so that the dynamical maps in \eqnref{eq:highTMap} and \eqnref{eq:highTMapSec} coincide. 

Moreover, the presence of NPC terms can enhance the value of the QFI maximized over all the initial conditions and hence enlarge the maximal achievable precision. This is explicitly shown by taking into account the states lying on the equatorial plane of the Bloch sphere, which as said maximize $F_{Q,\mathrm{PC}}^{(4)}$. For $\theta=\pi/2$, the second relation in \eqnref{eq:QFIst} reduces to
\begin{eqnarray}
F_{Q}^{(4)}=F_{Q,\rm{PC}}^{(4)}+\frac{\alpha t^4}{4}\left(\frac{\cos^4\Cang \cos^2\phi \sin^2\phi}{\sin^2\Cang \cos^2\phi+\sin^2\phi}\right),
\label{eq:fqfqpc}
\end{eqnarray}
which clearly shows that the maximum value of $F_{Q,\mathrm{PC}}^{(4)}$ can be actually overcome for any value of $\Cang\neq \pi/2$. In \figref{fig:QFIplot}(\textbf{b}), we plot the increase 
of the QFI due to the NPC terms for $\theta=\pi/2$, while varying the initial phase $\phi$ and mixing angle $\Cang$. The QFI with the NPC terms is always bigger or equal than $F_{Q,\rm{PC}}^{(4)}$ and the maximum enhancement occurs for $\phi$ close to $k \pi$ with $k=0,1,2$ and the pure transversal noise corresponding to $\Cang=0$. However, the latter condition depends on the specific choice of the initial state: for $\theta\neq \pi/2$ one can have the maximal amplification due to the NPC terms for non-zero values of $\Cang$. 
\begin{figure}[t!]
\includegraphics[width=\columnwidth]{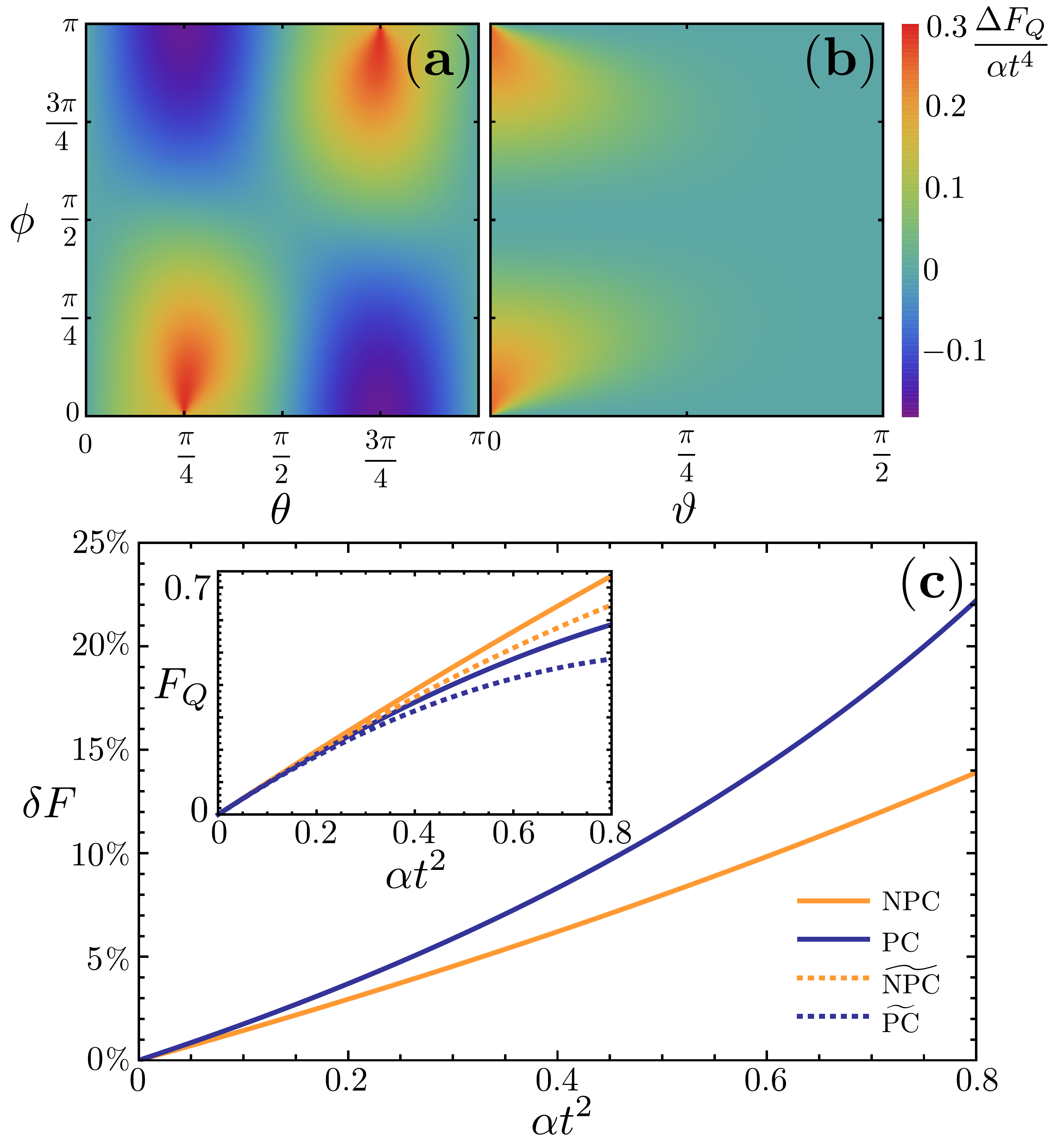}
\caption{
\textbf{Difference between NPC and PC QFI at short time scales and contributions due to the dependence of the master-equation rates on $\omega_0$}.
(\textbf{a}-\textbf{b}):~Difference (adimensional) between the QFI of NPC and the PC dynamics,
$\Delta F_{Q}/(\alpha t^4) =(F_{Q}^{(4)}-F_{Q,\rm{PC}}^{(4)})/(\alpha t^4)$,
as a function of:~ (\textbf{a}) $\phi$ and $\theta$ for a fixed coupling angle $\Cang=\pi/4$;
(\textbf{b}) of $\phi$ and $\Cang$ for states maximizing $F_{Q,\rm{PC}}^{(4)}$ in \eqnref{eq:fqfqpc} at $\theta =\pi/2$.
(\textbf{c}):~Increase of the QFI at short time-scales for PC and NPC dynamics when the $\omega_0$ dependence of the noise rates is taken into account, here $\delta F_Q=(F_Q-\tilde{F_Q})/F_Q$. The inset shows the QFI plotted exactly and after neglecting the dependence of the noise rates on $\omega_0$
(denoted by $\tilde{\bullet}$). 
}
\label{fig:QFIplot}
\end{figure}

\subsubsection*{Different contributions to the QFI.}

To get a more quantitative and general understanding of the different contributions fixing the QFI in PC and NPC dynamics,
let us move a step back and recall them explicitly.

First, the non-commutativity between the noise and the
free evolution will induce some specific contributions to the
QFI, typical of the NPC regime. For illustration, let us use the decomposition $\mathcal{L}(t) = \mathcal{H}(t)+\mathcal{D}(t)$, where $\mathcal{H}(t) = - i \left[H_0+H^{\rm{LS}}(t), \cdot \right]$
is the Hamiltonian term, while $\mathcal{D}(t) = \sum_{ij}b_{ij}(t)\left(\sigma_i \cdot \sigma_j^{\dag}-(1/2)\left\{\sigma^{\dag}_j \sigma_i, \cdot\right\}\right)$ represents the dissipator. In the PC case we have that $[\mathcal{H}(t),\mathcal{L}(t)]=0$ which does not hold for NPC dynamics}, as can be directly checked, for instance, by comparing \eqnref{eq:SecMasterEquation} and \eqnref{eq:masterEquation}. Recalling the Dyson expansion in \eqnref{eq:mtl}, we have to consider terms as $\mathcal{H}(t_1) \mathcal{D}(t_2) \ldots \mathcal{H}(t_k)$ to obtain the
dynamical maps fixing the evolution of the probes. If the Hamiltonian and the dissipative part do not commute, then the dependence on $\omega_0$ within $\mathcal{H}(t)$
will mix with the dissipative terms contained in $\mathcal{D}(t)$ and will be thus spread among more parameters of the dynamical
map at time $t$ or, equivalently, on more features of the Bloch vector at time $t$, possibly enhancing the QFI. In particular, this mechanism leads to the dependence of the QFI on the phase of the probes initial state in the NPC case, a feature which is not shared with the PC case, see \eqnref{eq:fq}.

Second, the noise terms themselves depend on $\omega_0$: As already pointed out in \secref{sec:som} 
the coefficients of the master equation will in general contain a dependence on the parameter to be estimated. To quantify explicitly such a phenomenon,  we compared, for both PC and NPC dynamics,
the QFI which is obtained including the dependence of the rates on $\omega_0$, with the QFI where such a dependence is disregarded. In particular, in the latter case we replace the dependence of the coefficients $b_{ij}(t), H^{\mathrm{LS}}_{ij}(t)$ on $\omega_0$
with the dependence on a generic frequency $\Omega$, and only after that the QFI has been evaluated, we set $\Omega=\omega_0$. Let us denote this auxiliary object as $\tilde{F}_Q$, contrary to the former calculations of the QFI which have been denoted by $F_{Q}$.
We stress that that $\tilde{F}_Q$
is actually the object utilized in more phenomenological approaches to quantum metrology, where the master equation is postulated to describe some specific kinds of noise, rather than microscopically derived so that the contributions due to the dependence of the rates on $\omega_0$ are not accounted for.
On the other hand, let us mention that in \cite{Szankowski2014} the role of the dependence of the emission and absorption rates on the free system frequency
for a qubit system coupled to a Gaussian classical noise has been investigated.

\figref{fig:QFIplot}(\textbf{c}) summarizes the effects of the two contributions described above. In the main panel we plot the percentage increase $\delta F_Q = 100(F_Q-\tilde{F_Q})/F_Q$ for both PC and NPC QFI. We
see that in both cases the dependence on $\omega_0$ of the noise terms non-negligible and the compliance of these noise terms can increase the QFI way beyond the value of the auxiliary QFI, e.g. reaching $10\%$ for NPC and $15\%$ for PC at $\alpha t^2 \approx 0.6$.

In the case of the PC dynamics, we can derive a very intuitive geometrical picture of the information encoding. In \appref{app:QFIan} we show that the auxiliary QFI $\tilde{F}_{Q,\mathrm{PC}}(t)$ is simply proportional to $t^2D_z(t)^2$ where $D_z(t)$ is the length of the projection of the Bloch vector into the $xy$ plane, see \eqnref{eq:qfipc1}. Hence the information about the frequency we want to estimate, i.e. the rotation speed about the $z$-axis, is fully enclosed into the distance of the Bloch vector from the rotation axis. Crucially, if we take the dependence of the rates on $\omega_0$ into account, some further contributions to the QFI will appear, see \eqnref{eq:qfipc2}. There is one additional term due to the dependence of $D_z(t)$ on $\omega_0$ and a second term in accordance with \eqnref{eq:PCmap}, which contains the noise parameters $v_z(t)$ and $d(t)$. By construction, these two terms are positive for any PC dynamics, so that the dependence of the rates on $\omega_0$ will always yield an improvement on the estimation precision, as already indicated in \figref{fig:QFIplot}(\textbf{c}).

The time course of the QFI provides us with direct access to the contribution of the non-commutativity by comparing $F_{Q,\mathrm{PC}}$ and $F_{Q,\mathrm{NPC}}$ in the inset of \figref{fig:QFIplot}(\textbf{c}). This effect is even
more relevant than the contribution due to the dependence of
the noise rates on $\omega_0$ and, in any case we can further confirm that the inclusion of nonsecular
terms modifies significantly the one-probe QFI, as already
discussed referring to \figref{fig:QFIplot}(\textbf{a}) and (\textbf{b}).

\begin{figure}[t!]
\includegraphics[width=\columnwidth]{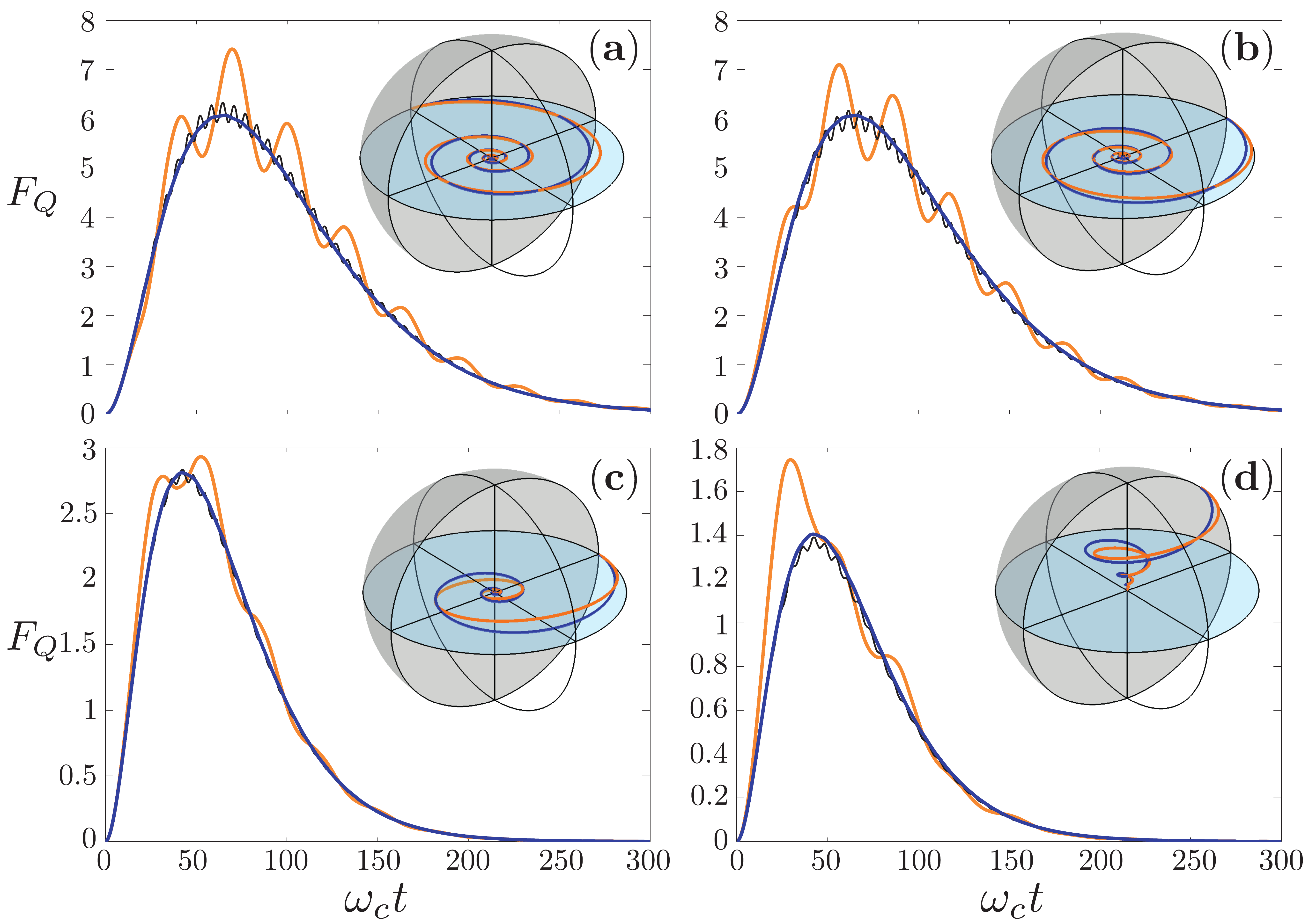}
\caption{
\textbf{Difference at finite times between the QFIs of NPC and PC dynamics for baths of Ohmic spectral density.}
Time evolution of the QFI for different initial states and values of $\Cang$. The NPC curves are shown in orange (light grey) and black for, respectively, $\omega_0 =1$ and $\omega_0=5$, 
while the PC curve after the secular approximation is shown in blue (dark grey) and it describes both the case of $\omega_0 =1$ and $\omega_0=5$; the noise parameter is $\lambda/\beta=0.1$. The insets show the corresponding evolution of the Bloch vector, here NPC in orange (light grey), PC in blue (dark grey). The initial conditions are the following: a) $\phi=\pi/2,\;\theta=\pi/2\;\Cang=0$, b) $\phi=0,\;\theta=\pi/2\;\Cang=0$, c) $\phi=0,\;\theta=\pi/2\;\Cang=\pi/4$, d) $\phi=0,\;\theta=\pi/4\;\Cang=\pi/4$.}
\label{fig:finiteQFI}
\end{figure}
%

\subsection{Finite-time analysis for the Ohmic spectral density}
\label{sec:ft}
In this paragraph we examine the behavior of the QFI for finite times, when the dynamics are dictated by the master equation expressed in Eqs.~\eqref{eq:finiteME} and \eqref{eq:masterEquationohmht}.
This will allow us to analyze more in detail the difference between the NPC and the PC
contributions to the QFI. 
The results presented in this section are numeric, calculated using \eqnref{eq:QFInori} and the same parametrization of the Bloch vector as before. \figref{fig:finiteQFI} contains the foundation of the following discussion. 

Let us first note that the dependence on the initial phase $\phi$ already mentioned above affects the whole time evolution of the NPC-QFI. \figsref{fig:finiteQFI}(\textbf{a}-\textbf{b}) show the evolution with $\Cang=0$ 
for an initial state in an equally weighted superposition, i.e., a state in the $\hat{x}-\hat{y}$ plane of the Bloch sphere ($\theta=\pi/2$), but with initial phases $\phi=\pi/2$ and $\phi=0$, respectively. 
Comparing the two figures, one observes that the initial phase is of no relevance for PC dynamics on the whole timescale, while the NPC dynamics introduces a dependence on $\phi$. 
The NPC contributions enhance the maximum value of the QFI and shift its position, depending on the value of the initial phase. 

For the Ohmic spectral density considered here, the noise terms do not depend on $\omega_0$, see \secref{sec:ohm},
so that $\tilde{F}_{Q}(t) = F_{Q}(t)$ and the same result holds for the PC case.
Hence, $F_{Q,\rm{PC}}(t)$ is directly fixed by the distance of the Bloch vector from the $z$-axis, along with the elapsed time $t$, see \eqnref{eq:qfipc1} in \appref{app:QFIan},
while the further contributions within the NPC-QFI $F_{Q}(t)$ can be fully ascribed to the non-commutativity of the Hamiltonian and dissipative part, see the discussion in the previous paragraph. 

While the independence of the QFI from the parameter to be estimated in the PC case can be readily shown \cite{Kolodynski2013,Smirne2016},
we can see from \figref{fig:finiteQFI} that the NPC-QFI depends on $\omega_0$. 
In particular, with growing values $\omega_0$, the NPC-QFI converges to its PC counterpart: higher values of $\omega_0$ imply a faster free dynamics of the system, 
which thus reduces the relevance of NPC terms and increases the validity of the secular approximation, see \secref{sec:blv}. 

We further observe that the overall effect of the NPC terms can yield an increase or a decrease of the QFI, depending on the time interval considered.
On the one hand, the NPC terms induce a contraction in the $x$-$y$ plane, which is no longer isotropic. Comparing the evolution of the QFIs in \figsref{fig:finiteQFI}({\textbf{a}-\textbf{b)} with the evolution of the Bloch vector in the insets, it is
clear how the non-isotropic contractions can bring the Bloch vector further or closer to the $z$-axis,
thus increasing or decreasing the QFI. 
On the other hand, as mentioned in the previous paragraph, due to the non-commutativity of the dynamics additional information about $\omega_0$ is enclosed in other features of the Bloch vector;
the action of decoherence itself adds 
some information about $\omega_0$ to the information imprinted by the rotation about the $z$-axis given by the Hamiltonian encoding.

The delicate interplay of the different mechanisms of production and annihilation of the QFI is also illustrated in \figsref{fig:finiteQFI}(\textbf{c}-\textbf{d}).
Here we consider values of $\Cang$ different from 0, so that the states initially on the equator of the Bloch sphere are no longer confined to the $xy$-plane. 
Comparing \figsref{fig:finiteQFI}(\textbf{b}) and (\textbf{c}), we see how the introduced NPC rotation partially counterbalances the oscillations due to the 
non-isotropic contraction. Furthermore, the role of the different NPC terms strongly depends on the initial state.
As an example, \figref{fig:finiteQFI}(\textbf{d}) shows the strongest (relative) enhancement of the maximum value of the QFI
due to the action of both the NPC rotations and contractions.

\section{$N$-probe quantum Fisher information and achievable metrological limits}
\label{sec:QFIBounds}
\begin{figure*}[t!]
 \includegraphics[width=1\textwidth]{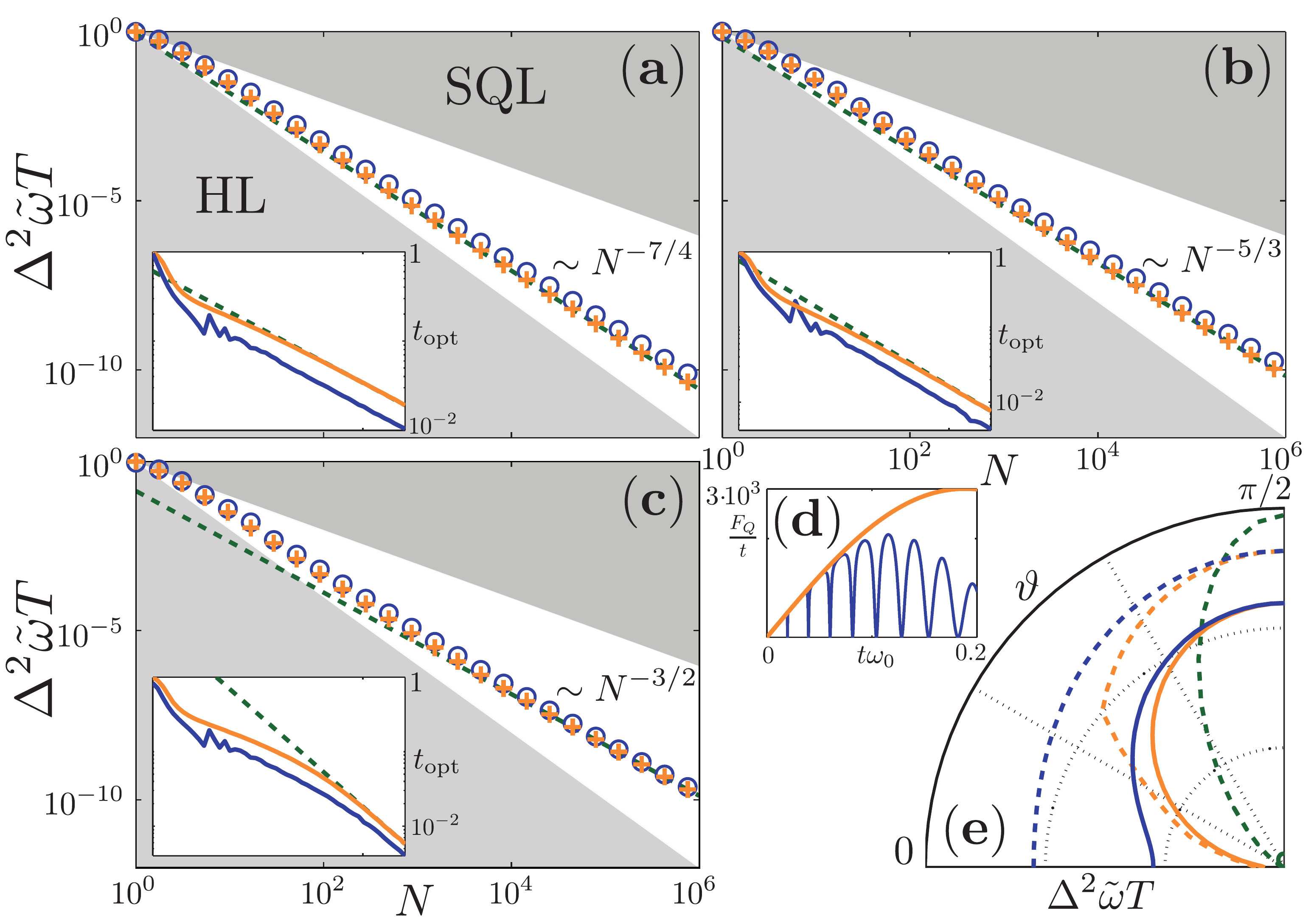}
\caption{%
\textbf{Sensing with N probes in parallel undergoing NPC Ohmic dynamics \eref{eq:finiteME}}. 
The panels (\textbf{a}) to (\textbf{c}) show the MSE as functions of $N$ that is 
attained with the parity measurement and GHZ inputs, $\Delta^2\tilde{\omega}_{0,P} T$ of \eqnref{eq:parityPrecision} (\emph{blue circles}),
in comparison with the general lower bound on the error, $t/F^{\uparrow}_{Q}\left[\rho^{(N)}(t)\right]$
with $F_Q^{\uparrow}$ defined in \eqnref{eq:channelQFI} (\emph{orange crosses});~%
both minimised over the round duration $t$ with corresponding optimal $t_\mathrm{opt}$
plotted within the insets (in matching colours). In cases, (\textbf{a}) $\Cang=0$ and $\omega_c=10$, (\textbf{b}, semigroup) for $\Cang=0$ and $\omega_c\rightarrow\infty$, 
and (\textbf{c}) $\Cang=\pi/100$ and $\omega_c=10$; in all the three cases $\alpha=1$. All the curves are normalized with respect to their
values at $N=1$; the \emph{grey} areas mark the regions below the HL and above the SQL scalings respectively, while the \emph{green (dashed)} line follows the 
scaling $N^{-\eta}$ , with the different $\eta$ denoting the asymptotic scaling observed.
The panel (\textbf{d}) shows the ratio $F_{Q}^{\uparrow}/t$ (\emph{orange}) for the upper bound on the QFI and 
the inverse error $1/(\Delta^2\tilde{\omega}_{0,P}  T)$ \emph{[blue (dark grey)]} for $N=160$ and the same parameters shown in (\textbf{a}). 
Panel (\textbf{e}) illustrates the dependency of the MSE on $\Cang$ for $N=160$ as a polar plot. 
\emph{Solid} lines correspond to $t/F^{\uparrow}_{Q}\left[\rho^{(N)}(t)\right]$, while \emph{dashed} lines represent
$\Delta^2\tilde{\omega}_{0,P}  T$. NPC noise is coloured in \emph{orange}, PC noise in \emph{blue} and the NPC semigroup limit in \emph{green}. 
Note that the lines for the semigroup cases are reduced by a factor $200$.}
\label{fig:NprobeFI}
\end{figure*}

In this final section, we want to explore the QFI for an estimation utilizing multiple probes, up to the asymptotic limit  $N \rightarrow  \infty$. 
In this way, we will also provide a complete picture for the model at hand of the different
scalings of the error in the presence of noise, including semigroup or non-semigroup noise,
as well as phase-covariant or non phase-covariant one.

As recalled in \secref{sec:qcrb}, evaluating the QFI becomes a more and more difficult task,
with the increasing of the dimensionality of the probing system. However, 
since we are assuming a non-interacting probe system subject to independent and identical noise,
we can exploit the finite-$N$ channel extension method \cite{Demkowicz2012,Kolodynski2013}.
Given the Kraus representation of the dynamical map $\Lambda(t)$ of a single probe, i.e.,
\begin{equation}
\Lambda(t)\left[\rho\right]=\sum_i K_i(t) \rho K^\dagger_i(t),
\end{equation}
the QFI of the resulting $N$-probe state can be bounded from above by the relation 
\begin{eqnarray}
	F_Q \left[\rho^{(N)}(t) \right] &\leq& F^{\uparrow}_Q \left[\rho^{(N)}(t) \right] \label{eq:channelQFI}\\
	&& \equiv 4 \min_{\tilde{K}} \left[N|| \alpha_{\tilde{K}}(t) ||+N(N-1) || \beta_{\tilde{K}}(t) ||^2 \right],\nonumber
\end{eqnarray}
which, along with the QCRB, directly provides us with a lower bound to the estimation error, i.e. the MSE of \eqnref{eq:QCRB}.
The minimum in \eqnref{eq:channelQFI} is taken over all Kraus representations, connected via 
a unitary transformation according to $\tilde{K_i}(t) = \sum_j	u_{ij}(t) K_j(t) $,
while the unitary transformation will generally depend on $\omega_0$ as well. We also introduced the quantities $\alpha_{\tilde{K}}(t) = \sum_j \dot{\tilde{K}}_j^\dagger(t)\dot{\tilde{K}}_j(t)$, $\beta_{\tilde{K}}=i \sum_j \dot{\tilde{K}}_j^\dagger\tilde{K}_j$ and recall that the dot notation represents a derivative with respect to the parameter $\omega_0$. We remark that this bound is already optimized over all possible input states and can hence be calculated without specifying both concrete preparation and measurement procedures. 
Furthermore, the optimization can be cast into a semidefinite programming task, which allows for an efficient numerical evaluation, see \refcite{Kolodynski2013}. 

In addition, to investigate the attainability of the bound, we will consider a measurement of the parity operator $P_x = \bigotimes_{k=1}^N \sp_x^{(k)}$
\cite{Ma2011}. 
Using the error propagation formula and since $\Delta^2 P_x = \langle P_x^2\rangle - \langle P_x \rangle ^2 = 1-\langle P_x \rangle^2$, 
the error $\Delta^2\tilde{\omega}_{0,P}$ under parity measurement reads \cite{Brask2015}:
\begin{eqnarray}
\Delta^2\tilde{\omega}_{P}  T = 
t \frac{1-\langle P_x(t)\rangle^2}{\left| \langle \dot{P}_x(t)\rangle \right|^2}.
\label{eq:parityPrecision}
\end{eqnarray}    
In particular, focusing on an initial GHZ state,
one finds
\begin{eqnarray}
\langle P_x(t) \rangle &=& \frac{1}{2} \left\lbrace\left[\xi(t)+i\chi(t)\right]^N \right.\nonumber\\
&&\left.+ \left[\xi(t)-i\chi(t)\right]^N + \left[1-(-1)^N\right] \varsigma(t)^N\right\rbrace,\label{eq:parityPrecision2}
\end{eqnarray}
where $\xi(t)$, $\chi(t)$ and $\varsigma(t)$ are proper time- and frequency-dependent functions obtained as in \cite{Brask2015},
from which \eqnref{eq:parityPrecision} can be evaluated.  
Note that the last term only contributes if $N$ is an even number and hence the precision may heavily change when $N$ is changed by one. 
However, for all the cases examined here, we have $\varsigma(t) = 0$. 

We focus on the case of an Ohmic spectral density, which
provides us with numerically easily solvable differential equations for any time $t$, cut-off frequency $\omega_C$ and coupling strength. Furthermore, by taking the limit $\omega_C\rightarrow\infty$ we recover the semigroup limit as mentioned in Sec.~\ref{sec:ohm} and \appref{app:semigroupLimit}, which will be useful to compare our results to those already known in the literature. 

\subsection{Asymptotic scaling of the ultimate estimation precision}\label{sec:tab}

\begin{table}[t!]
	\begin{center}
		\begin{tabular}{c||cccc}
			$\eta$	& NPC & PC & NPC, semigroup & PC, semigroup \\ 
			\hline \hline
			$\Cang = 0$ & $7/4$ & $3/2$ & $5/3$ & $1$ \\
			$\Cang \not= 0$ & $3/2$ & $3/2$ & $1$ & $1$ \\
		\end{tabular} 
	\end{center}
\caption{%
\textbf{Ultimate scaling exponent}, $\eta$ in \eqnref{eq:turbo_bound}, of the optimal estimation error $\Delta^2\tilde{\omega}_0T$ for different types of noise in the asymptotic limit of $N\rightarrow\infty$.
}
\label{tab:scalingExponentsOhmic}
\end{table}

The starting point is the master equation given by \eqnref{eq:finiteME}. In particular, 
we considered three different NPC noise scenarios:
the first two cases of a purely transversal noise, i.e., $\Cang=0$, for a non-semigroup [see \figref{fig:NprobeFI}(\textbf{a})]
and for a semigroup [\figref{fig:NprobeFI}(\textbf{b})] dynamics. As a third case, we chose noise with a (small) longitudinal component fixed by $\Cang=\pi/100$ for a non-semigroup dynamics [see \figref{fig:NprobeFI}(\textbf{c})].  
In \figref{fig:NprobeFI}(\textbf{a}-\textbf{c}) we report the numerical study of $t/F^{\uparrow}_{Q}\left[\rho^{(N)}(t)\right]$,
which fixes a lower bound to the estimation error, see \eqnsref{eq:QCRB}{eq:channelQFI},
along with the estimation error for the parity measurement,  $\Delta^2\tilde{\omega}_{0,P}  T$, see \eqnref{eq:parityPrecision}. 
As clearly observed in \figref{fig:NprobeFI}, the two quantities have
the same asymptotic scaling, therefore the bound is achievable, at most up to a constant factor. Hence we can infer the scaling with respect to $N$ of the error for the optimal estimation strategy. 
Denoting the latter as $\Delta^2\tilde{\omega}_{0}  T$, we can in fact write
the lower and upper bounds as
\begin{equation}
\Delta^2\tilde{\omega}_{0,P}  T\geq\Delta^2\tilde{\omega}_{0}  T\geq\frac{t}{F^{\uparrow}_{Q}\left[\rho^{(N)}(t)\right]}
\;\Rightarrow\;
\Delta^2\tilde{\omega}_{0} T \propto \frac{1}{N^{\eta}},
\label{eq:turbo_bound}
\end{equation}
where the implication follows from the fact that 
since both the lower and the upper bound approach 0 for $N \rightarrow  \infty$ as $N^{-\eta}$
with the same value of $\eta$, this will be the case also for $\Delta^2\tilde{\omega}_{0}  T$.

Table \ref{tab:scalingExponentsOhmic} contains the values of the optimal scaling $\eta$ for the different NPC noise scenarios
as inferred from our numerical
analysis, along with the corresponding PC scaling behavior (i.e., those for the dynamics after the secular approximation,
see \eqnref{eq:masterEquationohmht}) taken from \cite{Kolodynski2013,Smirne2016}.
The optimal scaling of the estimation error for the full NPC dynamics is fixed by two key features:
Whether we have a semigroup or a non-semigroup evolution
and the direction of the noise fixed by the angle $\Cang$.
The presence of a time-dependent rate $\gamma(t)$ as in \eqnref{eq:ohmf}
always leads to an improved scaling, with respect to the constant rate $\gamma$
of the semigroup evolution; in particular,
for any $\Cang\neq0$ we have the Zeno $\eta=3/2$ scaling, associated with the linear increase of the rate $\gamma(t)$
for short times \cite{Chin2012,Smirne2016}. 
Moreover, we numerically find the novel \textit{$\eta = 7/4$ scaling for a non-semigroup, purely transversal noise}.

We stress that for any value of $\Cang$ different from 0 the full NPC dynamics leads to the same scaling behavior
as in the corresponding PC case.
We can say that the transversal noise represents
a special case of NPC noise, which might be seen as a "purely
NPC noise". For any $\Cang\neq 0$, the dissipative part of the master equation given in \eqnref{eq:finiteME} together with the resulting dynamical maps,
will have a component longitudinal to the parameter imprinting, fixing the asymptotic scaling
to the less favorable one proper to PC dynamics and hence extending the Zeno regime recalled in \secref{sec:nfe} to the scenario governed by NPC noise. 
This result, already known for the semigroup regime \cite{Chaves2013} (see also \secref{sec:nfe}),
is here extended to the non-semigroup case.
Summarizing, we can conclude that the ultimate achievable estimation precision can overcome
the SQL whenever we have a non-semigroup (short-time) evolution, irrespective of the direction of the noise,
or in the cases where we have a purely transversal noise, irrespective of whether we have a semigroup or not.

Interestingly, similar results have been derived recently \cite{Jarzyna2017} for rather different, infinite dimensional probing systems. The
probes are prepared in Gaussian states and undergo a Gaussian dynamics,
possibly non-semigroup and NPC. The NPC contributions are induced by the presence of squeezing in the initial bath state.
Also there the optimal asymptotic scaling of the error is also found to be the same for PC and NPC 
dynamics, going from the SQL
for a semigroup to the Zeno limit for a linear increase of the dissipative rates.
Such a transition for a PC evolution of a Gaussian system has been shown also in \cite{Latune2016}.

\subsection{Finite-$N$ behavior}

The plots in \figref{fig:NprobeFI}(\textbf{a}-\textbf{c}) allow us to get some interesting information also
about the behavior of the estimation error for a finite number of probes,
showing that the asymptotic scaling is approached in a possibly non-trivial way.

First of all, we note that for smaller values of $N$, the lower bound to the estimation precision $t/F^{\uparrow}_{Q}\left[\rho^{(N)}(t)\right]$ 
and the error under parity $\Delta^2\tilde{\omega}_{0,P}  T$ seem to follow the SQL and then, only
for intermediate and high values of $N$, the two quantities converge to the asymptotic behavior,
approaching it always from above.
This was already shown for a semigroup NPC noise, also with a longitudinal component (see \cite{Chaves2013}, in particular \figref{fig:dynamicCompare})
and here we see how the same happens for a non-semigroup NPC noise.
Actually, the effect is even more pronounced for a non-semigroup non-transversal noise, where the 
asymptotic behavior emerges only if almost $10^4$ probes are used, see \figref{fig:NprobeFI}(\textbf{c}).
Additional numerical studies (not reported here) show that the asymptotic scaling is approached earlier when the coupling to the bath is increased.
Even if it is clear that the finite-$N$ behavior do not spoil the validity of the different scalings pointed out in the previous
paragraph, it should also be clear the relevance of such behavior in many experimental frameworks, when, indeed, the high-$N$
regime might be not achievable. 
In such situations, the experimental data would follow a scaling
which is different from the asymptotic one for all practical purposes.

In addition, the behavior of the estimation error for finite values of $N$ provides us with a more complete understanding of the specific role
played by the geometry of the noise, i.e. the coupling angle $\Cang$.
In \figref{fig:NprobeFI}(\textbf{e}) we study $t/F^{\uparrow}_{Q}\left[\rho^{(N)}(t)\right]$ and $\Delta^2\tilde{\omega}_{0,P}  T$, 
but
now for different values of $\Cang \in [0,\pi/2]$ and a fixed number of probes $N=160$.
For this value of $N$ and $\Cang=0$ the two quantities have essentially already reached their asymptotic values,
see \figref{fig:NprobeFI}(\textbf{a}), while this is not the case for $\Cang\neq0$, see \figref{fig:NprobeFI}(\textbf{c}).
Now, \figref{fig:NprobeFI}(\textbf{d}) shows how both
$t/F^{\uparrow}_{Q}\left[\rho^{(N)}(t)\right]$ and $\Delta^2\tilde{\omega}_{0,P}  T$
change continuously with the variation of $\Cang$. They increase from $\Cang=0$ up to $\Cang=\pi/2$, with the increment 
being more pronounced for values of $\Cang$ close to $0$. 
The sudden transition between different scalings for, respectively, $\Cang=0$ and $\Cang\neq0$
is a peculiarity of the asymptotic limit, $N\rightarrow\infty$. Furthermore, this also confirms that noise in the direction of the paramater imprinting is more detrimental than any other direction, if the absolute noise strength is kept identical.

As a final remark, note that the optimal time of the estimation error for a parity measurement as a function of $N$ has discrete jumps between smooth periods, see the lower insets in \figsref{fig:NprobeFI}(\textbf{a}-\textbf{c}). These jumps originate from the fact that $\Delta^2\tilde{\omega}_{0,P}  T$ does possess multiple local maxima instead of one global maxima as $t/F^{\uparrow}_{Q}\left[\rho^{(N)}(t)\right]$ does, see \figref{fig:NprobeFI}(\textbf{d}). The jump occurs when the global maximum of $\Delta^2\tilde{\omega}_{0,P}  T$ changes to a different peak, which was only a local maximum before. 
On the other hand, for large values of $N$ $\Delta^2\tilde{\omega}_{0,P}  T$ will converge to a function with only one local maximum, 
as the following ones have been damped off, so that the optimal time will stay a smooth function of $N$. 
The jumps in the optimal evaluation time for a parity measurement can
be observed also in the polar plot in \figref{fig:NprobeFI}(\textbf{e}), in terms of non-smooth
variation as a function of $\Cang$.

\section{Conclusions}

We have exploited a detailed analysis of the spin-boson model, which is a general, well-known and widely used noise model, to investigate how the 
ultimate achievable limits to frequency estimation are affected by the
different microscopic features of the interaction between the quantum probes and their environment.   
Hence, we used common tools of the theory of open quantum systems
to extend the characterization of noisy quantum metrology beyond the common framework,
where the description of the noise is usually postulated on a phenomenological basis.

First, we derived the master equation fixing the dynamics of the probes, employing the second order TCL expansion. 
Thereby, we clarified that the distinction between phase-covariant and non-phase-covariant noise,
which plays a key role in frequency estimation \cite{Smirne2016}, corresponds to the distinction between secular and non-secular dynamics. Moreover, we characterized explicitly the dependence of the noise rates, as well as of the correction
to the system Hamiltonian, on the free frequency of the probes, i.e., on the parameter to be estimated.
This is another aspect commonly overlooked in phenomenological approaches to noisy metrology.

Then, employing a solution to the master equation in the short time regime,
valid for any spectral density,
and a solution on the whole time scale for an Ohmic spectral density,  
we investigated the single probe QFI and hence how the microscopic details
of the model influence the estimation precision. 
In particular, we compared the differences between the effects of, respectively, phase-covariant and non-phase-covariant
dynamics. The non-secular contributions can both increase or decrease the QFI,
also depending on the initial condition, as they lead to a dependence of the QFI
on the initial phase of the probes state. However, in general, the maximum (over time)
QFI is higher in the non-phase covariant case, due to the positive contributions induced
by the non-commutativity of the noise and the free Hamiltonian.
Furthermore, we examined the mentioned dependence of the noise terms on the estimated frequency. While for non-secular dynamics no definite statement can be made, we found that this dependence is always beneficial for secular dynamics.

In the last part of the paper, 
we moved to the regime of multiple probes and gave a complete characterization of the possible asymptotic scalings of the estimation precision, putting results already existing in the literature onto a common ground, as well as exploring new regimes. In particular, we extended the validity of the super-classical Zeno scaling $N^{-3/2}$ onto non-phase-covariant, non-semigroup dynamics, as long as $\Cang \ne 0$. Furthermore, we identified the novel $N^{-7/4}$ scaling for  $\Cang \ne 0$, i.e., for a non-phase-covariant and non-semigroup dynamics, due to a coupling with the environment fully orthogonal to the direction of the encoding of the parameter. 

Concluding, our analysis offers a complete and physically motivated characterization of the scenarios where one can actually achieve super-classical precision in frequency estimation in the presence of (independent) noise. 
In addition, the microscopic characterization of the probes dynamics enabled us to present an in depth study of the influence of the microscopic details 
of the probe-environment interaction on the precision. 
The adopted scheme can be directly linked to widely used sensing scenarios as exploited with color-centers in diamond, superconducting qubits or optomechanical setups.

\acknowledgments AS would like to thank Walter Strunz for very useful discussions and for pointing out the ubiquitous dependence of the 
noise terms on the system frequency. This work was supported by the EU through STREP project QUCHIP and the ERC through the Synergy grant BioQ. J.K. acknowledges support from the Spanish MINECO (QIBEQI FIS2016-80773-P 
and Severo Ochoa SEV-2015-0522), Fundacio Cellex and Generalitat de Catalunya 
(SGR875 and CERCA Program). R.D.D acknowledges support from National Science Center (Poland) grant No. 2016/22/E/ST2/00559.

%

%
\onecolumngrid
\begin{appendix}
\section{Equivalence with and engineered coupling Hamiltonian}
\label{app:HamEngine}
Despite the fact that the Hamiltonian given in \eqnref{eq:Hamiltonian} can arise as the natural model for specific systems, we can also engineer this type of coupling out of a pure dephasing spin boson Hamiltonian by a continuous driving of the central spin. Therefore consider the Hamiltonian
\begin{eqnarray}
\tilde{H} = \frac{\omega\sp_z}{2} + \frac{\Omega \sp_x \cos\omega_L t}{2} - \frac{\sp_z}{2} \otimes \sum_{n}\left(g_n a_n + g_n^* a_n^\dagger\right) + H_B, 
\end{eqnarray}
where $\Omega$ is the associated Rabi frequency of the driving with the frequency $\omega_L$, e.g. these correspond to amplitude and frequency of a driving laser. In a frame rotating with the frequency $\omega_L$ we employ the rotating wave approximation with respect to that frequency and arrive at
\begin{eqnarray}
\tilde{H}^\prime = \frac{\omega-\omega_L}{2}\sp_z + \frac{\Omega}{2}\sp_x - \frac{\sp_z}{2} \otimes \sum_{n}\left(g_n a_n + g_n^* a_n^\dagger\right) + H_B.
\end{eqnarray}
Inserting the substitutions $\omega-\omega_L = -\omega_0 \sin \Cang$, $\Omega = - \omega_0 \cos \Cang$ and transforming the Hamiltonian with the help of the unitary matrix 
\begin{eqnarray}
U=\left(
\begin{array}{cc}
 \frac{\sec (\Cang ) (\sin (\Cang )-1) \sqrt{\sin (\Cang )+1}}{\sqrt{2}} & \frac{\sqrt{1-\sin (\Cang )} (\sec (\Cang )+\tan (\Cang ))}{\sqrt{2}} \\
 \frac{\sqrt{\sin (\Cang )+1}}{\sqrt{2}} & \frac{1}{\sqrt{2} \sqrt{\sec (\Cang ) (\sec (\Cang )+\tan (\Cang ))}} \\
\end{array}
\right),
\end{eqnarray}
directly yields the Hamiltonian \eqref{eq:Hamiltonian}, $H=U^\dagger\tilde{H}^\prime U$, described in the main text. Note that due to the linearity of the substitutions and the parameter-independent unitary transformation,
if one knows the driving laser frequency and amplitude, $\omega_L$ and $\Omega$, the
parameter estimation of $\omega_0$ is fully equivalent to the estimation of $\omega$.

\section{Derivation of the weak-coupling master equation}\label{app:1}
In this Appendix, we briefly sketch the derivation of the weak-coupling master equation in \eqnref{eq:masterEquation} and we provide the expression of the coefficients $b_{j k}(t)$ where $j, k =\{+,-,z\}$, as well as the correction to the Hamiltonian $H^{\mathrm{LS}}(t)$ in terms of the bath correlation function $C(t)$.

Recall that we start from \eqnref{eq:TCLmasterEquation}, which is obtained as the second order term in the expansion of the TCL master equation in the interaction picture, assuming an initial product state but without any assumption about the form of the global state at time $t$ \cite{Rivas2010}. The master equation is then readily obtained following the derivation described at pages 128-129 in \cite{Breuer2001}, the only difference being that we keep the integration at the r.h.s.~of \eqnref{eq:TCLmasterEquation} from $0$ to $t$, since we are 
\emph{not} making the Born-Markov approximation. Hence, following \cite{Breuer2001}, we expand the system operator in the interaction Hamiltonian in \eqnref{eq:Hamiltonian}, i.e.,
\begin{eqnarray}
A =\left(\cos\Cang \frac{\sp_x}{2}+\sin\Cang \frac{\sp_z}{2}\right),
\label{eq:A}
\end{eqnarray} 
via the projectors in the eigenspaces of the system free Hamiltonian,
\begin{eqnarray}
H_0 = \sum_{\epsilon} \epsilon \Pi(\epsilon),
\end{eqnarray}
where $\epsilon_1= \omega_0/2$, $\epsilon_2=-\omega_0/2$, $\Pi_1= \ketbra{1}{1}$ and $\Pi_2= \ketbra{0}{0}$. Thus, we define
\begin{eqnarray}
A(\varsigma) = \sum_{\epsilon'-\epsilon = \varsigma} \Pi(\epsilon) A \Pi(\epsilon'),
\label{eq:defasigm}
\end{eqnarray}
and we have 
\begin{eqnarray}
A = \sum_{\varsigma} A(\varsigma) = \sum_{\varsigma} A^{\dag}(\varsigma).
\label{eq:using}
\end{eqnarray}
Note that in our case $\varsigma$ can take the values $\pm \omega_0$ and $0$. Explicitly, 
\begin{eqnarray}
A(0) &=& \frac{\sin\Cang}{2} \sp_z; \nonumber\\
A(-\omega_0) &=& \frac{\cos\Cang}{2} \sp_+ \nonumber\\
A(\omega_0) &=& \frac{\cos\Cang}{2} \sp_-.
\label{eq:defasigm2}
\end{eqnarray}

The decomposition of the interaction operator in \eqnref{eq:using} allows us to express the interaction Hamiltonian $H_I(t)$ as
\begin{eqnarray}
H_I(t) = \sum_{\varsigma} e^{-i \varsigma t} A(\varsigma)\otimes B(t) = \sum_{\varsigma} e^{i \varsigma t} A^{\dag}(\varsigma)\otimes B(t).
\end{eqnarray}
Replacing these expansions in \eqnref{eq:TCLmasterEquation}, using  \eqnref{eq:defasigm2} and replacing the integration variable $\tau$ with $t-\tau$, one arrives at \cite{Breuer2001}
\begin{eqnarray}
\frac{\mathrm{d}}{\mathrm{d} t} \tilde{\rho}(t) &=& \sum_{\varsigma \varsigma'} e^{i(\varsigma'-\varsigma) t} \Gamma(\varsigma, t) \left(A(\varsigma) \tilde{\rho}(t) A^{\dag}(\varsigma')
- A^{\dag}(\varsigma') A(\varsigma) \tilde{\rho}(t)\right) + \mbox{h.c} \nonumber\\
&=& \Gamma(0, t) \left[\frac{\sin\Cang\cos\Cang}{4} \left(e^{-i \omega_0 t}  \left(\sp_z \tilde{\rho}(t) \sp_- - \sp_- \sp_z \tilde{\rho}(t)\right) 
+e^{i \omega_0 t}  \left(\sp_z \tilde{\rho}(t) \sp_+ - \sp_+ \sp_z \tilde{\rho}(t)\right)\right)+
\frac{\sin^2\Cang}{4}  \left(\sp_z \tilde{\rho}(t) \sp_z - \tilde{\rho}(t)\right)\right]\nonumber\\
&&+ \Gamma(-\omega_0, t) \left[\frac{\cos^2\Cang}{4} \left(\sp_+ \tilde{\rho}(t) \sp_- - \sp_- \sp_+ \tilde{\rho}(t)
+e^{2 i \omega_0 t}  \sp_+ \tilde{\rho}(t) \sp_+ \right)+
\frac{\sin\Cang\cos\Cang}{4}e^{i \omega_0 t}  \left(\sp_+ \tilde{\rho}(t) \sp_z - \sp_z\sp_+\tilde{\rho}(t)\right)\right]\nonumber\\
&&+ \Gamma(\omega_0, t) \left[\frac{\cos^2\Cang}{4} \left(e^{-2 i \omega_0 t}\sp_-\tilde{\rho}(t) \sp_- 
+  \sp_- \tilde{\rho}(t) \sp_+- \sp_+ \sp_- \tilde{\rho}(t) \right)+
\frac{\sin\Cang\cos\Cang}{4}e^{-i \omega_0 t}  \left(\sp_- \tilde{\rho}(t) \sp_z - \sp_z\sp_-\tilde{\rho}(t)\right)\right] + \mbox{h. c.}.
\label{eq:3132}
\end{eqnarray}
Here $\mbox{h.c.}$ stands for hermitian conjugate and we introduced the functions
\begin{eqnarray}
\Gamma(\varsigma, t) = \int^t_{0} \mathrm{d} \tau e^{i \varsigma \tau} C(\tau).
\label{eq:gamm}
\end{eqnarray}
Recall that $C(t)$ is defined in \eqnref{eq:ctt}. We still need to separate the Hamiltonian and the dissipative contributions of the dynamics. Before doing so, we go back to the Schr{\"o}dinger picture via $\tilde{\rho}(t)=e^{i H_0 t} \rho(t) e^{-i H_0 t}$, which adds a contribution $-i \left[H_0, \rho(t)\right]$ and removes all the phase terms $e^{\pm i \omega_0 t}$ and $e^{\pm 2 i \omega_0 t}$ in the previous equation (since $e^{-i H_0 t} \sp_{\pm} e^{i H_0 t} = e^{\mp i \omega_0 t} \sp_{\pm}$, while $e^{-i H_0 t} \sp_{z} e^{i H_0 t} = \sp_{z}$). If we now define
\begin{eqnarray}
c_{\varsigma \varsigma'}(t) &=& \Gamma(\varsigma, t) + \Gamma^*(\varsigma',t) \nonumber\\
d_{\varsigma \varsigma'}(t) &= &\frac{1}{2 i} \left( \Gamma(\varsigma, t) - \Gamma^*(\varsigma',t)\right),
\label{eq:cdcd}
\end{eqnarray}
we can write \eqnref{eq:3132} in the Schr{\"o}dinger picture as
\begin{eqnarray}
\frac{\mathrm{d}}{\mathrm{d} t} \rho(t) &=& -i \left[H_0 + \sum_{\varsigma, \varsigma'}d_{\varsigma \varsigma'}(t) A^{\dag}(\varsigma')A(\varsigma), \rho(t)\right] 
+ \sum_{\varsigma, \varsigma'}c_{\varsigma, \varsigma'}(t) \left(A(\varsigma) \rho(t) A^{\dag}(\varsigma') -\frac{1}{2}\left\{A^{\dag}(\varsigma')A(\varsigma), \rho(t)\right\}\right),
\label{eq:3132bis}
\end{eqnarray}
which can be written as the master equation~\eqref{eq:masterEquation} in the main text, when \eqnref{eq:defasigm2} is used. Thereby, exploiting Eq. \eqref{eq:cdcd}, the coefficients are fixed as in \eqnref{eq:coeffmain}, i.e., 
\begin{eqnarray}
b_{zz}(t)&=&\frac{\sin^2\Cang}{2}\Re{\Gamma(0,t)}=\frac{\sin^2\Cang}{2}\int_0^t \mathrm{d}\tau\; \Re{C(\tau)} \nonumber\\
b_{++}(t)&=&\frac{\cos^2\Cang}{2}\Re{\Gamma(-\omega_0,t)}=\frac{\cos^2\Cang}{2}\int_0^t \mathrm{d}\tau \;\Re{C(\tau)e^{-i\omega_0\tau}} \nonumber \\
b_{--}(t)&=&\frac{\cos^2\Cang}{2}\Re{\Gamma(\omega_0,t)}=\frac{\cos^2\Cang}{2}\int_0^t \mathrm{d}\tau\;\Re{C(\tau)e^{i\omega_0\tau}} \nonumber \\
b_{+-}(t)=b_{-+}^*(t)&=& \frac{\cos^2\Cang}{4}\left(\Gamma(-\omega_0,t)+\Gamma^*(\omega_0,t)\right) \nonumber \\ 
&=&\frac{\cos^2\Cang}{2}\int_0^t \mathrm{d}\tau\;\Re{C(\tau)}e^{-i\omega_0\tau}\nonumber \\
b_{z+}(t)=b_{+z}^*(t)&=&\frac{\sin\Cang\cos\Cang}{4}\left(\Gamma(0,t)+\Gamma^*(-\omega_0,t)\right) \nonumber \\ 
&=&\frac{\sin\Cang\cos\Cang}{4}\int_0^t\mathrm{d}\tau \left[C(\tau)+e^{i\omega_0\tau}C^*(\tau)\right] \nonumber \\
b_{z-}(t)=b_{-z}^*(t)&=&\frac{\sin\Cang\cos\Cang}{4}\left(\Gamma(0,t)+\Gamma^*(\omega_0,t)\right) \nonumber \\ 
&=&\frac{\sin\Cang\cos\Cang}{4}\int_0^t\mathrm{d}\tau \left[C(\tau)+e^{-i\omega_0\tau}C^*(\tau)\right],
\label{eq:coeff}
\end{eqnarray}
while the Hamiltonian contribution due to the interaction with the environment is given by $H^{\mathrm{LS}}(t)= \sum_{\varsigma, \varsigma'}d_{\varsigma \varsigma'}(t)A^{\dag}(\varsigma')A(\varsigma)$, which corresponds to \eqnref{eq:hhhmain} in the main text, that is,
\begin{eqnarray}
H^{\rm{LS}}(t)&=&\left(\begin{array}{cc} H^{\rm{LS}}_{11}(t) & H^{\rm{LS}}_{01}(t)\\
H^{\rm{LS}}_{01}(t) & H^{\rm{LS}}_{00}(t)
\end{array} \right),
\label{eq:hhh}
\end{eqnarray}
where
\begin{eqnarray}
H^{\text{LS}}(t)&=&\left(\begin{array}{cc} \frac{\cos^2\Cang}{4}\int_0^t  \mathrm{d}\tau \;\Im{e^{i\omega_0 \tau}C(\tau)} &  i \frac{\sin\Cang \cos\Cang}{4}\int_0^t  \mathrm{d}\tau\; \Re{C(\tau)} (1-e^{-i\omega_0\tau})  \\
- i\frac{\sin\Cang \cos\Cang}{4}\int_0^t  \mathrm{d}\tau \;\Re{C(\tau)} (1-e^{i\omega_0\tau})&    \frac{\cos^2\Cang}{4}\int_0^t  \mathrm{d}\tau\; \Im{e^{-i\omega_0 \tau}C(\tau)}
\end{array} \right).
\label{eq:hhh}
\end{eqnarray}
Summarizing, starting from the global Hamiltonian in \eqnref{eq:Hamiltonian}, after introducing the environmental correlation function $C(t)$ in \eqnref{eq:ctt} and the system's operators in \eqnref{eq:defasigm}, one directly gets the weak-coupling master equation via the Eqs.~\eqref{eq:gamm}-\eqref{eq:3132bis}.

\section{Solutions of the master equation in the high temperature limit}
\label{app:htl}
As said in the main text, we can use the approximation $j(\omega) \approx j(-\omega)$ to simplify the structure of the master equation in the high temperature regime. First, note that since $\mathcal{L}(t)$ is a linear map acting on the space of linear operators in $\mathbbm{C^2}$, we can represent it via a $4 \times 4$ matrix, using the same representation recalled in \secref{sec:blv}, see \eqnref{eq:matr}. In particular, the coefficients in the dissipative part of the generator as in \eqnref{eq:coefhight} imply the matrix representation of $\mathcal{L}(t)$ as
\begin{eqnarray}
\dynmat^{\mathcal{L}(t)} = \left( \begin{array}{cc} 
0 & \vec{0}^T \\
\vec{0} & L(t) 
\end{array}\right).
\label{eq:matrixGeneratorL}
\end{eqnarray}
Explicitly, using the definition of $\Gamma(\omega_0, t)$ and $\Gamma(0,t)$ in \eqnref{eq:gammmain}, as well as $H^{\rm{LS}}(t)$ in \eqnref{eq:hhh} and $j(\omega) \approx j(-\omega)$, we end up with 
\begin{eqnarray}
\dynmat^{\mathcal{L}(t)} &=&\left(
\begin{array}{cccc}
 0 & 0 & 0 & 0 \\
 0 & -   \sin^2(\Cang ) f_1(0,t)  &-\omega_0& 
 \cos (\Cang)\sin(\Cang) f_1(\omega_0,t)  \\
 0 &\omega_0+ \cos^2(\Cang)f_2(\omega_0,t) 
 & \begin{array}{l} - \sin ^2(\Cang )f_1(0,t) \\ - \cos^2(\Cang )f_1(\omega_0,t) \end{array}
 & \cos(\Cang)\sin(\Cang)f_2(\omega_0,t)  \\
 0 &\sin(\Cang)\cos(\Cang)f_1(0,t)  & 0
 &- \cos^2(\Cang )f_1(\omega_0,t),  \\
\end{array}
\right).\label{eq:genhight}
\end{eqnarray}
where
\begin{eqnarray}
f_1(\omega_0,t) &=&\int_{-\infty}^{\infty}\mathrm{d}\omega j(\omega) \frac{\sin (t (\omega-\omega_0))}{\omega-\omega_0} \nonumber\\
f_2(\omega_0,t) &=&\int_{-\infty}^{\infty}\mathrm{d}\omega j(\omega)\frac{\cos(t(\omega-\omega_0))-1}{\omega-\omega_0}.
\end{eqnarray}
Indeed, applying the same constraint on the secular master equation in \eqnref{eq:SecMasterEquation}, we get the PC master equation, where the coefficients in the last line of \eqnref{eq:coefhight} are set to 0, along with $H^{\rm{LS}}_{10}(t) = H^{\rm{LS}}_{01}(t)^*$ in \eqnref{eq:hhh}. The corresponding time-local generator is hence given by
\begin{eqnarray}
\dynmat^{\mathcal{L}(t)}_{\rm{PC}} &=&\left(
\begin{array}{cccc}
 0 & 0 & 0 & 0 \\
 0 & \begin{array}{l}-\sin^2(\Cang)f_1(0,t)\\-\frac{\cos^2(\Cang)}{2}f_1(\omega_0,t)\end{array}&-\omega_0-\frac{\cos^2(\Cang)}{2}f_2(\omega_0,t)& 
 0  \\
 0 &\omega_0+\frac{\cos^2(\Cang)}{2}f_2(\omega_0,t)
 & \begin{array}{l}-\sin^2(\Cang)f_1(0,t)\\-\frac{\cos^2(\Cang)}{2}f_1(\omega_0,t)\end{array}
 &0 \\
 0 &0& 0
 &- \cos^2(\Cang )f_1(\omega_0,t)  \\
\end{array}
\right).\label{eq:genhight2}
\end{eqnarray}
Now, the form of the time-local generator as in \eqnref{eq:matrixGeneratorL} implies the form for the dynamical map as in \eqnref{eq:matrixGenerator}. By means of, e.g., 
\eqnref{eq:mtl} we see that the block-diagonal structure of the generator directly implies the same block-diagonal structure of the dynamical map. Thus, we get \eqnref{eq:matrixGenerator} with
\begin{eqnarray}
V(t) = T_{\leftarrow} e^{\int_0^t d \tau L(\tau)}.
\end{eqnarray}
%

\subsection{Ohmic spectral density}\label{sec:appc2}
\subsubsection{Differential equations for the density matrix elements}
Due to the simple master equations in the Ohmic regime described in \secref{sec:ohm}, \eqnsref{eq:ohmf}{eq:masterEquationohmht}, it is more convenient to solve the dynamics taking into account the evolution of the elements of the system's density matrix, $\rho_{ij}(t) = \bra{i}\rho(t)\ket{j}$ for $i,j=1,2$.

For the NPC dynamics, the master equation in  \eqnref{eq:ohmf} is equivalent to the following system of equations (of course, $\rho_{01}(t)=\rho^{*}_{10}(t)$ and $\rho_{00}(t)=1-\rho_{11}(t)$):
\begin{eqnarray}
\frac{\mathrm{d}}{\mathrm{d}t}\rho_{11}(t) &=& \cos^2\Cang \gamma(t)\left(1-2\rho_{11}(t)\right) + 2 \cos\Cang \sin\Cang \gamma(t)\Re{\rho_{10}(t)} \nonumber \\
\frac{\mathrm{d}}{\mathrm{d}t}\rho_{10}(t) &=& -i \omega_0 \rho_{10}(t)-\sin\Cang \cos\Cang \gamma(t)(1-2 \rho_{11}(t))\nonumber \\
&&-(1+\sin^2\Cang)\gamma(t)\rho_{10}(t)+\cos^2\Cang \gamma(t)\rho^*_{10}(t),
\end{eqnarray}
which can be easily solved numerically. For the PC dynamics, \eqnref{eq:masterEquationohmht} leads us to
\begin{eqnarray}
\frac{\mathrm{d}}{\mathrm{d}t}\rho_{11}(t) &=& \cos^2\Cang \gamma(t)(1-2\rho_{11}(t)) \nonumber \\
\frac{\mathrm{d}}{\mathrm{d}t}\rho_{10}(t) &=& \left(-i \omega_0-(2-\cos^2\Cang) \gamma(t)\right)\rho_{10}(t).
\end{eqnarray}
Contrary to the NPC case, populations and coherences are decoupled. Indeed, the solution of this system of equations reads
\begin{eqnarray}
\frac{\mathrm{d}}{\mathrm{d}t}\rho_{11}(t) &=& e^{-2\cos^2\Cang\int_0^t \mathrm{d}\tau \gamma(\tau)}\rho_{11}(0)
\nonumber \\ &&+\cos^2\Cang\int_{0}^t\mathrm{d}\tau e^{-2\cos^2\Cang\int_\tau^t \mathrm{d}\tau' \gamma(\tau')}\gamma(\tau) 
\nonumber \\
\frac{\mathrm{d}}{\mathrm{d}t}\rho_{10}(t) &=& e^{-i \omega_0 t-(2-\cos^2\Cang)\int_0^t \mathrm{d}\tau \gamma(\tau)}\rho_{10}(0).
\end{eqnarray}
In \figref{fig:Blochsphere} we reported the evolution of the Bloch vector $\Bvec(t)$ for different initial conditions for $\rho(0)$. Indeed, the components of the vector $\Bvec(t)$ are directly related to the matrix elements of the corresponding state, see \secref{sec:blv}. Finally, the CP of the dynamics is guaranteed by the master equations themselves, as mentioned in the main text.

\subsubsection{Semigroup limit}
\label{app:semigroupLimit}
Taking the limit $\omega_C \rightarrow \infty$, the decay rate given by \eqnref{eq:ohmf} becomes time independent, 
\begin{equation}
\gamma_s = \lim_{\omega_C \rightarrow \infty} \frac{\lambda}{\beta}\arctan(\omega_C t) = \frac{\pi}{2} \frac{\lambda}{\beta}.
\end{equation}
In the NPC case, this yields the generator
\begin{equation}
\dynmat^{\mathcal{L}}_{\rm{NPC}}(t) = \left(
\begin{array}{cccc}
 0 & 0 & 0 & 0 \\
 0 & -2 \gamma_s  \sin ^2(\Cang ) & -\omega_0 & \gamma_s  \sin (2 \Cang ) \\
 0 & \omega_0 & -2 \gamma  & 0 \\
 0 & \gamma_s  \sin (2 \Cang ) & 0 & -2 \gamma_s  \cos ^2(\Cang ) \\
\end{array}
\right).
\end{equation}

\section{One-probe QFI: maximum for the short-time NPC expression and some general formulas for PC dynamics}\label{app:QFIan}

First, we would like to provide an approximate evaluation of the maximum of the short-time expression of the NPC-QFI, see \eqnref{eq:QFIst}, 
to gain some understanding of the dependence of the optimal QFI on the initial state also for NPC dynamics.
To second order in time, we have  $F_Q^{(4)}=F_{Q,\mathrm{PC}}^{(4)}$ and the QFI is maximal for $\theta=\pi/2$. 
We assume this to be around the optimal input even if the fourth order is considered. Taking the derivative of $F_Q^{(4)}$ with respect to $\phi$ we obtain 
\begin{equation}
\left.\frac{\partial F_{Q,}^{(4)}}{\partial \phi}\right|_{\theta = \pi/2}=\frac{\alpha  \tau ^4 \cos ^4\Cang \sin\phi \cos\phi \left(\sin ^2\Cang \cos ^4\phi-\sin ^4\phi\right)}{2 \left(\sin ^2\Cang \cos ^2\phi+\sin ^2\phi\right)^2}.
\end{equation}
This equation can be numerically solved for $\phi$ and yields:
\begin{eqnarray}
\theta_{\rm{opt}} \approx  \frac{\pi}{2} \qquad \t{and}\qquad \phi_{\rm{opt}} = \arctan \left\{\sqrt{\sin \Cang}\right\}.  
\end{eqnarray}
The (quasi-) optimality of this choice has been checked numerically, confirming that the value of the optimal $\phi_\mathrm{opt}$ is more sensitive to changes in tha bath-coupling angle $\Cang$.

Now, using the characterization of PC dynamics presented in \secref{sec:blv} and the formula for the QFI in \eqnref{eq:QFInori},
we will provide some analytical formulas for the one-probe QFI of a PC dynamics; for the sake of generality, we will not
restrict to the unital case (i.e., to the $T\rightarrow \infty$ regime for the spin-boson model, see \figref{fig:dynamicCompare}).
Any PC dynamical map can be written as in \eqnref{eq:PCmap}, where $\xi = \omega_0 t +\varphi$ and, in general,
also the other coefficients $v_z, d_z, d$ will depend both on $\omega_0$ and on $t$. However, if we neglect for a moment the dependence
of the noise rates (for a PC dynamics $b_{ii}(t), i= \pm, z$) on $\omega_0$, it is easy to see that the dependence on $\omega_0$
will be enclosed only in $\xi$, that is the coefficient due to the unitary component of the map. In this case the QFI,
which we denote as $\tilde{F}_{Q,PC}(t)$, will be simply given by
\begin{equation}\label{eq:qfipc1}
\tilde{F}_{Q,\rm{PC}}(t) = \frac{t^2}{2} D_z(t)^2,
\end{equation}
where $D_z(t) = \sqrt{x(t)^2+y(t)^2}=|d(t)|\sqrt{x(0)^2+y(0)^2}$ is the distance of the state at time $t$ from the $z$-axis
and we have used Cartesian coordinates to define the Bloch vector $\vec{r}(t)=\{x(t),y(t),z(t)\}$.
Instead, if we include the dependence of the noise parameters on $\omega_0$, we obtain the 'full' QFI
\begin{align}
F_{Q,PC}(t) &=\frac{1}{2}\left(t^2 D_z(t,\omega_0)^2+\dot{D}_z(t,\omega_0)^2+\dot{z}(t, \omega_0)^2\right)+\frac{1}{2}\frac{\left(D_z(t,\omega_0)^2 \dot{d}(t,\omega_0)/d(t,\omega_0)
+z(t,\omega_0)\dot{z}(t)\right)^2}
{1-D_z(t,\omega_0)^2-z^2(t,\omega_0)},\label{eq:qfipc2}
\end{align}
where for the sake of compactness we used explicitly that 
$z(t,\omega_0) = v_z(t,\omega_0)+z(0) d_z(t,\omega_0)$
Note that all the contributions are positive; in particular 
$1-D_z(t,\omega_0)^2-z^2(t)\ge0$ due to the positivity of the dynamics.


\end{appendix}

\begin{thebibliography}{104}%
\makeatletter
\providecommand \@ifxundefined [1]{%
 \@ifx{#1\undefined}
}%
\providecommand \@ifnum [1]{%
 \ifnum #1\expandafter \@firstoftwo
 \else \expandafter \@secondoftwo
 \fi
}%
\providecommand \@ifx [1]{%
 \ifx #1\expandafter \@firstoftwo
 \else \expandafter \@secondoftwo
 \fi
}%
\providecommand \natexlab [1]{#1}%
\providecommand \enquote  [1]{``#1''}%
\providecommand \bibnamefont  [1]{#1}%
\providecommand \bibfnamefont [1]{#1}%
\providecommand \citenamefont [1]{#1}%
\providecommand \href@noop [0]{\@secondoftwo}%
\providecommand \href [0]{\begingroup \@sanitize@url \@href}%
\providecommand \@href[1]{\@@startlink{#1}\@@href}%
\providecommand \@@href[1]{\endgroup#1\@@endlink}%
\providecommand \@sanitize@url [0]{\catcode `\\12\catcode `\$12\catcode
  `\&12\catcode `\#12\catcode `\^12\catcode `\_12\catcode `\%12\relax}%
\providecommand \@@startlink[1]{}%
\providecommand \@@endlink[0]{}%
\providecommand \url  [0]{\begingroup\@sanitize@url \@url }%
\providecommand \@url [1]{\endgroup\@href {#1}{\urlprefix }}%
\providecommand \urlprefix  [0]{URL }%
\providecommand \Eprint [0]{\href }%
\providecommand \doibase [0]{http://dx.doi.org/}%
\providecommand \selectlanguage [0]{\@gobble}%
\providecommand \bibinfo  [0]{\@secondoftwo}%
\providecommand \bibfield  [0]{\@secondoftwo}%
\providecommand \translation [1]{[#1]}%
\providecommand \BibitemOpen [0]{}%
\providecommand \bibitemStop [0]{}%
\providecommand \bibitemNoStop [0]{.\EOS\space}%
\providecommand \EOS [0]{\spacefactor3000\relax}%
\providecommand \BibitemShut  [1]{\csname bibitem#1\endcsname}%
\let\auto@bib@innerbib\@empty
\bibitem [{\citenamefont {Dowling}\ and\ \citenamefont
  {Seshadreesan}(2015)}]{Dowling2015}%
  \BibitemOpen
  \bibfield  {author} {\bibinfo {author} {\bibfnamefont {Jonathan~P.}\
  \bibnamefont {Dowling}}\ and\ \bibinfo {author} {\bibfnamefont {Kaushik~P.}\
  \bibnamefont {Seshadreesan}},\ }\bibfield  {title} {\enquote {\bibinfo
  {title} {{Quantum Optical Technologies for Metrology, Sensing, and
  Imaging}},}\ }\href {\doibase 10.1109/JLT.2014.2386795} {\bibfield  {journal}
  {\bibinfo  {journal} {J. Lightwave Technol.}\ }\textbf {\bibinfo {volume}
  {33}},\ \bibinfo {pages} {2359--2370} (\bibinfo {year} {2015})}\BibitemShut
  {NoStop}%
\bibitem [{\citenamefont {Schleich}\ \emph {et~al.}(2016)\citenamefont
  {Schleich}, \citenamefont {Ranade}, \citenamefont {Anton}, \citenamefont
  {Arndt}, \citenamefont {Aspelmeyer}, \citenamefont {Bayer}, \citenamefont
  {Berg}, \citenamefont {Calarco}, \citenamefont {Fuchs}, \citenamefont
  {Giacobino}, \citenamefont {Grassl}, \citenamefont {H{\"a}nggi},
  \citenamefont {Heckl}, \citenamefont {Hertel}, \citenamefont {Huelga},
  \citenamefont {Jelezko}, \citenamefont {Keimer}, \citenamefont {Kotthaus},
  \citenamefont {Leuchs}, \citenamefont {L{\"u}tkenhaus}, \citenamefont
  {Maurer}, \citenamefont {Pfau}, \citenamefont {Plenio}, \citenamefont
  {Rasel}, \citenamefont {Renn}, \citenamefont {Silberhorn}, \citenamefont
  {Schiedmayer}, \citenamefont {Schmitt-Landsiedel}, \citenamefont
  {Sch{\"o}nhammer}, \citenamefont {Ustinov}, \citenamefont {Walther},
  \citenamefont {Weinfurter}, \citenamefont {Welzl}, \citenamefont
  {Wiesendanger}, \citenamefont {Wolf}, \citenamefont {Zeilinger},\ and\
  \citenamefont {Zoller}}]{Schleich2016}%
  \BibitemOpen
  \bibfield  {author} {\bibinfo {author} {\bibfnamefont {Wolfgang~P.}\
  \bibnamefont {Schleich}}, \bibinfo {author} {\bibfnamefont {Kedar~S.}\
  \bibnamefont {Ranade}}, \bibinfo {author} {\bibfnamefont {Christian}\
  \bibnamefont {Anton}}, \bibinfo {author} {\bibfnamefont {Markus}\
  \bibnamefont {Arndt}}, \bibinfo {author} {\bibfnamefont {Markus}\
  \bibnamefont {Aspelmeyer}}, \bibinfo {author} {\bibfnamefont {Manfred}\
  \bibnamefont {Bayer}}, \bibinfo {author} {\bibfnamefont {Gunnar}\
  \bibnamefont {Berg}}, \bibinfo {author} {\bibfnamefont {Tommaso}\
  \bibnamefont {Calarco}}, \bibinfo {author} {\bibfnamefont {Harald}\
  \bibnamefont {Fuchs}}, \bibinfo {author} {\bibfnamefont {Elisabeth}\
  \bibnamefont {Giacobino}}, \bibinfo {author} {\bibfnamefont {Markus}\
  \bibnamefont {Grassl}}, \bibinfo {author} {\bibfnamefont {Peter}\
  \bibnamefont {H{\"a}nggi}}, \bibinfo {author} {\bibfnamefont {Wolfgang~M.}\
  \bibnamefont {Heckl}}, \bibinfo {author} {\bibfnamefont {Ingolf-Volker}\
  \bibnamefont {Hertel}}, \bibinfo {author} {\bibfnamefont {Susana}\
  \bibnamefont {Huelga}}, \bibinfo {author} {\bibfnamefont {Fedor}\
  \bibnamefont {Jelezko}}, \bibinfo {author} {\bibfnamefont {Bernhard}\
  \bibnamefont {Keimer}}, \bibinfo {author} {\bibfnamefont {J{\"o}rg~P.}\
  \bibnamefont {Kotthaus}}, \bibinfo {author} {\bibfnamefont {Gerd}\
  \bibnamefont {Leuchs}}, \bibinfo {author} {\bibfnamefont {Norbert}\
  \bibnamefont {L{\"u}tkenhaus}}, \bibinfo {author} {\bibfnamefont {Ueli}\
  \bibnamefont {Maurer}}, \bibinfo {author} {\bibfnamefont {Tilman}\
  \bibnamefont {Pfau}}, \bibinfo {author} {\bibfnamefont {Martin~B.}\
  \bibnamefont {Plenio}}, \bibinfo {author} {\bibfnamefont {Ernst~Maria}\
  \bibnamefont {Rasel}}, \bibinfo {author} {\bibfnamefont {Ortwin}\
  \bibnamefont {Renn}}, \bibinfo {author} {\bibfnamefont {Christine}\
  \bibnamefont {Silberhorn}}, \bibinfo {author} {\bibfnamefont {J{\"o}rg}\
  \bibnamefont {Schiedmayer}}, \bibinfo {author} {\bibfnamefont {Doris}\
  \bibnamefont {Schmitt-Landsiedel}}, \bibinfo {author} {\bibfnamefont {Kurt}\
  \bibnamefont {Sch{\"o}nhammer}}, \bibinfo {author} {\bibfnamefont {Alexey}\
  \bibnamefont {Ustinov}}, \bibinfo {author} {\bibfnamefont {Philip}\
  \bibnamefont {Walther}}, \bibinfo {author} {\bibfnamefont {Harald}\
  \bibnamefont {Weinfurter}}, \bibinfo {author} {\bibfnamefont {Emo}\
  \bibnamefont {Welzl}}, \bibinfo {author} {\bibfnamefont {Roland}\
  \bibnamefont {Wiesendanger}}, \bibinfo {author} {\bibfnamefont {Stefan}\
  \bibnamefont {Wolf}}, \bibinfo {author} {\bibfnamefont {Anton}\ \bibnamefont
  {Zeilinger}}, \ and\ \bibinfo {author} {\bibfnamefont {Peter}\ \bibnamefont
  {Zoller}},\ }\bibfield  {title} {\enquote {\bibinfo {title} {Quantum
  technology: from research to application},}\ }\href {\doibase
  10.1007/s00340-016-6353-8} {\bibfield  {journal} {\bibinfo  {journal} {Appl.
  Phys. B}\ }\textbf {\bibinfo {volume} {122}},\ \bibinfo {pages} {130}
  (\bibinfo {year} {2016})}\BibitemShut {NoStop}%
\bibitem [{\citenamefont {T\'{o}th}\ and\ \citenamefont
  {Apellaniz}(2014)}]{Toth2014}%
  \BibitemOpen
  \bibfield  {author} {\bibinfo {author} {\bibfnamefont {G\'{e}za}\
  \bibnamefont {T\'{o}th}}\ and\ \bibinfo {author} {\bibfnamefont {Iagoba}\
  \bibnamefont {Apellaniz}},\ }\bibfield  {title} {\enquote {\bibinfo {title}
  {Quantum metrology from a quantum information science perspective},}\ }\href
  {\doibase 10.1088/1751-8113/47/42/424006} {\bibfield  {journal} {\bibinfo
  {journal} {J. Phys. A: Math. Theor.}\ }\textbf {\bibinfo {volume} {47}},\
  \bibinfo {pages} {424006} (\bibinfo {year} {2014})}\BibitemShut {NoStop}%
\bibitem [{\citenamefont {Demkowicz-Dobrza\'{n}ski}\ \emph
  {et~al.}(2015)\citenamefont {Demkowicz-Dobrza\'{n}ski}, \citenamefont
  {Jarzyna},\ and\ \citenamefont {Ko\l{}ody\'{n}ski}}]{Demkowicz2015}%
  \BibitemOpen
  \bibfield  {author} {\bibinfo {author} {\bibfnamefont {R.}~\bibnamefont
  {Demkowicz-Dobrza\'{n}ski}}, \bibinfo {author} {\bibfnamefont
  {M.}~\bibnamefont {Jarzyna}}, \ and\ \bibinfo {author} {\bibfnamefont
  {J.}~\bibnamefont {Ko\l{}ody\'{n}ski}},\ }\bibfield  {title} {\enquote
  {\bibinfo {title} {Quantum limits in optical interferometry},}\ }in\ \href
  {\doibase 10.1016/bs.po.2015.02.003} {\emph {\bibinfo {booktitle} {Progress
  in Optics}}},\ Vol.~\bibinfo {volume} {60},\ \bibinfo {editor} {edited by\
  \bibinfo {editor} {\bibfnamefont {Emil}\ \bibnamefont {Wolf}}}\ (\bibinfo
  {publisher} {Elsevier},\ \bibinfo {year} {2015})\ pp.\ \bibinfo {pages}
  {345--435},\ \Eprint {http://arxiv.org/abs/1405.7703} {arXiv:1405.7703
  [quant-ph]} \BibitemShut {NoStop}%
\bibitem [{\citenamefont {Pezz{\`e}}\ \emph {et~al.}(2016)\citenamefont
  {Pezz{\`e}}, \citenamefont {Smerzi}, \citenamefont {Oberthaler},
  \citenamefont {Schmied},\ and\ \citenamefont {Treutlein}}]{Pezze2016}%
  \BibitemOpen
  \bibfield  {author} {\bibinfo {author} {\bibfnamefont {Luca}\ \bibnamefont
  {Pezz{\`e}}}, \bibinfo {author} {\bibfnamefont {Augusto}\ \bibnamefont
  {Smerzi}}, \bibinfo {author} {\bibfnamefont {Markus~K}\ \bibnamefont
  {Oberthaler}}, \bibinfo {author} {\bibfnamefont {Roman}\ \bibnamefont
  {Schmied}}, \ and\ \bibinfo {author} {\bibfnamefont {Philipp}\ \bibnamefont
  {Treutlein}},\ }\bibfield  {title} {\enquote {\bibinfo {title} {Non-classical
  states of atomic ensembles: fundamentals and applications in quantum
  metrology},}\ }\href@noop {} {\bibfield  {journal} {\bibinfo  {journal}
  {arXiv e-print}\ } (\bibinfo {year} {2016})},\ \Eprint
  {http://arxiv.org/abs/1609.01609} {arXiv:1609.01609 [quant-ph]} \BibitemShut
  {NoStop}%
\bibitem [{\citenamefont {Degen}\ \emph {et~al.}(2017)\citenamefont {Degen},
  \citenamefont {Reinhard},\ and\ \citenamefont {Cappellaro}}]{Degen2017}%
  \BibitemOpen
  \bibfield  {author} {\bibinfo {author} {\bibfnamefont {C.~L.}\ \bibnamefont
  {Degen}}, \bibinfo {author} {\bibfnamefont {F.}~\bibnamefont {Reinhard}}, \
  and\ \bibinfo {author} {\bibfnamefont {P.}~\bibnamefont {Cappellaro}},\
  }\bibfield  {title} {\enquote {\bibinfo {title} {Quantum sensing},}\ }\href
  {\doibase 10.1103/RevModPhys.89.035002} {\bibfield  {journal} {\bibinfo
  {journal} {Rev. Mod. Phys.}\ }\textbf {\bibinfo {volume} {89}},\ \bibinfo
  {pages} {035002} (\bibinfo {year} {2017})}\BibitemShut {NoStop}%
\bibitem [{\citenamefont {Braun}\ \emph {et~al.}(2017)\citenamefont {Braun},
  \citenamefont {Adesso}, \citenamefont {Benatti}, \citenamefont {Floreanini},
  \citenamefont {Marzolino}, \citenamefont {Mitchell},\ and\ \citenamefont
  {Pirandola}}]{Braun2017}%
  \BibitemOpen
  \bibfield  {author} {\bibinfo {author} {\bibfnamefont {Daniel}\ \bibnamefont
  {Braun}}, \bibinfo {author} {\bibfnamefont {Gerardo}\ \bibnamefont {Adesso}},
  \bibinfo {author} {\bibfnamefont {Fabio}\ \bibnamefont {Benatti}}, \bibinfo
  {author} {\bibfnamefont {Roberto}\ \bibnamefont {Floreanini}}, \bibinfo
  {author} {\bibfnamefont {Ugo}\ \bibnamefont {Marzolino}}, \bibinfo {author}
  {\bibfnamefont {Morgan~W}\ \bibnamefont {Mitchell}}, \ and\ \bibinfo {author}
  {\bibfnamefont {Stefano}\ \bibnamefont {Pirandola}},\ }\bibfield  {title}
  {\enquote {\bibinfo {title} {Quantum enhanced measurements without
  entanglement},}\ }\href@noop {} {\bibfield  {journal} {\bibinfo  {journal}
  {arXiv e-print}\ } (\bibinfo {year} {2017})},\ \Eprint
  {http://arxiv.org/abs/1701.05152} {1701.05152 [quant-ph]} \BibitemShut
  {NoStop}%
\bibitem [{\citenamefont {Giovannetti}\ \emph {et~al.}(2004)\citenamefont
  {Giovannetti}, \citenamefont {Lloyd},\ and\ \citenamefont
  {Maccone}}]{Giovannetti2004}%
  \BibitemOpen
  \bibfield  {author} {\bibinfo {author} {\bibfnamefont {Vittorio}\
  \bibnamefont {Giovannetti}}, \bibinfo {author} {\bibfnamefont {Seth}\
  \bibnamefont {Lloyd}}, \ and\ \bibinfo {author} {\bibfnamefont {Lorenzo}\
  \bibnamefont {Maccone}},\ }\bibfield  {title} {\enquote {\bibinfo {title}
  {Quantum-enhanced measurements: Beating the {Standard Quantum Limit}},}\
  }\href {\doibase 10.1126/science.1104149} {\bibfield  {journal} {\bibinfo
  {journal} {Science}\ }\textbf {\bibinfo {volume} {306}},\ \bibinfo {pages}
  {1330--1336} (\bibinfo {year} {2004})}\BibitemShut {NoStop}%
\bibitem [{\citenamefont {Schnabel}(2017)}]{Schnabel2017}%
  \BibitemOpen
  \bibfield  {author} {\bibinfo {author} {\bibfnamefont {Roman}\ \bibnamefont
  {Schnabel}},\ }\bibfield  {title} {\enquote {\bibinfo {title} {Squeezed
  states of light and their applications in laser interferometers},}\ }\href
  {\doibase http://dx.doi.org/10.1016/j.physrep.2017.04.001} {\bibfield
  {journal} {\bibinfo  {journal} {Physics Reports}\ }\textbf {\bibinfo {volume}
  {684}},\ \bibinfo {pages} {1 -- 51} (\bibinfo {year} {2017})}\BibitemShut
  {NoStop}%
\bibitem [{\citenamefont {LIGO{\,}Collaboration}(2011)}]{LIGO2011}%
  \BibitemOpen
  \bibfield  {author} {\bibinfo {author} {\bibnamefont
  {LIGO{\,}Collaboration}},\ }\bibfield  {title} {\enquote {\bibinfo {title} {A
  gravitational wave observatory operating beyond the quantum shot-noise
  limit},}\ }\href {\doibase 10.1038/nphys2083} {\bibfield  {journal} {\bibinfo
   {journal} {Nat. Phys.}\ }\textbf {\bibinfo {volume} {7}},\ \bibinfo {pages}
  {962--965} (\bibinfo {year} {2011})}\BibitemShut {NoStop}%
\bibitem [{\citenamefont {LIGO{\,}Collaboration}(2013)}]{LIGO2013}%
  \BibitemOpen
  \bibfield  {author} {\bibinfo {author} {\bibnamefont
  {LIGO{\,}Collaboration}},\ }\bibfield  {title} {\enquote {\bibinfo {title}
  {Enhanced sensitivity of the {LIGO} gravitational wave detector by using
  squeezed states of light},}\ }\href {\doibase 10.1038/nphoton.2013.177}
  {\bibfield  {journal} {\bibinfo  {journal} {Nat. Photonics}\ }\textbf
  {\bibinfo {volume} {7}},\ \bibinfo {pages} {613--619} (\bibinfo {year}
  {2013})}\BibitemShut {NoStop}%
\bibitem [{\citenamefont {Ma}\ \emph {et~al.}(2011)\citenamefont {Ma},
  \citenamefont {Wang}, \citenamefont {Sun},\ and\ \citenamefont
  {Nori}}]{Ma2011}%
  \BibitemOpen
  \bibfield  {author} {\bibinfo {author} {\bibfnamefont {Jian}\ \bibnamefont
  {Ma}}, \bibinfo {author} {\bibfnamefont {Xiaoguang}\ \bibnamefont {Wang}},
  \bibinfo {author} {\bibfnamefont {CP}~\bibnamefont {Sun}}, \ and\ \bibinfo
  {author} {\bibfnamefont {Franco}\ \bibnamefont {Nori}},\ }\bibfield  {title}
  {\enquote {\bibinfo {title} {Quantum spin squeezing},}\ }\href {\doibase
  10.1016/j.physrep.2011.08.003} {\bibfield  {journal} {\bibinfo  {journal}
  {Phys. Rep.}\ }\textbf {\bibinfo {volume} {509}},\ \bibinfo {pages} {89--165}
  (\bibinfo {year} {2011})}\BibitemShut {NoStop}%
\bibitem [{\citenamefont {Greenberger}\ \emph {et~al.}(1989)\citenamefont
  {Greenberger}, \citenamefont {Horne},\ and\ \citenamefont
  {Zeilinger}}]{Greenberger1989}%
  \BibitemOpen
  \bibfield  {author} {\bibinfo {author} {\bibfnamefont {D.~M.}\ \bibnamefont
  {Greenberger}}, \bibinfo {author} {\bibfnamefont {M.A.}\ \bibnamefont
  {Horne}}, \ and\ \bibinfo {author} {\bibfnamefont {A.}~\bibnamefont
  {Zeilinger}},\ }\bibfield  {title} {\enquote {\bibinfo {title} {Going beyond
  {B}ell's theorem},}\ }in\ \href {\doibase 10.1007/978-94-017-0849-4} {\emph
  {\bibinfo {booktitle} {Bell's Theorem, Quantum Theory and Conceptions of the
  Universe}}},\ \bibinfo {series} {Fundamental Theories of Physics},
  Vol.~\bibinfo {volume} {37},\ \bibinfo {editor} {edited by\ \bibinfo {editor}
  {\bibfnamefont {M.}~\bibnamefont {Kafatos}}}\ (\bibinfo  {publisher}
  {Springer Netherlands},\ \bibinfo {year} {1989})\ pp.\ \bibinfo {pages}
  {69--72}\BibitemShut {NoStop}%
\bibitem [{\citenamefont {Wineland}\ \emph {et~al.}(1992)\citenamefont
  {Wineland}, \citenamefont {Bollinger}, \citenamefont {Itano},\ and\
  \citenamefont {Moore}}]{Wineland1992}%
  \BibitemOpen
  \bibfield  {author} {\bibinfo {author} {\bibfnamefont {D.~J.}\ \bibnamefont
  {Wineland}}, \bibinfo {author} {\bibfnamefont {J.~J.}\ \bibnamefont
  {Bollinger}}, \bibinfo {author} {\bibfnamefont {W.~M.}\ \bibnamefont
  {Itano}}, \ and\ \bibinfo {author} {\bibfnamefont {F.~L.}\ \bibnamefont
  {Moore}},\ }\bibfield  {title} {\enquote {\bibinfo {title} {Spin squeezing
  and reduced quantum noise in spectroscopy},}\ }\href {\doibase
  10.1103/PhysRevA.46.R6797} {\bibfield  {journal} {\bibinfo  {journal} {Phys.
  Rev. A}\ }\textbf {\bibinfo {volume} {46}},\ \bibinfo {pages} {R6797--R6800}
  (\bibinfo {year} {1992})}\BibitemShut {NoStop}%
\bibitem [{\citenamefont {Wineland}\ \emph {et~al.}(1994)\citenamefont
  {Wineland}, \citenamefont {Bollinger}, \citenamefont {Itano},\ and\
  \citenamefont {Heinzen}}]{Wineland1994}%
  \BibitemOpen
  \bibfield  {author} {\bibinfo {author} {\bibfnamefont {D.~J.}\ \bibnamefont
  {Wineland}}, \bibinfo {author} {\bibfnamefont {J.~J.}\ \bibnamefont
  {Bollinger}}, \bibinfo {author} {\bibfnamefont {W.~M.}\ \bibnamefont
  {Itano}}, \ and\ \bibinfo {author} {\bibfnamefont {D.~J.}\ \bibnamefont
  {Heinzen}},\ }\bibfield  {title} {\enquote {\bibinfo {title} {Squeezed atomic
  states and projection noise in spectroscopy},}\ }\href {\doibase
  10.1103/PhysRevA.50.67} {\bibfield  {journal} {\bibinfo  {journal} {Phys.
  Rev. A}\ }\textbf {\bibinfo {volume} {50}},\ \bibinfo {pages} {67--88}
  (\bibinfo {year} {1994})}\BibitemShut {NoStop}%
\bibitem [{\citenamefont {Leibfried}\ \emph {et~al.}(2004)\citenamefont
  {Leibfried}, \citenamefont {Barrett}, \citenamefont {Schaetz}, \citenamefont
  {Britton}, \citenamefont {Chiaverini}, \citenamefont {Itano}, \citenamefont
  {Jost}, \citenamefont {Langer},\ and\ \citenamefont
  {Wineland}}]{Leibfried2004}%
  \BibitemOpen
  \bibfield  {author} {\bibinfo {author} {\bibfnamefont {D.}~\bibnamefont
  {Leibfried}}, \bibinfo {author} {\bibfnamefont {M.~D.}\ \bibnamefont
  {Barrett}}, \bibinfo {author} {\bibfnamefont {T.}~\bibnamefont {Schaetz}},
  \bibinfo {author} {\bibfnamefont {J.}~\bibnamefont {Britton}}, \bibinfo
  {author} {\bibfnamefont {J.}~\bibnamefont {Chiaverini}}, \bibinfo {author}
  {\bibfnamefont {W.~M.}\ \bibnamefont {Itano}}, \bibinfo {author}
  {\bibfnamefont {J.~D.}\ \bibnamefont {Jost}}, \bibinfo {author}
  {\bibfnamefont {C.}~\bibnamefont {Langer}}, \ and\ \bibinfo {author}
  {\bibfnamefont {D.~J.}\ \bibnamefont {Wineland}},\ }\bibfield  {title}
  {\enquote {\bibinfo {title} {Toward {H}eisenberg-limited spectroscopy with
  multiparticle entangled states},}\ }\href {\doibase 10.1126/science.1097576}
  {\bibfield  {journal} {\bibinfo  {journal} {Science}\ }\textbf {\bibinfo
  {volume} {304}},\ \bibinfo {pages} {1476--1478} (\bibinfo {year}
  {2004})}\BibitemShut {NoStop}%
\bibitem [{\citenamefont {Budker}\ and\ \citenamefont
  {Romalis}(2007)}]{Budker2007}%
  \BibitemOpen
  \bibfield  {author} {\bibinfo {author} {\bibfnamefont {Dmitry}\ \bibnamefont
  {Budker}}\ and\ \bibinfo {author} {\bibfnamefont {Michael}\ \bibnamefont
  {Romalis}},\ }\bibfield  {title} {\enquote {\bibinfo {title} {Optical
  magnetometry},}\ }\href {http://dx.doi.org/10.1038/nphys566} {\bibfield
  {journal} {\bibinfo  {journal} {Nat. Phys.}\ }\textbf {\bibinfo {volume}
  {3}},\ \bibinfo {pages} {227--234} (\bibinfo {year} {2007})}\BibitemShut
  {NoStop}%
\bibitem [{\citenamefont {Taylor}\ \emph {et~al.}(2008)\citenamefont {Taylor},
  \citenamefont {Cappellaro}, \citenamefont {Childress}, \citenamefont {Jiang},
  \citenamefont {Budker}, \citenamefont {Hemmer}, \citenamefont {Yacoby},
  \citenamefont {Walsworth},\ and\ \citenamefont {Lukin}}]{Taylor2008}%
  \BibitemOpen
  \bibfield  {author} {\bibinfo {author} {\bibfnamefont {J.~M.}\ \bibnamefont
  {Taylor}}, \bibinfo {author} {\bibfnamefont {P.}~\bibnamefont {Cappellaro}},
  \bibinfo {author} {\bibfnamefont {L.}~\bibnamefont {Childress}}, \bibinfo
  {author} {\bibfnamefont {L.}~\bibnamefont {Jiang}}, \bibinfo {author}
  {\bibfnamefont {D.}~\bibnamefont {Budker}}, \bibinfo {author} {\bibfnamefont
  {P.~R.}\ \bibnamefont {Hemmer}}, \bibinfo {author} {\bibfnamefont
  {A.}~\bibnamefont {Yacoby}}, \bibinfo {author} {\bibfnamefont
  {R.}~\bibnamefont {Walsworth}}, \ and\ \bibinfo {author} {\bibfnamefont
  {M.~D.}\ \bibnamefont {Lukin}},\ }\bibfield  {title} {\enquote {\bibinfo
  {title} {High-sensitivity diamond magnetometer with nanoscale resolution},}\
  }\href {http://dx.doi.org/10.1038/nphys1075} {\bibfield  {journal} {\bibinfo
  {journal} {Nat. Phys.}\ }\textbf {\bibinfo {volume} {4}},\ \bibinfo {pages}
  {810--816} (\bibinfo {year} {2008})}\BibitemShut {NoStop}%
\bibitem [{\citenamefont {Clerk}\ \emph {et~al.}(2010)\citenamefont {Clerk},
  \citenamefont {Devoret}, \citenamefont {Girvin}, \citenamefont {Marquardt},\
  and\ \citenamefont {Schoelkopf}}]{Clerk2010}%
  \BibitemOpen
  \bibfield  {author} {\bibinfo {author} {\bibfnamefont {A.~A.}\ \bibnamefont
  {Clerk}}, \bibinfo {author} {\bibfnamefont {M.~H.}\ \bibnamefont {Devoret}},
  \bibinfo {author} {\bibfnamefont {S.~M.}\ \bibnamefont {Girvin}}, \bibinfo
  {author} {\bibfnamefont {Florian}\ \bibnamefont {Marquardt}}, \ and\ \bibinfo
  {author} {\bibfnamefont {R.~J.}\ \bibnamefont {Schoelkopf}},\ }\bibfield
  {title} {\enquote {\bibinfo {title} {Introduction to quantum noise,
  measurement, and amplification},}\ }\href {\doibase
  10.1103/RevModPhys.82.1155} {\bibfield  {journal} {\bibinfo  {journal} {Rev.
  Mod. Phys.}\ }\textbf {\bibinfo {volume} {82}},\ \bibinfo {pages}
  {1155--1208} (\bibinfo {year} {2010})}\BibitemShut {NoStop}%
\bibitem [{\citenamefont {Ludlow}\ \emph {et~al.}(2015)\citenamefont {Ludlow},
  \citenamefont {Boyd}, \citenamefont {Ye}, \citenamefont {Peik},\ and\
  \citenamefont {Schmidt}}]{Ludlow2015}%
  \BibitemOpen
  \bibfield  {author} {\bibinfo {author} {\bibfnamefont {Andrew~D.}\
  \bibnamefont {Ludlow}}, \bibinfo {author} {\bibfnamefont {Martin~M.}\
  \bibnamefont {Boyd}}, \bibinfo {author} {\bibfnamefont {Jun}\ \bibnamefont
  {Ye}}, \bibinfo {author} {\bibfnamefont {E.}~\bibnamefont {Peik}}, \ and\
  \bibinfo {author} {\bibfnamefont {P.~O.}\ \bibnamefont {Schmidt}},\
  }\bibfield  {title} {\enquote {\bibinfo {title} {Optical atomic clocks},}\
  }\href {\doibase 10.1103/RevModPhys.87.637} {\bibfield  {journal} {\bibinfo
  {journal} {Rev. Mod. Phys.}\ }\textbf {\bibinfo {volume} {87}},\ \bibinfo
  {pages} {637--701} (\bibinfo {year} {2015})}\BibitemShut {NoStop}%
\bibitem [{\citenamefont {Helstrom}(1967)}]{Helstrom1967}%
  \BibitemOpen
  \bibfield  {author} {\bibinfo {author} {\bibfnamefont {C.W.}\ \bibnamefont
  {Helstrom}},\ }\bibfield  {title} {\enquote {\bibinfo {title} {Minimum
  mean-squared error of estimates in quantum statistics},}\ }\href {\doibase
  http://dx.doi.org/10.1016/0375-9601(67)90366-0} {\bibfield  {journal}
  {\bibinfo  {journal} {Phys. Lett. A}\ }\textbf {\bibinfo {volume} {25}},\
  \bibinfo {pages} {101 -- 102} (\bibinfo {year} {1967})}\BibitemShut {NoStop}%
\bibitem [{\citenamefont {Barndorff-Nielsen}\ \emph {et~al.}(2003)\citenamefont
  {Barndorff-Nielsen}, \citenamefont {Gill},\ and\ \citenamefont
  {Jupp}}]{Barndorff2003}%
  \BibitemOpen
  \bibfield  {author} {\bibinfo {author} {\bibfnamefont {Ole~E.}\ \bibnamefont
  {Barndorff-Nielsen}}, \bibinfo {author} {\bibfnamefont {Richard~D.}\
  \bibnamefont {Gill}}, \ and\ \bibinfo {author} {\bibfnamefont {Peter~E.}\
  \bibnamefont {Jupp}},\ }\bibfield  {title} {\enquote {\bibinfo {title} {On
  quantum statistical inference},}\ }\href {\doibase 10.1111/1467-9868.00415}
  {\bibfield  {journal} {\bibinfo  {journal} {J. Royal Stat. Soc. B}\ }\textbf
  {\bibinfo {volume} {65}},\ \bibinfo {pages} {775--804} (\bibinfo {year}
  {2003})}\BibitemShut {NoStop}%
\bibitem [{\citenamefont {Braunstein}\ and\ \citenamefont
  {Caves}(1994)}]{Braunstein1994}%
  \BibitemOpen
  \bibfield  {author} {\bibinfo {author} {\bibfnamefont {Samuel~L.}\
  \bibnamefont {Braunstein}}\ and\ \bibinfo {author} {\bibfnamefont
  {Carlton~M.}\ \bibnamefont {Caves}},\ }\bibfield  {title} {\enquote {\bibinfo
  {title} {Statistical distance and the geometry of quantum states},}\ }\href
  {\doibase 10.1103/PhysRevLett.72.3439} {\bibfield  {journal} {\bibinfo
  {journal} {Phys. Rev. Lett.}\ }\textbf {\bibinfo {volume} {72}},\ \bibinfo
  {pages} {3439--3443} (\bibinfo {year} {1994})}\BibitemShut {NoStop}%
\bibitem [{\citenamefont {Bouten}\ \emph {et~al.}(2007)\citenamefont {Bouten},
  \citenamefont {Van~Handel},\ and\ \citenamefont {James}}]{Bouten2007}%
  \BibitemOpen
  \bibfield  {author} {\bibinfo {author} {\bibfnamefont {Luc}\ \bibnamefont
  {Bouten}}, \bibinfo {author} {\bibfnamefont {Ramon}\ \bibnamefont
  {Van~Handel}}, \ and\ \bibinfo {author} {\bibfnamefont {Matthew~R}\
  \bibnamefont {James}},\ }\bibfield  {title} {\enquote {\bibinfo {title} {An
  introduction to quantum filtering},}\ }\href {\doibase 10.1137/060651239}
  {\bibfield  {journal} {\bibinfo  {journal} {SIAM J. Control Optim.}\ }\textbf
  {\bibinfo {volume} {46}},\ \bibinfo {pages} {2199--2241} (\bibinfo {year}
  {2007})}\BibitemShut {NoStop}%
\bibitem [{\citenamefont {Tsang}\ \emph {et~al.}(2011)\citenamefont {Tsang},
  \citenamefont {Wiseman},\ and\ \citenamefont {Caves}}]{Tsang2011}%
  \BibitemOpen
  \bibfield  {author} {\bibinfo {author} {\bibfnamefont {Mankei}\ \bibnamefont
  {Tsang}}, \bibinfo {author} {\bibfnamefont {Howard~M.}\ \bibnamefont
  {Wiseman}}, \ and\ \bibinfo {author} {\bibfnamefont {Carlton~M.}\
  \bibnamefont {Caves}},\ }\bibfield  {title} {\enquote {\bibinfo {title}
  {Fundamental quantum limit to waveform estimation},}\ }\href {\doibase
  10.1103/PhysRevLett.106.090401} {\bibfield  {journal} {\bibinfo  {journal}
  {Phys. Rev. Lett.}\ }\textbf {\bibinfo {volume} {106}},\ \bibinfo {pages}
  {090401} (\bibinfo {year} {2011})}\BibitemShut {NoStop}%
\bibitem [{\citenamefont {Huelga}\ \emph {et~al.}(1997)\citenamefont {Huelga},
  \citenamefont {Macchiavello}, \citenamefont {Pellizzari}, \citenamefont
  {Ekert}, \citenamefont {Plenio},\ and\ \citenamefont {Cirac}}]{Huelga1997}%
  \BibitemOpen
  \bibfield  {author} {\bibinfo {author} {\bibfnamefont {S.~F.}\ \bibnamefont
  {Huelga}}, \bibinfo {author} {\bibfnamefont {C.}~\bibnamefont
  {Macchiavello}}, \bibinfo {author} {\bibfnamefont {T.}~\bibnamefont
  {Pellizzari}}, \bibinfo {author} {\bibfnamefont {A.~K.}\ \bibnamefont
  {Ekert}}, \bibinfo {author} {\bibfnamefont {M.~B.}\ \bibnamefont {Plenio}}, \
  and\ \bibinfo {author} {\bibfnamefont {J.~I.}\ \bibnamefont {Cirac}},\
  }\bibfield  {title} {\enquote {\bibinfo {title} {Improvement of frequency
  standards with quantum entanglement},}\ }\href {\doibase
  10.1103/PhysRevLett.79.3865} {\bibfield  {journal} {\bibinfo  {journal}
  {Phys. Rev. Lett.}\ }\textbf {\bibinfo {volume} {79}},\ \bibinfo {pages}
  {3865--3868} (\bibinfo {year} {1997})}\BibitemShut {NoStop}%
\bibitem [{\citenamefont {Escher}\ \emph {et~al.}(2011)\citenamefont {Escher},
  \citenamefont {de~Matos~Filho},\ and\ \citenamefont
  {Davidovich}}]{Escher2011}%
  \BibitemOpen
  \bibfield  {author} {\bibinfo {author} {\bibfnamefont {B.~M.}\ \bibnamefont
  {Escher}}, \bibinfo {author} {\bibfnamefont {R.~L.}\ \bibnamefont
  {de~Matos~Filho}}, \ and\ \bibinfo {author} {\bibfnamefont {L.}~\bibnamefont
  {Davidovich}},\ }\bibfield  {title} {\enquote {\bibinfo {title} {General
  framework for estimating the ultimate precision limit in noisy
  quantum-enhanced metrology},}\ }\href {http://dx.doi.org/10.1038/nphys1958}
  {\bibfield  {journal} {\bibinfo  {journal} {Nat. Phys.}\ }\textbf {\bibinfo
  {volume} {7}},\ \bibinfo {pages} {406--411} (\bibinfo {year}
  {2011})}\BibitemShut {NoStop}%
\bibitem [{\citenamefont {Demkowicz-Dobrza\'{n}ski}\ \emph
  {et~al.}(2012)\citenamefont {Demkowicz-Dobrza\'{n}ski}, \citenamefont
  {Ko\l{}ody\'{n}ski},\ and\ \citenamefont {{Gu{\c t}{\u a}}}}]{Demkowicz2012}%
  \BibitemOpen
  \bibfield  {author} {\bibinfo {author} {\bibfnamefont {Rafa\l{}}\
  \bibnamefont {Demkowicz-Dobrza\'{n}ski}}, \bibinfo {author} {\bibfnamefont
  {Jan}\ \bibnamefont {Ko\l{}ody\'{n}ski}}, \ and\ \bibinfo {author}
  {\bibfnamefont {M{\u{a}}d{\u{a}}lin}\ \bibnamefont {{Gu{\c t}{\u a}}}},\
  }\bibfield  {title} {\enquote {\bibinfo {title} {The elusive {H}eisenberg
  limit in quantum-enhanced metrology},}\ }\href {\doibase 10.1038/ncomms2067}
  {\bibfield  {journal} {\bibinfo  {journal} {Nat. Commun.}\ }\textbf {\bibinfo
  {volume} {3}},\ \bibinfo {pages} {1063} (\bibinfo {year} {2012})}\BibitemShut
  {NoStop}%
\bibitem [{\citenamefont {Tsang}(2013)}]{Tsang2013}%
  \BibitemOpen
  \bibfield  {author} {\bibinfo {author} {\bibfnamefont {Mankei}\ \bibnamefont
  {Tsang}},\ }\bibfield  {title} {\enquote {\bibinfo {title} {Quantum metrology
  with open dynamical systems},}\ }\href
  {http://stacks.iop.org/1367-2630/15/i=7/a=073005} {\bibfield  {journal}
  {\bibinfo  {journal} {New J. Phys.}\ }\textbf {\bibinfo {volume} {15}},\
  \bibinfo {pages} {073005} (\bibinfo {year} {2013})}\BibitemShut {NoStop}%
\bibitem [{\citenamefont {Leggett}\ \emph {et~al.}(1987)\citenamefont
  {Leggett}, \citenamefont {Chakravarty}, \citenamefont {Dorsey}, \citenamefont
  {Fisher}, \citenamefont {Garg},\ and\ \citenamefont {Zwerger}}]{Leggett1987}%
  \BibitemOpen
  \bibfield  {author} {\bibinfo {author} {\bibfnamefont {A.~J.}\ \bibnamefont
  {Leggett}}, \bibinfo {author} {\bibfnamefont {S.}~\bibnamefont
  {Chakravarty}}, \bibinfo {author} {\bibfnamefont {A.~T.}\ \bibnamefont
  {Dorsey}}, \bibinfo {author} {\bibfnamefont {Matthew P.~A.}\ \bibnamefont
  {Fisher}}, \bibinfo {author} {\bibfnamefont {Anupam}\ \bibnamefont {Garg}}, \
  and\ \bibinfo {author} {\bibfnamefont {W.}~\bibnamefont {Zwerger}},\
  }\bibfield  {title} {\enquote {\bibinfo {title} {Dynamics of the dissipative
  two-state system},}\ }\href {\doibase 10.1103/RevModPhys.59.1} {\bibfield
  {journal} {\bibinfo  {journal} {Rev. Mod. Phys.}\ }\textbf {\bibinfo {volume}
  {59}},\ \bibinfo {pages} {1--85} (\bibinfo {year} {1987})}\BibitemShut
  {NoStop}%
\bibitem [{\citenamefont {Garg}\ \emph {et~al.}(1985)\citenamefont {Garg},
  \citenamefont {Onuchic},\ and\ \citenamefont {Ambegaokar}}]{Garg1985}%
  \BibitemOpen
  \bibfield  {author} {\bibinfo {author} {\bibfnamefont {Anupam}\ \bibnamefont
  {Garg}}, \bibinfo {author} {\bibfnamefont {Jos{\'e}~Nelson}\ \bibnamefont
  {Onuchic}}, \ and\ \bibinfo {author} {\bibfnamefont {Vinay}\ \bibnamefont
  {Ambegaokar}},\ }\bibfield  {title} {\enquote {\bibinfo {title} {Effect of
  friction on electron transfer in biomolecules},}\ }\href {\doibase
  10.1063/1.449017} {\bibfield  {journal} {\bibinfo  {journal} {The Journal of
  chemical physics}\ }\textbf {\bibinfo {volume} {83}},\ \bibinfo {pages}
  {4491--4503} (\bibinfo {year} {1985})}\BibitemShut {NoStop}%
\bibitem [{\citenamefont {Golding}\ \emph {et~al.}(1992)\citenamefont
  {Golding}, \citenamefont {Zimmerman},\ and\ \citenamefont
  {Coppersmith}}]{Golding1992}%
  \BibitemOpen
  \bibfield  {author} {\bibinfo {author} {\bibfnamefont {Brage}\ \bibnamefont
  {Golding}}, \bibinfo {author} {\bibfnamefont {M.~Neil}\ \bibnamefont
  {Zimmerman}}, \ and\ \bibinfo {author} {\bibfnamefont {S.~N.}\ \bibnamefont
  {Coppersmith}},\ }\bibfield  {title} {\enquote {\bibinfo {title} {Dissipative
  quantum tunneling of a single microscopic defect in a mesoscopic metal},}\
  }\href {\doibase 10.1103/PhysRevLett.68.998} {\bibfield  {journal} {\bibinfo
  {journal} {Phys. Rev. Lett.}\ }\textbf {\bibinfo {volume} {68}},\ \bibinfo
  {pages} {998--1001} (\bibinfo {year} {1992})}\BibitemShut {NoStop}%
\bibitem [{\citenamefont {Makhlin}\ \emph {et~al.}(2001)\citenamefont
  {Makhlin}, \citenamefont {Sch\"on},\ and\ \citenamefont
  {Shnirman}}]{Makhlin2001}%
  \BibitemOpen
  \bibfield  {author} {\bibinfo {author} {\bibfnamefont {Yuriy}\ \bibnamefont
  {Makhlin}}, \bibinfo {author} {\bibfnamefont {Gerd}\ \bibnamefont {Sch\"on}},
  \ and\ \bibinfo {author} {\bibfnamefont {Alexander}\ \bibnamefont
  {Shnirman}},\ }\bibfield  {title} {\enquote {\bibinfo {title} {Quantum-state
  engineering with josephson-junction devices},}\ }\href {\doibase
  10.1103/RevModPhys.73.357} {\bibfield  {journal} {\bibinfo  {journal} {Rev.
  Mod. Phys.}\ }\textbf {\bibinfo {volume} {73}},\ \bibinfo {pages} {357--400}
  (\bibinfo {year} {2001})}\BibitemShut {NoStop}%
\bibitem [{\citenamefont {Ko{\l}ody{\'n}ski}\ and\ \citenamefont
  {Demkowicz-Dobrza{\'n}ski}(2013)}]{Kolodynski2013}%
  \BibitemOpen
  \bibfield  {author} {\bibinfo {author} {\bibfnamefont {Jan}\ \bibnamefont
  {Ko{\l}ody{\'n}ski}}\ and\ \bibinfo {author} {\bibfnamefont {Rafa{\l}}\
  \bibnamefont {Demkowicz-Dobrza{\'n}ski}},\ }\bibfield  {title} {\enquote
  {\bibinfo {title} {Efficient tools for quantum metrology with uncorrelated
  noise},}\ }\href {http://stacks.iop.org/1367-2630/15/i=7/a=073043} {\bibfield
   {journal} {\bibinfo  {journal} {New J. Phys.}\ }\textbf {\bibinfo {volume}
  {15}},\ \bibinfo {pages} {073043} (\bibinfo {year} {2013})}\BibitemShut
  {NoStop}%
\bibitem [{\citenamefont {Sza\ifmmode~\acute{n}\else \'{n}\fi{}kowski}\ \emph
  {et~al.}(2014)\citenamefont {Sza\ifmmode~\acute{n}\else \'{n}\fi{}kowski},
  \citenamefont {Trippenbach},\ and\ \citenamefont
  {Chwede\ifmmode~\acute{n}\else \'{n}\fi{}czuk}}]{Szankowski2014}%
  \BibitemOpen
  \bibfield  {author} {\bibinfo {author} {\bibfnamefont {P.}~\bibnamefont
  {Sza\ifmmode~\acute{n}\else \'{n}\fi{}kowski}}, \bibinfo {author}
  {\bibfnamefont {M.}~\bibnamefont {Trippenbach}}, \ and\ \bibinfo {author}
  {\bibfnamefont {J.}~\bibnamefont {Chwede\ifmmode~\acute{n}\else
  \'{n}\fi{}czuk}},\ }\bibfield  {title} {\enquote {\bibinfo {title} {Parameter
  estimation in memory-assisted noisy quantum interferometry},}\ }\href
  {\doibase 10.1103/PhysRevA.90.063619} {\bibfield  {journal} {\bibinfo
  {journal} {Phys. Rev. A}\ }\textbf {\bibinfo {volume} {90}},\ \bibinfo
  {pages} {063619} (\bibinfo {year} {2014})}\BibitemShut {NoStop}%
\bibitem [{\citenamefont {Sekatski}\ \emph {et~al.}(2017)\citenamefont
  {Sekatski}, \citenamefont {Skotiniotis}, \citenamefont
  {Ko{\l{}}ody{\'{n}}ski},\ and\ \citenamefont {D{\"{u}}r}}]{Sekatski2017}%
  \BibitemOpen
  \bibfield  {author} {\bibinfo {author} {\bibfnamefont {Pavel}\ \bibnamefont
  {Sekatski}}, \bibinfo {author} {\bibfnamefont {Michalis}\ \bibnamefont
  {Skotiniotis}}, \bibinfo {author} {\bibfnamefont {Janek}\ \bibnamefont
  {Ko{\l{}}ody{\'{n}}ski}}, \ and\ \bibinfo {author} {\bibfnamefont {Wolfgang}\
  \bibnamefont {D{\"{u}}r}},\ }\bibfield  {title} {\enquote {\bibinfo {title}
  {Quantum metrology with full and fast quantum control},}\ }\href {\doibase
  10.22331/q-2017-09-06-27} {\bibfield  {journal} {\bibinfo  {journal}
  {{Quantum}}\ }\textbf {\bibinfo {volume} {1}},\ \bibinfo {pages} {27}
  (\bibinfo {year} {2017})}\BibitemShut {NoStop}%
\bibitem [{\citenamefont {Chaves}\ \emph {et~al.}(2013)\citenamefont {Chaves},
  \citenamefont {Brask}, \citenamefont {Markiewicz}, \citenamefont
  {Ko\l{}ody\ifmmode~\acute{n}\else \'{n}\fi{}ski},\ and\ \citenamefont
  {Ac\'{\i}n}}]{Chaves2013}%
  \BibitemOpen
  \bibfield  {author} {\bibinfo {author} {\bibfnamefont {R.}~\bibnamefont
  {Chaves}}, \bibinfo {author} {\bibfnamefont {J.~B.}\ \bibnamefont {Brask}},
  \bibinfo {author} {\bibfnamefont {M.}~\bibnamefont {Markiewicz}}, \bibinfo
  {author} {\bibfnamefont {J.}~\bibnamefont {Ko\l{}ody\ifmmode~\acute{n}\else
  \'{n}\fi{}ski}}, \ and\ \bibinfo {author} {\bibfnamefont {A.}~\bibnamefont
  {Ac\'{\i}n}},\ }\bibfield  {title} {\enquote {\bibinfo {title} {Noisy
  metrology beyond the standard quantum limit},}\ }\href {\doibase
  10.1103/PhysRevLett.111.120401} {\bibfield  {journal} {\bibinfo  {journal}
  {Phys. Rev. Lett.}\ }\textbf {\bibinfo {volume} {111}},\ \bibinfo {pages}
  {120401} (\bibinfo {year} {2013})}\BibitemShut {NoStop}%
\bibitem [{\citenamefont {Holevo}(1993)}]{Holevo1993}%
  \BibitemOpen
  \bibfield  {author} {\bibinfo {author} {\bibfnamefont {A.~S.}\ \bibnamefont
  {Holevo}},\ }\bibfield  {title} {\enquote {\bibinfo {title} {A note on
  covariant dynamical semigroups},}\ }\href {\doibase
  http://dx.doi.org/10.1016/0034-4877(93)90014-6} {\bibfield  {journal}
  {\bibinfo  {journal} {Rep. Math. Phys.}\ }\textbf {\bibinfo {volume} {32}},\
  \bibinfo {pages} {211 -- 216} (\bibinfo {year} {1993})}\BibitemShut {NoStop}%
\bibitem [{\citenamefont {Holevo}(1996)}]{Holevo1996}%
  \BibitemOpen
  \bibfield  {author} {\bibinfo {author} {\bibfnamefont {A.~S.}\ \bibnamefont
  {Holevo}},\ }\bibfield  {title} {\enquote {\bibinfo {title} {Covariant
  quantum markovian evolutions},}\ }\href {\doibase 10.1063/1.531481}
  {\bibfield  {journal} {\bibinfo  {journal} {J. Math. Phys.}\ }\textbf
  {\bibinfo {volume} {37}},\ \bibinfo {pages} {1812--1832} (\bibinfo {year}
  {1996})}\BibitemShut {NoStop}%
\bibitem [{\citenamefont {Vacchini}(2010)}]{Vacchini2010}%
  \BibitemOpen
  \bibfield  {author} {\bibinfo {author} {\bibfnamefont {B}~\bibnamefont
  {Vacchini}},\ }\bibfield  {title} {\enquote {\bibinfo {title} {Covariant
  mappings for the description of measurement, dissipation and decoherence in
  quantum mechanics},}\ }\href@noop {} {\bibfield  {journal} {\bibinfo
  {journal} {Lect. Notes Phys.}\ }\textbf {\bibinfo {volume} {787}},\ \bibinfo
  {pages} {39--77} (\bibinfo {year} {2010})}\BibitemShut {NoStop}%
\bibitem [{\citenamefont {Breuer}\ and\ \citenamefont
  {Petruccione}(2002)}]{BreuerPetruccione}%
  \BibitemOpen
  \bibfield  {author} {\bibinfo {author} {\bibfnamefont {Heinz-Peter}\
  \bibnamefont {Breuer}}\ and\ \bibinfo {author} {\bibfnamefont {Francesco}\
  \bibnamefont {Petruccione}},\ }\href@noop {} {\emph {\bibinfo {title} {The
  theory of open quantum systems}}}\ (\bibinfo  {publisher} {Oxford University
  Press on Demand},\ \bibinfo {year} {2002})\BibitemShut {NoStop}%
\bibitem [{\citenamefont {Matsuzaki}\ \emph {et~al.}(2011)\citenamefont
  {Matsuzaki}, \citenamefont {Benjamin},\ and\ \citenamefont
  {Fitzsimons}}]{Matsuzaki2011}%
  \BibitemOpen
  \bibfield  {author} {\bibinfo {author} {\bibfnamefont {Yuichiro}\
  \bibnamefont {Matsuzaki}}, \bibinfo {author} {\bibfnamefont {Simon~C.}\
  \bibnamefont {Benjamin}}, \ and\ \bibinfo {author} {\bibfnamefont {Joseph}\
  \bibnamefont {Fitzsimons}},\ }\bibfield  {title} {\enquote {\bibinfo {title}
  {Magnetic field sensing beyond the standard quantum limit under the effect of
  decoherence},}\ }\href {\doibase 10.1103/PhysRevA.84.012103} {\bibfield
  {journal} {\bibinfo  {journal} {Phys. Rev. A}\ }\textbf {\bibinfo {volume}
  {84}},\ \bibinfo {pages} {012103} (\bibinfo {year} {2011})}\BibitemShut
  {NoStop}%
\bibitem [{\citenamefont {Chin}\ \emph {et~al.}(2012)\citenamefont {Chin},
  \citenamefont {Huelga},\ and\ \citenamefont {Plenio}}]{Chin2012}%
  \BibitemOpen
  \bibfield  {author} {\bibinfo {author} {\bibfnamefont {Alex~W.}\ \bibnamefont
  {Chin}}, \bibinfo {author} {\bibfnamefont {Susana~F.}\ \bibnamefont
  {Huelga}}, \ and\ \bibinfo {author} {\bibfnamefont {Martin~B.}\ \bibnamefont
  {Plenio}},\ }\bibfield  {title} {\enquote {\bibinfo {title} {Quantum
  metrology in non-markovian environments},}\ }\href {\doibase
  10.1103/PhysRevLett.109.233601} {\bibfield  {journal} {\bibinfo  {journal}
  {Phys. Rev. Lett.}\ }\textbf {\bibinfo {volume} {109}},\ \bibinfo {pages}
  {233601} (\bibinfo {year} {2012})}\BibitemShut {NoStop}%
\bibitem [{\citenamefont {Smirne}\ \emph {et~al.}(2016)\citenamefont {Smirne},
  \citenamefont {Ko\l{}ody\ifmmode~\acute{n}\else \'{n}\fi{}ski}, \citenamefont
  {Huelga},\ and\ \citenamefont {Demkowicz-Dobrza\ifmmode~\acute{n}\else
  \'{n}\fi{}ski}}]{Smirne2016}%
  \BibitemOpen
  \bibfield  {author} {\bibinfo {author} {\bibfnamefont {Andrea}\ \bibnamefont
  {Smirne}}, \bibinfo {author} {\bibfnamefont {Jan}\ \bibnamefont
  {Ko\l{}ody\ifmmode~\acute{n}\else \'{n}\fi{}ski}}, \bibinfo {author}
  {\bibfnamefont {Susana~F.}\ \bibnamefont {Huelga}}, \ and\ \bibinfo {author}
  {\bibfnamefont {Rafa\l{}}\ \bibnamefont
  {Demkowicz-Dobrza\ifmmode~\acute{n}\else \'{n}\fi{}ski}},\ }\bibfield
  {title} {\enquote {\bibinfo {title} {Ultimate precision limits for noisy
  frequency estimation},}\ }\href {\doibase 10.1103/PhysRevLett.116.120801}
  {\bibfield  {journal} {\bibinfo  {journal} {Phys. Rev. Lett.}\ }\textbf
  {\bibinfo {volume} {116}},\ \bibinfo {pages} {120801} (\bibinfo {year}
  {2016})}\BibitemShut {NoStop}%
\bibitem [{\citenamefont {Kay}(1993)}]{Kay1993}%
  \BibitemOpen
  \bibfield  {author} {\bibinfo {author} {\bibfnamefont {Steven~M.}\
  \bibnamefont {Kay}},\ }\href@noop {} {\emph {\bibinfo {title} {Fundamentals
  of Statistical Signal Processing: Estimation Theory}}}\ (\bibinfo
  {publisher} {Prentice Hall},\ \bibinfo {year} {1993})\BibitemShut {NoStop}%
\bibitem [{Note1()}]{Note1}%
  \BibitemOpen
  \bibinfo {note} {For instance, when maximising the sensitivity of slope
  detection in external-field sensing scenarios \cite {Degen2017}.}\BibitemShut
  {Stop}%
\bibitem [{Note2()}]{Note2}%
  \BibitemOpen
  \bibinfo {note} {Without loss of generality, we also require for convenience
  that maximal variance of $\protect \mathaccentV {hat}05Eh$ is $\protect
  \qopname \relax m{max}_{\psi } \left .\Delta ^2 \protect \mathaccentV
  {hat}05Eh\right |_\psi =1/4$.}\BibitemShut {Stop}%
\bibitem [{\citenamefont {Bollinger}\ \emph {et~al.}(1996)\citenamefont
  {Bollinger}, \citenamefont {Itano}, \citenamefont {Wineland},\ and\
  \citenamefont {Heinzen}}]{Bollinger1996}%
  \BibitemOpen
  \bibfield  {author} {\bibinfo {author} {\bibfnamefont {J.~J~.}\ \bibnamefont
  {Bollinger}}, \bibinfo {author} {\bibfnamefont {Wayne~M.}\ \bibnamefont
  {Itano}}, \bibinfo {author} {\bibfnamefont {D.~J.}\ \bibnamefont {Wineland}},
  \ and\ \bibinfo {author} {\bibfnamefont {D.~J.}\ \bibnamefont {Heinzen}},\
  }\bibfield  {title} {\enquote {\bibinfo {title} {Optimal frequency
  measurements with maximally correlated states},}\ }\href {\doibase
  10.1103/PhysRevA.54.R4649} {\bibfield  {journal} {\bibinfo  {journal} {Phys.
  Rev. A}\ }\textbf {\bibinfo {volume} {54}},\ \bibinfo {pages} {R4649--R4652}
  (\bibinfo {year} {1996})}\BibitemShut {NoStop}%
\bibitem [{\citenamefont {Pang}\ and\ \citenamefont {Jordan}(2017)}]{Pang2017}%
  \BibitemOpen
  \bibfield  {author} {\bibinfo {author} {\bibfnamefont {Shengshi}\
  \bibnamefont {Pang}}\ and\ \bibinfo {author} {\bibfnamefont {Andrew~N.}\
  \bibnamefont {Jordan}},\ }\bibfield  {title} {\enquote {\bibinfo {title}
  {Optimal adaptive control for quantum metrology with time-dependent
  {Hamiltonians}},}\ }\href {\doibase 10.1038/ncomms14695} {\bibfield
  {journal} {\bibinfo  {journal} {Nat. Commun.}\ }\textbf {\bibinfo {volume}
  {8}},\ \bibinfo {pages} {14695} (\bibinfo {year} {2017})}\BibitemShut
  {NoStop}%
\bibitem [{\citenamefont {Dooley}\ \emph {et~al.}(2016)\citenamefont {Dooley},
  \citenamefont {Munro},\ and\ \citenamefont {Nemoto}}]{Dooley2016}%
  \BibitemOpen
  \bibfield  {author} {\bibinfo {author} {\bibfnamefont {Shane}\ \bibnamefont
  {Dooley}}, \bibinfo {author} {\bibfnamefont {William~J.}\ \bibnamefont
  {Munro}}, \ and\ \bibinfo {author} {\bibfnamefont {Kae}\ \bibnamefont
  {Nemoto}},\ }\bibfield  {title} {\enquote {\bibinfo {title} {Quantum
  metrology including state preparation and readout times},}\ }\href {\doibase
  10.1103/PhysRevA.94.052320} {\bibfield  {journal} {\bibinfo  {journal} {Phys.
  Rev. A}\ }\textbf {\bibinfo {volume} {94}},\ \bibinfo {pages} {052320}
  (\bibinfo {year} {2016})}\BibitemShut {NoStop}%
\bibitem [{\citenamefont {Fujiwara}(2001)}]{Fujiwara2001}%
  \BibitemOpen
  \bibfield  {author} {\bibinfo {author} {\bibfnamefont {Akio}\ \bibnamefont
  {Fujiwara}},\ }\bibfield  {title} {\enquote {\bibinfo {title} {Quantum
  channel identification problem},}\ }\href {\doibase
  10.1103/PhysRevA.63.042304} {\bibfield  {journal} {\bibinfo  {journal} {Phys.
  Rev. A}\ }\textbf {\bibinfo {volume} {63}},\ \bibinfo {pages} {042304}
  (\bibinfo {year} {2001})}\BibitemShut {NoStop}%
\bibitem [{\citenamefont {Fujiwara}\ and\ \citenamefont
  {Imai}(2008)}]{Fujiwara2008}%
  \BibitemOpen
  \bibfield  {author} {\bibinfo {author} {\bibfnamefont {Akio}\ \bibnamefont
  {Fujiwara}}\ and\ \bibinfo {author} {\bibfnamefont {Hiroshi}\ \bibnamefont
  {Imai}},\ }\bibfield  {title} {\enquote {\bibinfo {title} {A fibre bundle
  over manifolds of quantum channels and its application to quantum
  statistics},}\ }\href {\doibase 10.1088/1751-8113/41/25/255304} {\bibfield
  {journal} {\bibinfo  {journal} {J. Phys. A: Math. Theor.}\ }\textbf {\bibinfo
  {volume} {41}},\ \bibinfo {pages} {255304} (\bibinfo {year}
  {2008})}\BibitemShut {NoStop}%
\bibitem [{\citenamefont {Nielsen}\ and\ \citenamefont
  {Chuang}(2010)}]{Nielsen2010}%
  \BibitemOpen
  \bibfield  {author} {\bibinfo {author} {\bibfnamefont {Michael~A.}\
  \bibnamefont {Nielsen}}\ and\ \bibinfo {author} {\bibfnamefont {Issaac~L.}\
  \bibnamefont {Chuang}},\ }\href@noop {} {\emph {\bibinfo {title} {Quantum
  Computation and Quantum Information}}}\ (\bibinfo  {publisher} {Cambridge
  University Press},\ \bibinfo {year} {2010})\BibitemShut {NoStop}%
\bibitem [{\citenamefont {Bengtsson}\ and\ \citenamefont
  {{\.Z}yczkowski}(2006)}]{Bengtsson2006}%
  \BibitemOpen
  \bibfield  {author} {\bibinfo {author} {\bibfnamefont {Ingemar}\ \bibnamefont
  {Bengtsson}}\ and\ \bibinfo {author} {\bibfnamefont {Karol}\ \bibnamefont
  {{\.Z}yczkowski}},\ }\href@noop {} {\emph {\bibinfo {title} {Geometry of
  quantum states: An introduction to quantum entanglement}}}\ (\bibinfo
  {publisher} {Cambridge University Press},\ \bibinfo {year}
  {2006})\BibitemShut {NoStop}%
\bibitem [{\citenamefont {Dorner}(2012)}]{Dorner2012}%
  \BibitemOpen
  \bibfield  {author} {\bibinfo {author} {\bibfnamefont {U}~\bibnamefont
  {Dorner}},\ }\bibfield  {title} {\enquote {\bibinfo {title} {Quantum
  frequency estimation with trapped ions and atoms},}\ }\href
  {http://stacks.iop.org/1367-2630/14/i=4/a=043011} {\bibfield  {journal}
  {\bibinfo  {journal} {New J. Phys.}\ }\textbf {\bibinfo {volume} {14}},\
  \bibinfo {pages} {043011} (\bibinfo {year} {2012})}\BibitemShut {NoStop}%
\bibitem [{\citenamefont {Jeske}\ \emph {et~al.}(2014)\citenamefont {Jeske},
  \citenamefont {Cole},\ and\ \citenamefont {Huelga}}]{Jeske2014}%
  \BibitemOpen
  \bibfield  {author} {\bibinfo {author} {\bibfnamefont {Jan}\ \bibnamefont
  {Jeske}}, \bibinfo {author} {\bibfnamefont {Jared~H}\ \bibnamefont {Cole}}, \
  and\ \bibinfo {author} {\bibfnamefont {Susana~F}\ \bibnamefont {Huelga}},\
  }\bibfield  {title} {\enquote {\bibinfo {title} {Quantum metrology subject to
  spatially correlated markovian noise: restoring the heisenberg limit},}\
  }\href {http://stacks.iop.org/1367-2630/16/i=7/a=073039} {\bibfield
  {journal} {\bibinfo  {journal} {New J. Phys.}\ }\textbf {\bibinfo {volume}
  {16}},\ \bibinfo {pages} {073039} (\bibinfo {year} {2014})}\BibitemShut
  {NoStop}%
\bibitem [{\citenamefont {Knysh}\ and\ \citenamefont
  {Durkin}(2013)}]{Knysh2013}%
  \BibitemOpen
  \bibfield  {author} {\bibinfo {author} {\bibfnamefont {Sergey~I.}\
  \bibnamefont {Knysh}}\ and\ \bibinfo {author} {\bibfnamefont {Gabriel~A.}\
  \bibnamefont {Durkin}},\ }\bibfield  {title} {\enquote {\bibinfo {title}
  {Estimation of {Phase} and {Diffusion}: {Combining} {Quantum} {Statistics}
  and {Classical} {Noise}},}\ }\href@noop {} {\bibfield  {journal} {\bibinfo
  {journal} {arXiv e-print}\ } (\bibinfo {year} {2013})},\ \Eprint
  {http://arxiv.org/abs/1307.0470} {arXiv:1307.0470 [quant-ph]} \BibitemShut
  {NoStop}%
\bibitem [{\citenamefont {Knysh}\ \emph {et~al.}(2014)\citenamefont {Knysh},
  \citenamefont {Chen},\ and\ \citenamefont {Durkin}}]{Knysh2014}%
  \BibitemOpen
  \bibfield  {author} {\bibinfo {author} {\bibfnamefont {Sergey~I.}\
  \bibnamefont {Knysh}}, \bibinfo {author} {\bibfnamefont {Edward~H.}\
  \bibnamefont {Chen}}, \ and\ \bibinfo {author} {\bibfnamefont {Gabriel~A.}\
  \bibnamefont {Durkin}},\ }\bibfield  {title} {\enquote {\bibinfo {title}
  {True limits to precision via unique quantum probe},}\ }\href@noop {}
  {\bibfield  {journal} {\bibinfo  {journal} {arXiv e-print}\ } (\bibinfo
  {year} {2014})},\ \Eprint {http://arxiv.org/abs/1402.0495} {arXiv:1402.0495
  [quant-ph]} \BibitemShut {NoStop}%
\bibitem [{\citenamefont {Berry}\ \emph {et~al.}(2009)\citenamefont {Berry},
  \citenamefont {Higgins}, \citenamefont {Bartlett}, \citenamefont {Mitchell},
  \citenamefont {Pryde},\ and\ \citenamefont {Wiseman}}]{Berry2009}%
  \BibitemOpen
  \bibfield  {author} {\bibinfo {author} {\bibfnamefont {D.~W.}\ \bibnamefont
  {Berry}}, \bibinfo {author} {\bibfnamefont {B.~L.}\ \bibnamefont {Higgins}},
  \bibinfo {author} {\bibfnamefont {S.~D.}\ \bibnamefont {Bartlett}}, \bibinfo
  {author} {\bibfnamefont {M.~W.}\ \bibnamefont {Mitchell}}, \bibinfo {author}
  {\bibfnamefont {G.~J.}\ \bibnamefont {Pryde}}, \ and\ \bibinfo {author}
  {\bibfnamefont {H.~M.}\ \bibnamefont {Wiseman}},\ }\bibfield  {title}
  {\enquote {\bibinfo {title} {How to perform the most accurate possible phase
  measurements},}\ }\href {\doibase 10.1103/PhysRevA.80.052114} {\bibfield
  {journal} {\bibinfo  {journal} {Phys. Rev. A}\ }\textbf {\bibinfo {volume}
  {80}},\ \bibinfo {pages} {052114} (\bibinfo {year} {2009})}\BibitemShut
  {NoStop}%
\bibitem [{Note3()}]{Note3}%
  \BibitemOpen
  \bibinfo {note} {A situation that naturally applies in the so-called ``slope
  detection''' scenarios (e.g., in Ramsey spectroscopy) of quantum sensing
  \cite {Degen2017}.}\BibitemShut {Stop}%
\bibitem [{\citenamefont {Gu{\c{t}}{\u{a}}}\ and\ \citenamefont
  {Jen{\v{c}}ov{\'a}}(2007)}]{Guta2007}%
  \BibitemOpen
  \bibfield  {author} {\bibinfo {author} {\bibfnamefont {M{\u{a}}d{\u{a}}lin}\
  \bibnamefont {Gu{\c{t}}{\u{a}}}}\ and\ \bibinfo {author} {\bibfnamefont
  {Anna}\ \bibnamefont {Jen{\v{c}}ov{\'a}}},\ }\bibfield  {title} {\enquote
  {\bibinfo {title} {Local asymptotic normality in quantum statistics},}\
  }\href {\doibase 10.1007/s00220-007-0340-1} {\bibfield  {journal} {\bibinfo
  {journal} {Commun. Math. Phys.}\ }\textbf {\bibinfo {volume} {276}},\
  \bibinfo {pages} {341--379} (\bibinfo {year} {2007})}\BibitemShut {NoStop}%
\bibitem [{\citenamefont {Fujiwara}(2006)}]{Fujiwara2006}%
  \BibitemOpen
  \bibfield  {author} {\bibinfo {author} {\bibfnamefont {Akio}\ \bibnamefont
  {Fujiwara}},\ }\bibfield  {title} {\enquote {\bibinfo {title} {Strong
  consistency and asymptotic efficiency for adaptive quantum estimation
  problems},}\ }\href {\doibase 10.1088/0305-4470/39/40/014} {\bibfield
  {journal} {\bibinfo  {journal} {J. Phys. A: Math. Gen.}\ }\textbf {\bibinfo
  {volume} {39}},\ \bibinfo {pages} {12489} (\bibinfo {year}
  {2006})}\BibitemShut {NoStop}%
\bibitem [{\citenamefont {Macieszczak}\ \emph {et~al.}(2014)\citenamefont
  {Macieszczak}, \citenamefont {Fraas},\ and\ \citenamefont
  {Demkowicz-Dobrza{\'n}ski}}]{Macieszczak2014}%
  \BibitemOpen
  \bibfield  {author} {\bibinfo {author} {\bibfnamefont {Katarzyna}\
  \bibnamefont {Macieszczak}}, \bibinfo {author} {\bibfnamefont {Martin}\
  \bibnamefont {Fraas}}, \ and\ \bibinfo {author} {\bibfnamefont {Rafa\l{}}\
  \bibnamefont {Demkowicz-Dobrza{\'n}ski}},\ }\bibfield  {title} {\enquote
  {\bibinfo {title} {Bayesian quantum frequency estimation in presence of
  collective dephasing},}\ }\href {\doibase 10.1088/1367-2630/16/11/113002}
  {\bibfield  {journal} {\bibinfo  {journal} {New J. Phys.}\ }\textbf {\bibinfo
  {volume} {16}},\ \bibinfo {pages} {113002} (\bibinfo {year}
  {2014})}\BibitemShut {NoStop}%
\bibitem [{\citenamefont {Jarzyna}\ and\ \citenamefont
  {Demkowicz-Dobrza\'{n}ski}(2015)}]{Jarzyna2015}%
  \BibitemOpen
  \bibfield  {author} {\bibinfo {author} {\bibfnamefont {M.}~\bibnamefont
  {Jarzyna}}\ and\ \bibinfo {author} {\bibfnamefont {R.}~\bibnamefont
  {Demkowicz-Dobrza\'{n}ski}},\ }\bibfield  {title} {\enquote {\bibinfo {title}
  {True precision limits in quantum metrology},}\ }\href {\doibase
  10.1088/1367-2630/17/1/013010} {\bibfield  {journal} {\bibinfo  {journal}
  {New J. Phys.}\ }\textbf {\bibinfo {volume} {17}},\ \bibinfo {pages} {013010}
  (\bibinfo {year} {2015})}\BibitemShut {NoStop}%
\bibitem [{Note4()}]{Note4}%
  \BibitemOpen
  \bibinfo {note} {Pure initial states may always be considered optimal due
  convexity of the QFI \protect \textup {\hbox {\mathsurround \z@ \protect
  \normalfont (\ignorespaces \ref {eq:QFI}\unskip \@@italiccorr )}} on states,
  $F_Q[\DOTSB \sum@ \slimits@ _i p_i\rho _{\omega }^{(i)}]\le \DOTSB \sum@
  \slimits@ _i p_i F_Q[\rho _{\omega }^{(i)}]$ \cite
  {Alipour2015}.}\BibitemShut {Stop}%
\bibitem [{\citenamefont {Giovannetti}\ \emph {et~al.}(2006)\citenamefont
  {Giovannetti}, \citenamefont {Lloyd},\ and\ \citenamefont
  {Maccone}}]{Giovannetti2006}%
  \BibitemOpen
  \bibfield  {author} {\bibinfo {author} {\bibfnamefont {Vittorio}\
  \bibnamefont {Giovannetti}}, \bibinfo {author} {\bibfnamefont {Seth}\
  \bibnamefont {Lloyd}}, \ and\ \bibinfo {author} {\bibfnamefont {Lorenzo}\
  \bibnamefont {Maccone}},\ }\bibfield  {title} {\enquote {\bibinfo {title}
  {Quantum metrology},}\ }\href {\doibase 10.1103/PhysRevLett.96.010401}
  {\bibfield  {journal} {\bibinfo  {journal} {Phys. Rev. Lett.}\ }\textbf
  {\bibinfo {volume} {96}},\ \bibinfo {pages} {010401} (\bibinfo {year}
  {2006})}\BibitemShut {NoStop}%
\bibitem [{\citenamefont {Lindblad}(1976)}]{Lindblad1976}%
  \BibitemOpen
  \bibfield  {author} {\bibinfo {author} {\bibfnamefont {G.}~\bibnamefont
  {Lindblad}},\ }\bibfield  {title} {\enquote {\bibinfo {title} {On the
  generators of quantum dynamical semigroups},}\ }\href {\doibase
  10.1007/BF01608499} {\bibfield  {journal} {\bibinfo  {journal} {Commun. Math.
  Phys.}\ }\textbf {\bibinfo {volume} {48}},\ \bibinfo {pages} {119--130}
  (\bibinfo {year} {1976})}\BibitemShut {NoStop}%
\bibitem [{\citenamefont {Gorini}\ \emph {et~al.}(1976)\citenamefont {Gorini},
  \citenamefont {Kossakowski},\ and\ \citenamefont {Sudarshan}}]{Gorini1976}%
  \BibitemOpen
  \bibfield  {author} {\bibinfo {author} {\bibfnamefont {Vittorio}\
  \bibnamefont {Gorini}}, \bibinfo {author} {\bibfnamefont {Andrzej}\
  \bibnamefont {Kossakowski}}, \ and\ \bibinfo {author} {\bibfnamefont
  {E.~C.~G.}\ \bibnamefont {Sudarshan}},\ }\bibfield  {title} {\enquote
  {\bibinfo {title} {Completely positive dynamical semigroups of n-level
  systems},}\ }\href {\doibase 10.1063/1.522979} {\bibfield  {journal}
  {\bibinfo  {journal} {J. Math. Phys.}\ }\textbf {\bibinfo {volume} {17}},\
  \bibinfo {pages} {821--825} (\bibinfo {year} {1976})}\BibitemShut {NoStop}%
\bibitem [{Note5()}]{Note5}%
  \BibitemOpen
  \bibinfo {note} {Note that in \cite {Demkowicz2012, Kolodynski2013}, as well
  as in \cite {Smirne2016}, an $\omega _0$-independent noise term $\Gamma (t)$
  was considered. Here, instead, we will take into account a possible
  dependence on $\omega _0$ also in $\Gamma _{\omega _0}(t)$.}\BibitemShut
  {Stop}%
\bibitem [{\citenamefont {Macieszczak}(2015)}]{Macieszczak2015}%
  \BibitemOpen
  \bibfield  {author} {\bibinfo {author} {\bibfnamefont {Katarzyna}\
  \bibnamefont {Macieszczak}},\ }\bibfield  {title} {\enquote {\bibinfo {title}
  {Zeno limit in frequency estimation with non-markovian environments},}\
  }\href {\doibase 10.1103/PhysRevA.92.010102} {\bibfield  {journal} {\bibinfo
  {journal} {Phys. Rev. A}\ }\textbf {\bibinfo {volume} {92}},\ \bibinfo
  {pages} {010102} (\bibinfo {year} {2015})}\BibitemShut {NoStop}%
\bibitem [{\citenamefont {Misra}\ and\ \citenamefont
  {Sudarshan}(1977)}]{Misra1977}%
  \BibitemOpen
  \bibfield  {author} {\bibinfo {author} {\bibfnamefont {B.}~\bibnamefont
  {Misra}}\ and\ \bibinfo {author} {\bibfnamefont {E.~C.~G.}\ \bibnamefont
  {Sudarshan}},\ }\bibfield  {title} {\enquote {\bibinfo {title} {The zeno's
  paradox in quantum theory},}\ }\href {\doibase 10.1063/1.523304} {\bibfield
  {journal} {\bibinfo  {journal} {J. Math. Phys.}\ }\textbf {\bibinfo {volume}
  {18}},\ \bibinfo {pages} {756--763} (\bibinfo {year} {1977})}\BibitemShut
  {NoStop}%
\bibitem [{\citenamefont {Facchi}\ and\ \citenamefont
  {Pascazio}(2008)}]{Facchi2008}%
  \BibitemOpen
  \bibfield  {author} {\bibinfo {author} {\bibfnamefont {P}~\bibnamefont
  {Facchi}}\ and\ \bibinfo {author} {\bibfnamefont {S}~\bibnamefont
  {Pascazio}},\ }\bibfield  {title} {\enquote {\bibinfo {title} {Quantum zeno
  dynamics: mathematical and physical aspects},}\ }\href
  {http://stacks.iop.org/1751-8121/41/i=49/a=493001} {\bibfield  {journal}
  {\bibinfo  {journal} {J. Phys. A: Math. Theor.}\ }\textbf {\bibinfo {volume}
  {41}},\ \bibinfo {pages} {493001} (\bibinfo {year} {2008})}\BibitemShut
  {NoStop}%
\bibitem [{\citenamefont {Rivas}\ \emph {et~al.}(2014)\citenamefont {Rivas},
  \citenamefont {Huelga},\ and\ \citenamefont {Plenio}}]{Rivas2014}%
  \BibitemOpen
  \bibfield  {author} {\bibinfo {author} {\bibfnamefont {{\'A}ngel}\
  \bibnamefont {Rivas}}, \bibinfo {author} {\bibfnamefont {Susana~F}\
  \bibnamefont {Huelga}}, \ and\ \bibinfo {author} {\bibfnamefont {Martin~B}\
  \bibnamefont {Plenio}},\ }\bibfield  {title} {\enquote {\bibinfo {title}
  {Quantum non-markovianity: characterization, quantification and detection},}\
  }\href {http://stacks.iop.org/0034-4885/77/i=9/a=094001} {\bibfield
  {journal} {\bibinfo  {journal} {Rep. Prog. Phys.}\ }\textbf {\bibinfo
  {volume} {77}},\ \bibinfo {pages} {094001} (\bibinfo {year}
  {2014})}\BibitemShut {NoStop}%
\bibitem [{\citenamefont {Breuer}\ \emph {et~al.}(2016)\citenamefont {Breuer},
  \citenamefont {Laine}, \citenamefont {Piilo},\ and\ \citenamefont
  {Vacchini}}]{Breuer2016}%
  \BibitemOpen
  \bibfield  {author} {\bibinfo {author} {\bibfnamefont {Heinz-Peter}\
  \bibnamefont {Breuer}}, \bibinfo {author} {\bibfnamefont {Elsi-Mari}\
  \bibnamefont {Laine}}, \bibinfo {author} {\bibfnamefont {Jyrki}\ \bibnamefont
  {Piilo}}, \ and\ \bibinfo {author} {\bibfnamefont {Bassano}\ \bibnamefont
  {Vacchini}},\ }\bibfield  {title} {\enquote {\bibinfo {title} {Colloquium:
  Non-markovian dynamics in open quantum systems},}\ }\href {\doibase
  10.1103/RevModPhys.88.021002} {\bibfield  {journal} {\bibinfo  {journal}
  {Rev. Mod. Phys.}\ }\textbf {\bibinfo {volume} {88}},\ \bibinfo {pages}
  {021002} (\bibinfo {year} {2016})}\BibitemShut {NoStop}%
\bibitem [{\citenamefont {King}\ and\ \citenamefont {Ruskai}(2001)}]{King2001}%
  \BibitemOpen
  \bibfield  {author} {\bibinfo {author} {\bibfnamefont {Christopher}\
  \bibnamefont {King}}\ and\ \bibinfo {author} {\bibfnamefont {Mary~Beth}\
  \bibnamefont {Ruskai}},\ }\bibfield  {title} {\enquote {\bibinfo {title}
  {Minimal entropy of states emerging from noisy quantum channels},}\ }\href
  {\doibase 10.1109/18.904522} {\bibfield  {journal} {\bibinfo  {journal} {IEEE
  Trans. Inf. Theory}\ }\textbf {\bibinfo {volume} {47}},\ \bibinfo {pages}
  {192--209} (\bibinfo {year} {2001})}\BibitemShut {NoStop}%
\bibitem [{\citenamefont {{Andersson}}\ \emph {et~al.}(2007)\citenamefont
  {{Andersson}}, \citenamefont {{Cresser}},\ and\ \citenamefont
  {{Hall}}}]{Andersson2007}%
  \BibitemOpen
  \bibfield  {author} {\bibinfo {author} {\bibfnamefont {E.}~\bibnamefont
  {{Andersson}}}, \bibinfo {author} {\bibfnamefont {J.~D.}\ \bibnamefont
  {{Cresser}}}, \ and\ \bibinfo {author} {\bibfnamefont {M.~J.~W.}\
  \bibnamefont {{Hall}}},\ }\bibfield  {title} {\enquote {\bibinfo {title}
  {{Finding the Kraus decomposition from a master equation and vice versa}},}\
  }\href {\doibase 10.1080/09500340701352581} {\bibfield  {journal} {\bibinfo
  {journal} {J. Mod. Opt.}\ }\textbf {\bibinfo {volume} {54}},\ \bibinfo
  {pages} {1695--1716} (\bibinfo {year} {2007})}\BibitemShut {NoStop}%
\bibitem [{\citenamefont {Asorey}\ \emph {et~al.}(2009)\citenamefont {Asorey},
  \citenamefont {Kossakowski}, \citenamefont {Marmo},\ and\ \citenamefont
  {Sudarshan}}]{Asorey2009}%
  \BibitemOpen
  \bibfield  {author} {\bibinfo {author} {\bibfnamefont {M}~\bibnamefont
  {Asorey}}, \bibinfo {author} {\bibfnamefont {A}~\bibnamefont {Kossakowski}},
  \bibinfo {author} {\bibfnamefont {G}~\bibnamefont {Marmo}}, \ and\ \bibinfo
  {author} {\bibfnamefont {E~C~G}\ \bibnamefont {Sudarshan}},\ }\bibfield
  {title} {\enquote {\bibinfo {title} {Dynamical maps and density matrices},}\
  }\href {http://stacks.iop.org/1742-6596/196/i=1/a=012023} {\bibfield
  {journal} {\bibinfo  {journal} {J. Phys. Conf. Ser.}\ }\textbf {\bibinfo
  {volume} {196}},\ \bibinfo {pages} {012023} (\bibinfo {year}
  {2009})}\BibitemShut {NoStop}%
\bibitem [{\citenamefont {Chru\ifmmode \acute{s}\else
  \'{s}\fi{}ci\ifmmode~\acute{n}\else \'{n}\fi{}ski}\ and\ \citenamefont
  {Kossakowski}(2010)}]{Chruscinski2010}%
  \BibitemOpen
  \bibfield  {author} {\bibinfo {author} {\bibfnamefont {Dariusz}\ \bibnamefont
  {Chru\ifmmode \acute{s}\else \'{s}\fi{}ci\ifmmode~\acute{n}\else
  \'{n}\fi{}ski}}\ and\ \bibinfo {author} {\bibfnamefont {Andrzej}\
  \bibnamefont {Kossakowski}},\ }\bibfield  {title} {\enquote {\bibinfo {title}
  {Non-markovian quantum dynamics: Local versus nonlocal},}\ }\href {\doibase
  10.1103/PhysRevLett.104.070406} {\bibfield  {journal} {\bibinfo  {journal}
  {Phys. Rev. Lett.}\ }\textbf {\bibinfo {volume} {104}},\ \bibinfo {pages}
  {070406} (\bibinfo {year} {2010})}\BibitemShut {NoStop}%
\bibitem [{\citenamefont {Chru\ifmmode \acute{s}\else
  \'{s}\fi{}ci\ifmmode~\acute{n}\else \'{n}\fi{}ski}\ \emph
  {et~al.}(2010)\citenamefont {Chru\ifmmode \acute{s}\else
  \'{s}\fi{}ci\ifmmode~\acute{n}\else \'{n}\fi{}ski}, \citenamefont
  {Kossakowski}, \citenamefont {Aniello}, \citenamefont {Marmo},\ and\
  \citenamefont {Ventriglia}}]{Chruscinski2010_2}%
  \BibitemOpen
  \bibfield  {author} {\bibinfo {author} {\bibfnamefont {Dariusz}\ \bibnamefont
  {Chru\ifmmode \acute{s}\else \'{s}\fi{}ci\ifmmode~\acute{n}\else
  \'{n}\fi{}ski}}, \bibinfo {author} {\bibfnamefont {Andrzej}\ \bibnamefont
  {Kossakowski}}, \bibinfo {author} {\bibfnamefont {Paolo}\ \bibnamefont
  {Aniello}}, \bibinfo {author} {\bibfnamefont {Giuseppe}\ \bibnamefont
  {Marmo}}, \ and\ \bibinfo {author} {\bibfnamefont {Franco}\ \bibnamefont
  {Ventriglia}},\ }\bibfield  {title} {\enquote {\bibinfo {title} {A class of
  commutative dynamics of open quantum systems},}\ }\href {\doibase
  10.1142/S1230161210000163} {\bibfield  {journal} {\bibinfo  {journal} {Open
  Syst. Inf. Dyn.}\ }\textbf {\bibinfo {volume} {17}},\ \bibinfo {pages}
  {255--277} (\bibinfo {year} {2010})}\BibitemShut {NoStop}%
\bibitem [{\citenamefont {Smirne}\ and\ \citenamefont
  {Vacchini}(2010)}]{Smirne2010}%
  \BibitemOpen
  \bibfield  {author} {\bibinfo {author} {\bibfnamefont {Andrea}\ \bibnamefont
  {Smirne}}\ and\ \bibinfo {author} {\bibfnamefont {Bassano}\ \bibnamefont
  {Vacchini}},\ }\bibfield  {title} {\enquote {\bibinfo {title}
  {Nakajima-zwanzig versus time-convolutionless master equation for the
  non-markovian dynamics of a two-level system},}\ }\href {\doibase
  10.1103/PhysRevA.82.022110} {\bibfield  {journal} {\bibinfo  {journal} {Phys.
  Rev. A}\ }\textbf {\bibinfo {volume} {82}},\ \bibinfo {pages} {022110}
  (\bibinfo {year} {2010})}\BibitemShut {NoStop}%
\bibitem [{Note6()}]{Note6}%
  \BibitemOpen
  \bibinfo {note} {The global rotation about $z$ will be given by the encoding
  rotation by $\omega *t$ plus possibly a further contribution, i.e., $\phi =
  \omega t + {\vartheta }$.}\BibitemShut {Stop}%
\bibitem [{\citenamefont {Clos}\ and\ \citenamefont {Breuer}(2012)}]{Clos2012}%
  \BibitemOpen
  \bibfield  {author} {\bibinfo {author} {\bibfnamefont {Govinda}\ \bibnamefont
  {Clos}}\ and\ \bibinfo {author} {\bibfnamefont {Heinz-Peter}\ \bibnamefont
  {Breuer}},\ }\bibfield  {title} {\enquote {\bibinfo {title} {Quantification
  of memory effects in the spin-boson model},}\ }\href {\doibase
  10.1103/PhysRevA.86.012115} {\bibfield  {journal} {\bibinfo  {journal} {Phys.
  Rev. A}\ }\textbf {\bibinfo {volume} {86}},\ \bibinfo {pages} {012115}
  (\bibinfo {year} {2012})}\BibitemShut {NoStop}%
\bibitem [{\citenamefont {Breuer}\ \emph {et~al.}(2001)\citenamefont {Breuer},
  \citenamefont {Kappler},\ and\ \citenamefont {Petruccione}}]{Breuer2001}%
  \BibitemOpen
  \bibfield  {author} {\bibinfo {author} {\bibfnamefont {Heinz-Peter}\
  \bibnamefont {Breuer}}, \bibinfo {author} {\bibfnamefont {Bernd}\
  \bibnamefont {Kappler}}, \ and\ \bibinfo {author} {\bibfnamefont {Francesco}\
  \bibnamefont {Petruccione}},\ }\bibfield  {title} {\enquote {\bibinfo {title}
  {The time-convolutionless projection operator technique in the quantum theory
  of dissipation and decoherence},}\ }\href {\doibase
  http://dx.doi.org/10.1006/aphy.2001.6152} {\bibfield  {journal} {\bibinfo
  {journal} {Ann. Phys.}\ }\textbf {\bibinfo {volume} {291}},\ \bibinfo {pages}
  {36 -- 70} (\bibinfo {year} {2001})}\BibitemShut {NoStop}%
\bibitem [{\citenamefont {{Gasbarri}}\ and\ \citenamefont
  {{Ferialdi}}(2017)}]{Gasbarri2017}%
  \BibitemOpen
  \bibfield  {author} {\bibinfo {author} {\bibfnamefont {G.}~\bibnamefont
  {{Gasbarri}}}\ and\ \bibinfo {author} {\bibfnamefont {L.}~\bibnamefont
  {{Ferialdi}}},\ }\bibfield  {title} {\enquote {\bibinfo {title} {{Recursive
  approach for non-Markovian maps and their time convolutionless master
  equations}},}\ }\href@noop {} {\bibfield  {journal} {\bibinfo  {journal}
  {arXiv e-print}\ } (\bibinfo {year} {2017})},\ \Eprint
  {http://arxiv.org/abs/1707.06540} {arXiv:1707.06540 [quant-ph]} \BibitemShut
  {NoStop}%
\bibitem [{\citenamefont {Maniscalco}\ \emph {et~al.}(2004)\citenamefont
  {Maniscalco}, \citenamefont {Intravaia}, \citenamefont {Piilo},\ and\
  \citenamefont {Messina}}]{Maniscalco2004}%
  \BibitemOpen
  \bibfield  {author} {\bibinfo {author} {\bibfnamefont {S}~\bibnamefont
  {Maniscalco}}, \bibinfo {author} {\bibfnamefont {F}~\bibnamefont
  {Intravaia}}, \bibinfo {author} {\bibfnamefont {J}~\bibnamefont {Piilo}}, \
  and\ \bibinfo {author} {\bibfnamefont {A}~\bibnamefont {Messina}},\
  }\bibfield  {title} {\enquote {\bibinfo {title} {Misbeliefs and
  misunderstandings about the non-markovian dynamics of a damped harmonic
  oscillator},}\ }\href {http://stacks.iop.org/1464-4266/6/i=3/a=016}
  {\bibfield  {journal} {\bibinfo  {journal} {J. Opt. B: Quantum Semiclassical
  Opt.}\ }\textbf {\bibinfo {volume} {6}},\ \bibinfo {pages} {S98} (\bibinfo
  {year} {2004})}\BibitemShut {NoStop}%
\bibitem [{\citenamefont {Fleming}\ \emph {et~al.}(2010)\citenamefont
  {Fleming}, \citenamefont {Cummings}, \citenamefont {Anastopoulos},\ and\
  \citenamefont {Hu}}]{Fleming2010}%
  \BibitemOpen
  \bibfield  {author} {\bibinfo {author} {\bibfnamefont {C.}~\bibnamefont
  {Fleming}}, \bibinfo {author} {\bibfnamefont {N.~I.}\ \bibnamefont
  {Cummings}}, \bibinfo {author} {\bibfnamefont {C.}~\bibnamefont
  {Anastopoulos}}, \ and\ \bibinfo {author} {\bibfnamefont {B.~L.}\
  \bibnamefont {Hu}},\ }\bibfield  {title} {\enquote {\bibinfo {title} {The
  rotating-wave approximation: consistency and applicability from an open
  quantum system analysis},}\ }\href@noop {} {\bibfield  {journal} {\bibinfo
  {journal} {Journal of Physics A: Mathematical and Theoretical}\ }\textbf
  {\bibinfo {volume} {43}},\ \bibinfo {pages} {405304} (\bibinfo {year}
  {2010})}\BibitemShut {NoStop}%
\bibitem [{\citenamefont {Lankinen}\ \emph {et~al.}(2016)\citenamefont
  {Lankinen}, \citenamefont {Lyyra}, \citenamefont {Sokolov}, \citenamefont
  {Teittinen}, \citenamefont {Ziaei},\ and\ \citenamefont
  {Maniscalco}}]{Lankinen2016}%
  \BibitemOpen
  \bibfield  {author} {\bibinfo {author} {\bibfnamefont {Juho}\ \bibnamefont
  {Lankinen}}, \bibinfo {author} {\bibfnamefont {Henri}\ \bibnamefont {Lyyra}},
  \bibinfo {author} {\bibfnamefont {Boris}\ \bibnamefont {Sokolov}}, \bibinfo
  {author} {\bibfnamefont {Jose}\ \bibnamefont {Teittinen}}, \bibinfo {author}
  {\bibfnamefont {Babak}\ \bibnamefont {Ziaei}}, \ and\ \bibinfo {author}
  {\bibfnamefont {Sabrina}\ \bibnamefont {Maniscalco}},\ }\bibfield  {title}
  {\enquote {\bibinfo {title} {Complete positivity, finite-temperature effects,
  and additivity of noise for time-local qubit dynamics},}\ }\href {\doibase
  10.1103/PhysRevA.93.052103} {\bibfield  {journal} {\bibinfo  {journal} {Phys.
  Rev. A}\ }\textbf {\bibinfo {volume} {93}},\ \bibinfo {pages} {052103}
  (\bibinfo {year} {2016})}\BibitemShut {NoStop}%
\bibitem [{\citenamefont {Oviedo-Casado}\ \emph {et~al.}(2016)\citenamefont
  {Oviedo-Casado}, \citenamefont {Prior}, \citenamefont {Chin}, \citenamefont
  {Rosenbach}, \citenamefont {Huelga},\ and\ \citenamefont
  {Plenio}}]{Oviedo2016}%
  \BibitemOpen
  \bibfield  {author} {\bibinfo {author} {\bibfnamefont {S.}~\bibnamefont
  {Oviedo-Casado}}, \bibinfo {author} {\bibfnamefont {J.}~\bibnamefont
  {Prior}}, \bibinfo {author} {\bibfnamefont {A.~W.}\ \bibnamefont {Chin}},
  \bibinfo {author} {\bibfnamefont {R.}~\bibnamefont {Rosenbach}}, \bibinfo
  {author} {\bibfnamefont {S.~F.}\ \bibnamefont {Huelga}}, \ and\ \bibinfo
  {author} {\bibfnamefont {M.~B.}\ \bibnamefont {Plenio}},\ }\bibfield  {title}
  {\enquote {\bibinfo {title} {Phase-dependent exciton transport and energy
  harvesting from thermal environments},}\ }\href {\doibase
  10.1103/PhysRevA.93.020102} {\bibfield  {journal} {\bibinfo  {journal} {Phys.
  Rev. A}\ }\textbf {\bibinfo {volume} {93}},\ \bibinfo {pages} {020102}
  (\bibinfo {year} {2016})}\BibitemShut {NoStop}%
\bibitem [{\citenamefont {Jeske}\ \emph {et~al.}(2015)\citenamefont {Jeske},
  \citenamefont {Ing}, \citenamefont {Plenio}, \citenamefont {Huelga},\ and\
  \citenamefont {Cole}}]{Jeske2015}%
  \BibitemOpen
  \bibfield  {author} {\bibinfo {author} {\bibfnamefont {Jan}\ \bibnamefont
  {Jeske}}, \bibinfo {author} {\bibfnamefont {David~J.}\ \bibnamefont {Ing}},
  \bibinfo {author} {\bibfnamefont {Martin~B.}\ \bibnamefont {Plenio}},
  \bibinfo {author} {\bibfnamefont {Susana~F.}\ \bibnamefont {Huelga}}, \ and\
  \bibinfo {author} {\bibfnamefont {Jared~H.}\ \bibnamefont {Cole}},\
  }\bibfield  {title} {\enquote {\bibinfo {title} {Bloch-redfield equations for
  modeling light-harvesting complexes},}\ }\href {\doibase 10.1063/1.4907370}
  {\bibfield  {journal} {\bibinfo  {journal} {The Journal of Chemical Physics}\
  }\textbf {\bibinfo {volume} {142}},\ \bibinfo {pages} {064104} (\bibinfo
  {year} {2015})},\ \Eprint
  {http://arxiv.org/abs/http://dx.doi.org/10.1063/1.4907370}
  {http://dx.doi.org/10.1063/1.4907370} \BibitemShut {NoStop}%
\bibitem [{\citenamefont {Sun}\ \emph {et~al.}(2015)\citenamefont {Sun},
  \citenamefont {Liu}, \citenamefont {Ma},\ and\ \citenamefont
  {Wang}}]{Sun2015}%
  \BibitemOpen
  \bibfield  {author} {\bibinfo {author} {\bibfnamefont {Zhe}\ \bibnamefont
  {Sun}}, \bibinfo {author} {\bibfnamefont {Jing}\ \bibnamefont {Liu}},
  \bibinfo {author} {\bibfnamefont {Jian}\ \bibnamefont {Ma}}, \ and\ \bibinfo
  {author} {\bibfnamefont {Xiaoguang}\ \bibnamefont {Wang}},\ }\bibfield
  {title} {\enquote {\bibinfo {title} {Quantum speed limits in open systems:
  Non-markovian dynamics without rotating-wave approximation},}\ }\href
  {http://dx.doi.org/10.1038/srep08444} {\bibfield  {journal} {\bibinfo
  {journal} {Sc. Rep}\ }\textbf {\bibinfo {volume} {5}},\ \bibinfo {pages}
  {8444} (\bibinfo {year} {2015})}\BibitemShut {NoStop}%
\bibitem [{\citenamefont {Zhang}\ \emph {et~al.}(2015)\citenamefont {Zhang},
  \citenamefont {Han}, \citenamefont {Xia}, \citenamefont {Cao},\ and\
  \citenamefont {Fan}}]{Zhang2015}%
  \BibitemOpen
  \bibfield  {author} {\bibinfo {author} {\bibfnamefont {Ying-Jie}\
  \bibnamefont {Zhang}}, \bibinfo {author} {\bibfnamefont {Wei}\ \bibnamefont
  {Han}}, \bibinfo {author} {\bibfnamefont {Yun-Jie}\ \bibnamefont {Xia}},
  \bibinfo {author} {\bibfnamefont {Jun-Peng}\ \bibnamefont {Cao}}, \ and\
  \bibinfo {author} {\bibfnamefont {Heng}\ \bibnamefont {Fan}},\ }\bibfield
  {title} {\enquote {\bibinfo {title} {Classical-driving-assisted quantum
  speed-up},}\ }\href {\doibase 10.1103/PhysRevA.91.032112} {\bibfield
  {journal} {\bibinfo  {journal} {Phys. Rev. A}\ }\textbf {\bibinfo {volume}
  {91}},\ \bibinfo {pages} {032112} (\bibinfo {year} {2015})}\BibitemShut
  {NoStop}%
\bibitem [{\citenamefont {Lostaglio}\ \emph {et~al.}(2017)\citenamefont
  {Lostaglio}, \citenamefont {Korzekwa},\ and\ \citenamefont
  {Milne}}]{Lostaglio2017}%
  \BibitemOpen
  \bibfield  {author} {\bibinfo {author} {\bibfnamefont {M.}~\bibnamefont
  {Lostaglio}}, \bibinfo {author} {\bibfnamefont {K.}~\bibnamefont {Korzekwa}},
  \ and\ \bibinfo {author} {\bibfnamefont {A.}~\bibnamefont {Milne}},\
  }\bibfield  {title} {\enquote {\bibinfo {title} {Markovian evolution of
  quantum coherence under symmetric dynamics},}\ }\href@noop {} {\bibfield
  {journal} {\bibinfo  {journal} {Phys. Rev. A}\ }\textbf {\bibinfo {volume}
  {96}},\ \bibinfo {pages} {032109} (\bibinfo {year} {2017})}\BibitemShut
  {NoStop}%
\bibitem [{\citenamefont {{\.{Z}yczkowski}}\ and\ \citenamefont
  {{Bengtsson}}(2004)}]{Zyczkowski2004}%
  \BibitemOpen
  \bibfield  {author} {\bibinfo {author} {\bibfnamefont {K.}~\bibnamefont
  {{\.{Z}yczkowski}}}\ and\ \bibinfo {author} {\bibfnamefont {I.}~\bibnamefont
  {{Bengtsson}}},\ }\bibfield  {title} {\enquote {\bibinfo {title} {{On duality
  between quantum maps and quantum states}},}\ }\href@noop {} {\bibfield
  {journal} {\bibinfo  {journal} {Open Syst. Inf. Dyn.}\ }\textbf {\bibinfo
  {volume} {11}},\ \bibinfo {pages} {3--42} (\bibinfo {year}
  {2004})}\BibitemShut {NoStop}%
\bibitem [{\citenamefont {Choi.}(1975)}]{Choi1975}%
  \BibitemOpen
  \bibfield  {author} {\bibinfo {author} {\bibfnamefont {M.-D.}\ \bibnamefont
  {Choi.}},\ }\bibfield  {title} {\enquote {\bibinfo {title} {Completely
  positive linear maps on complex matrices},}\ }\href@noop {} {\bibfield
  {journal} {\bibinfo  {journal} {Linear Algebra and its Applications}\
  }\textbf {\bibinfo {volume} {10}},\ \bibinfo {pages} {285} (\bibinfo {year}
  {1975})}\BibitemShut {NoStop}%
\bibitem [{\citenamefont {D\"ur}\ \emph {et~al.}(2014)\citenamefont {D\"ur},
  \citenamefont {Skotiniotis}, \citenamefont {Fr\"owis},\ and\ \citenamefont
  {Kraus}}]{Dur2014}%
  \BibitemOpen
  \bibfield  {author} {\bibinfo {author} {\bibfnamefont {W.}~\bibnamefont
  {D\"ur}}, \bibinfo {author} {\bibfnamefont {M.}~\bibnamefont {Skotiniotis}},
  \bibinfo {author} {\bibfnamefont {F.}~\bibnamefont {Fr\"owis}}, \ and\
  \bibinfo {author} {\bibfnamefont {B.}~\bibnamefont {Kraus}},\ }\bibfield
  {title} {\enquote {\bibinfo {title} {Improved quantum metrology using quantum
  error correction},}\ }\href@noop {} {\bibfield  {journal} {\bibinfo
  {journal} {Phys. Rev. Lett.}\ }\textbf {\bibinfo {volume} {112}},\ \bibinfo
  {pages} {080801} (\bibinfo {year} {2014})}\BibitemShut {NoStop}%
\bibitem [{\citenamefont {Brask}\ \emph {et~al.}(2015)\citenamefont {Brask},
  \citenamefont {Chaves},\ and\ \citenamefont {Ko\l{}ody\ifmmode~\acute{n}\else
  \'{n}\fi{}ski}}]{Brask2015}%
  \BibitemOpen
  \bibfield  {author} {\bibinfo {author} {\bibfnamefont {J.~B.}\ \bibnamefont
  {Brask}}, \bibinfo {author} {\bibfnamefont {R.}~\bibnamefont {Chaves}}, \
  and\ \bibinfo {author} {\bibfnamefont {J.}~\bibnamefont
  {Ko\l{}ody\ifmmode~\acute{n}\else \'{n}\fi{}ski}},\ }\bibfield  {title}
  {\enquote {\bibinfo {title} {Improved quantum magnetometry beyond the
  standard quantum limit},}\ }\href {\doibase 10.1103/PhysRevX.5.031010}
  {\bibfield  {journal} {\bibinfo  {journal} {Phys. Rev. X}\ }\textbf {\bibinfo
  {volume} {5}},\ \bibinfo {pages} {031010} (\bibinfo {year}
  {2015})}\BibitemShut {NoStop}%
\bibitem [{\citenamefont {{Zhou}}\ \emph {et~al.}(2017)\citenamefont {{Zhou}},
  \citenamefont {{Zhang}}, \citenamefont {{Preskill}},\ and\ \citenamefont
  {{Jiang}}}]{Zhou2017}%
  \BibitemOpen
  \bibfield  {author} {\bibinfo {author} {\bibfnamefont {S.}~\bibnamefont
  {{Zhou}}}, \bibinfo {author} {\bibfnamefont {M.}~\bibnamefont {{Zhang}}},
  \bibinfo {author} {\bibfnamefont {J.}~\bibnamefont {{Preskill}}}, \ and\
  \bibinfo {author} {\bibfnamefont {L.}~\bibnamefont {{Jiang}}},\ }\bibfield
  {title} {\enquote {\bibinfo {title} {{Achieving the Heisenberg limit in
  quantum metrology using quantum error correction}},}\ }\href@noop {}
  {\bibfield  {journal} {\bibinfo  {journal} {arXiv e-print}\ } (\bibinfo
  {year} {2017})},\ \Eprint {http://arxiv.org/abs/1706.02445} {arXiv:1706.02445
  [quant-ph]} \BibitemShut {NoStop}%
\bibitem [{\citenamefont {Unden}\ \emph {et~al.}(2016)\citenamefont {Unden},
  \citenamefont {Balasubramanian}, \citenamefont {Louzon}, \citenamefont
  {Vinkler}, \citenamefont {Plenio}, \citenamefont {Markham}, \citenamefont
  {Twitchen}, \citenamefont {Stacey}, \citenamefont {Lovchinsky}, \citenamefont
  {Sushkov}, \citenamefont {Lukin}, \citenamefont {Retzker}, \citenamefont
  {Naydenov}, \citenamefont {McGuinness},\ and\ \citenamefont
  {Jelezko}}]{Unden2016}%
  \BibitemOpen
  \bibfield  {author} {\bibinfo {author} {\bibfnamefont {Thomas}\ \bibnamefont
  {Unden}}, \bibinfo {author} {\bibfnamefont {Priya}\ \bibnamefont
  {Balasubramanian}}, \bibinfo {author} {\bibfnamefont {Daniel}\ \bibnamefont
  {Louzon}}, \bibinfo {author} {\bibfnamefont {Yuval}\ \bibnamefont {Vinkler}},
  \bibinfo {author} {\bibfnamefont {Martin~B.}\ \bibnamefont {Plenio}},
  \bibinfo {author} {\bibfnamefont {Matthew}\ \bibnamefont {Markham}}, \bibinfo
  {author} {\bibfnamefont {Daniel}\ \bibnamefont {Twitchen}}, \bibinfo {author}
  {\bibfnamefont {Alastair}\ \bibnamefont {Stacey}}, \bibinfo {author}
  {\bibfnamefont {Igor}\ \bibnamefont {Lovchinsky}}, \bibinfo {author}
  {\bibfnamefont {Alexander~O.}\ \bibnamefont {Sushkov}}, \bibinfo {author}
  {\bibfnamefont {Mikhail~D.}\ \bibnamefont {Lukin}}, \bibinfo {author}
  {\bibfnamefont {Alex}\ \bibnamefont {Retzker}}, \bibinfo {author}
  {\bibfnamefont {Boris}\ \bibnamefont {Naydenov}}, \bibinfo {author}
  {\bibfnamefont {Liam~P.}\ \bibnamefont {McGuinness}}, \ and\ \bibinfo
  {author} {\bibfnamefont {Fedor}\ \bibnamefont {Jelezko}},\ }\bibfield
  {title} {\enquote {\bibinfo {title} {Quantum metrology enhanced by repetitive
  quantum error correction},}\ }\href {\doibase 10.1103/PhysRevLett.116.230502}
  {\bibfield  {journal} {\bibinfo  {journal} {Phys. Rev. Lett.}\ }\textbf
  {\bibinfo {volume} {116}},\ \bibinfo {pages} {230502} (\bibinfo {year}
  {2016})}\BibitemShut {NoStop}%
\bibitem [{\citenamefont {Rivas}\ \emph {et~al.}(2010)\citenamefont {Rivas},
  \citenamefont {Huelga},\ and\ \citenamefont {Plenio}}]{Rivas2010}%
  \BibitemOpen
  \bibfield  {author} {\bibinfo {author} {\bibfnamefont {\'Angel}\ \bibnamefont
  {Rivas}}, \bibinfo {author} {\bibfnamefont {Susana~F.}\ \bibnamefont
  {Huelga}}, \ and\ \bibinfo {author} {\bibfnamefont {Martin~B.}\ \bibnamefont
  {Plenio}},\ }\bibfield  {title} {\enquote {\bibinfo {title} {Entanglement and
  non-markovianity of quantum evolutions},}\ }\href {\doibase
  10.1103/PhysRevLett.105.050403} {\bibfield  {journal} {\bibinfo  {journal}
  {Phys. Rev. Lett.}\ }\textbf {\bibinfo {volume} {105}},\ \bibinfo {pages}
  {050403} (\bibinfo {year} {2010})}\BibitemShut {NoStop}%
\bibitem [{Note7()}]{Note7}%
  \BibitemOpen
  \bibinfo {note} {Once again, this could be shown by exploiting a
  block-diagonal structure of the generator $\protect \mathcal {L}(t)$ and thus
  of the resulting dynamical maps; compare with \ref {app:htl}.}\BibitemShut
  {Stop}%
\bibitem [{\citenamefont {Zhong}\ \emph {et~al.}(2013)\citenamefont {Zhong},
  \citenamefont {Sun}, \citenamefont {Ma}, \citenamefont {Wang},\ and\
  \citenamefont {Nori}}]{Zhong2013}%
  \BibitemOpen
  \bibfield  {author} {\bibinfo {author} {\bibfnamefont {Wei}\ \bibnamefont
  {Zhong}}, \bibinfo {author} {\bibfnamefont {Zhe}\ \bibnamefont {Sun}},
  \bibinfo {author} {\bibfnamefont {Jian}\ \bibnamefont {Ma}}, \bibinfo
  {author} {\bibfnamefont {Xiaoguang}\ \bibnamefont {Wang}}, \ and\ \bibinfo
  {author} {\bibfnamefont {Franco}\ \bibnamefont {Nori}},\ }\bibfield  {title}
  {\enquote {\bibinfo {title} {Fisher information under decoherence in bloch
  representation},}\ }\href {\doibase 10.1103/PhysRevA.87.022337} {\bibfield
  {journal} {\bibinfo  {journal} {Phys. Rev. A}\ }\textbf {\bibinfo {volume}
  {87}},\ \bibinfo {pages} {022337} (\bibinfo {year} {2013})}\BibitemShut
  {NoStop}%
\bibitem [{\citenamefont {Jarzyna}\ and\ \citenamefont
  {Zwierz}(2017)}]{Jarzyna2017}%
  \BibitemOpen
  \bibfield  {author} {\bibinfo {author} {\bibfnamefont {Marcin}\ \bibnamefont
  {Jarzyna}}\ and\ \bibinfo {author} {\bibfnamefont {Marcin}\ \bibnamefont
  {Zwierz}},\ }\bibfield  {title} {\enquote {\bibinfo {title} {Parameter
  estimation in the presence of the most general gaussian dissipative
  reservoir},}\ }\href {\doibase 10.1103/PhysRevA.95.012109} {\bibfield
  {journal} {\bibinfo  {journal} {Phys. Rev. A}\ }\textbf {\bibinfo {volume}
  {95}},\ \bibinfo {pages} {012109} (\bibinfo {year} {2017})}\BibitemShut
  {NoStop}%
\bibitem [{\citenamefont {Latune}\ \emph {et~al.}(2016)\citenamefont {Latune},
  \citenamefont {Sinayskiy},\ and\ \citenamefont {Petruccione}}]{Latune2016}%
  \BibitemOpen
  \bibfield  {author} {\bibinfo {author} {\bibfnamefont {C.~L.}\ \bibnamefont
  {Latune}}, \bibinfo {author} {\bibfnamefont {I.}~\bibnamefont {Sinayskiy}}, \
  and\ \bibinfo {author} {\bibfnamefont {F.}~\bibnamefont {Petruccione}},\
  }\bibfield  {title} {\enquote {\bibinfo {title} {Quantum force estimation in
  arbitrary non-markovian gaussian baths},}\ }\href {\doibase
  10.1103/PhysRevA.94.052115} {\bibfield  {journal} {\bibinfo  {journal} {Phys.
  Rev. A}\ }\textbf {\bibinfo {volume} {94}},\ \bibinfo {pages} {052115}
  (\bibinfo {year} {2016})}\BibitemShut {NoStop}%
\bibitem [{\citenamefont {Alipour}\ and\ \citenamefont
  {Rezakhani}(2015)}]{Alipour2015}%
  \BibitemOpen
  \bibfield  {author} {\bibinfo {author} {\bibfnamefont {S.}~\bibnamefont
  {Alipour}}\ and\ \bibinfo {author} {\bibfnamefont {A.~T.}\ \bibnamefont
  {Rezakhani}},\ }\bibfield  {title} {\enquote {\bibinfo {title} {Extended
  convexity of quantum fisher information in quantum metrology},}\ }\href@noop
  {} {\bibfield  {journal} {\bibinfo  {journal} {Phys. Rev. A}\ }\textbf
  {\bibinfo {volume} {91}},\ \bibinfo {pages} {042104} (\bibinfo {year}
  {2015})}\BibitemShut {NoStop}%
\end{thebibliography}
\end{document}